\documentclass[12pt]{article}
\pdfoutput=1
\usepackage{authblk}
\newcommand{\SeriesTitle}{Persistent Misperception with Correctly Specified Models}
\newcommand{\PartIITitle}{\SeriesTitle, Part~II: Rational Under-learning}

\newcommand{\PartIThmLocal}{Theorem~7 of Part~I}%
\newcommand{\PartILemComplete}{Lemma~3 of Part~I}%
\newcommand{\PartIFigBeliefsAd}{Figure~1a of Part~I}%
\newcommand{\PartIFigBeliefsPricing}{Figure~1b of Part~I}%

\newcommand{\PartILemPosterior}{Lemma~S1.1 of Part~I}%
\newcommand{\PartILemDisintegration}{Lemma~S1.2 of Part~I}%
\newcommand{\PartIRemPredictable}{Remark~S1.3 of Part~I}%
\newcommand{\PartIDefTracking}{Definition~S2.1 of Part~I}%
\newcommand{\PartILemFiniteHistory}{Lemma~S2.2 of Part~I}%
\newcommand{\PartILemTrapping}{Lemma~S2.3 of Part~I}%
\newcommand{\PartIRemTrapping}{Remark~S2.4 of Part~I}%
\newcommand{\PartILemAnchored}{Lemma~S3.1 of Part~I}%
\newcommand{\PartIDefPackage}{Definition~S3.3 of Part~I}%
\newcommand{\PartIThmExclusion}{Theorem~S3.4 of Part~I}%
\newcommand{\PartIRemSeeded}{Remark~S3.5 of Part~I}%
\newcommand{\PartILemEventualConstant}{Lemma~C.4 of Part~I}%

\newcommand{\ArticleTitle}{\PartIITitle}
\newcommand{\AuthorOne}{Tetsuya Kaji}
\newcommand{\AuthorTwo}{Michihiro Kandori}
\newcommand{\AuthorThree}{Yuichi Yamamoto}
\newcommand{\AffiliationOne}{University of Chicago Booth School of Business}
\newcommand{\AffiliationTwo}{Department of Economics, University of Tokyo}
\newcommand{\AffiliationThree}{Department of Economics, University of Tokyo}

\author[1]{\AuthorOne}
\author[2]{\AuthorTwo}
\author[3]{\AuthorThree}
\affil[1]{\AffiliationOne}
\affil[2]{\AffiliationTwo}
\affil[3]{\AffiliationThree}
\title{\ArticleTitle%
\thanks{This work is supported by the Richard N.\ Rosett Faculty Fellowship at the University of Chicago Booth School of Business. Generative AI was used to assist with proofs and to polish the writing.}}
\date{\today}
\usepackage{amsmath,amssymb,amsthm,mathtools,mathrsfs,bm,bbm}
\usepackage{mleftright}
\mleftright
\usepackage{enumitem,graphicx,booktabs,longtable,array}
\usepackage[onehalfspacing,nodisplayskipstretch]{setspace}
\usepackage{microtype}
\usepackage[authoryear]{natbib}
\usepackage{xcolor}
\usepackage[colorlinks,citecolor=blue,urlcolor=blue,linkcolor=blue,hypertexnames=false]{hyperref}
\usepackage{cleveref}
\makeatletter
\newcommand*{\condtarget}[2]{%
  \begingroup\let\cond@savedHref\@currentHref
  \phantomsection\def\@currentlabel{#2}\label{#1}%
  \global\let\@currentHref\cond@savedHref\endgroup#2}
\newcommand*{\parstartref}[1]{\leavevmode\MT@get@prot{(}{left}\ref{#1}}
\makeatother
\usepackage{refcount}
\usepackage[justification=raggedright,singlelinecheck=false]{caption}
\usepackage{subcaption}
\usepackage[margin=1.25in]{geometry}
\allowdisplaybreaks[3]
\theoremstyle{plain}
\newtheorem{theorem}{Theorem}
\newtheorem{proposition}[theorem]{Proposition}
\newtheorem{lemma}[theorem]{Lemma}
\newtheorem{corollary}[theorem]{Corollary}

\newtheorem{auxlemma}[theorem]{Lemma}
\theoremstyle{definition}
\newtheorem{definition}[theorem]{Definition}
\newtheorem{assumption}[theorem]{Assumption}
\theoremstyle{remark}
\newtheorem{remark}[theorem]{Remark}
\usepackage{thmtools}
\newcommand{\exasymbol}{$\square$}
\declaretheoremstyle[
  spaceabove=6pt, spacebelow=6pt,
  headfont=\normalfont\bfseries,
  notefont=\mdseries, notebraces={(}{)},
  bodyfont=\normalfont, postheadspace=1em,
  qed=\exasymbol
]{exampleStyle}
\declaretheorem[style=exampleStyle,name=Example]{example}
\renewcommand\thmcontinues[1]{continued}
\crefname{example}{Example}{Examples}
\newcommand{\R}{\mathbb R}
\newcommand{\1}{\mathbf1}
\DeclareMathOperator{\Var}{Var}
\begin{document}
\pdfbookmark[0]{Article}{part2-article}
\maketitle
\begin{abstract}
A price is both a source of profit and an experiment about demand. We ask
whether a forward-looking Bayesian decision maker, who values the
information its actions generate, eventually learns the true response of
outcomes to actions. In a correctly specified Gaussian linear regression in which the decision maker chooses the regressor, the optimal policy need not be smooth. Near regular Berk--Nash equilibria and near complete learning, however, it admits a smooth approximation, and every policy with such an approximation inherits the incomplete-learning results for smooth decision rules: for every discount factor below one, incorrect Berk--Nash equilibria
persist with probability arbitrarily close to one from nearby concentrated
priors, and optimal policies yield incorrect learning on the regular
parameter set with exactly its prior probability. Only infinite patience justifies complete learning by a forward-looking decision maker. Advertising, monopoly pricing, and inflation stabilization
illustrate the results.
\par\medskip
\noindent\textbf{Keywords:} Bayesian learning, incomplete learning, Berk--Nash equilibrium, optimal experimentation, exploration--exploitation.
\end{abstract}

\section{Introduction}\label{sec:introduction}
A price determines current profit and what a seller learns about demand. A
campaign allocation determines current sales and what a firm learns about advertising.
An interest-rate decision changes inflation and the evidence available about
monetary transmission. In each setting, the decision maker learns from
outcomes generated by its own decisions.

There are two opposing forces. Experimentation improves future decisions by
separating the effects of alternative actions. Optimization directs the
decision maker toward actions that appear valuable given what is already
known. A forward-looking decision maker weighs both forces. Does its
incentive to experiment ensure that it eventually learns the true response of
outcomes to actions?

We study this question in a two-parameter linear regression model. The decision
maker chooses the regressor and learns its intercept and slope by Bayesian
updating from a nondegenerate Gaussian prior. The observation model is
correctly specified. Actions lie in a compact interval, and the objective is
expected discounted payoff. Correct specification and rational updating
remove two familiar explanations of persistent belief errors, so the errors
studied here arise from the actions chosen to pursue the economic objective.

A companion paper \citep{kky2026parti} shows that decision rules that respond
smoothly to beliefs generate too little variation in the regressor for the
slope to be learned. The failure is generic. For every rule in a residual set of
rules that are smooth near complete information, complete learning has
probability zero under every nondegenerate Gaussian prior. Its results cover myopic optimization and rules of thumb,
which do not value information. However, they do not settle the present question, for
two reasons. First, a forward-looking decision maker values the information
its actions generate and experiments when that information is worth the cost,
so one might expect experimentation to restore learning. Second, the results
of the companion paper require the decision rule to be smooth, and the
smoothness of an optimal policy cannot be taken for granted. Establishing
smoothness of optimal policies is a long-standing difficulty in dynamic
programming. \citet{santos1991} establishes differentiability of the optimal
policy in concave discrete-time problems, and \citet{araujo1991} constructs
\(C^3\), strongly concave return functions whose optimal policies are not
twice differentiable. In continuous time, \citet{struloviciszydlowski2015}
establish smoothness of value functions and the existence of optimal controls
in scalar diffusion models, where optimal controls can be discontinuous. Our
Bellman problem is not concave. The action affects both the current expected
payoff and the information that shapes future beliefs. The optimal action may
also switch between several maximizers, and an optimal policy may randomize
among them.

We devise a method that bypasses this difficulty. We first show that every
policy whose actions lie within a distance quadratic in the current action gap
of a smooth map with the properties that drive incomplete learning inherits
the conclusions for smooth rules (Theorem~\ref{thm:transfer}). We then show
that the optimal policy admits such an approximation where it matters, that is, near
regular Berk--Nash equilibria and near complete learning
(Theorem~\ref{thm:approximation}). The economic reason
for the approximation is simple. Near a sufficiently concentrated belief, a
small deviation from the locally optimal action has a second-order current
payoff cost, and the additional slope information it generates also has a
second-order value. For a fixed discount factor below one, the value of
further experimentation becomes too small to sustain the variation needed for
learning. Optimality is not itself the mechanism; it is the economic reason
why the optimal policy acquires the local structure that prevents persistent
experimentation.

Two impossibility results in the spirit of the companion paper then follow
for optimal policies. Both are implications of the approximation and transfer
theorems, except that the global result uses a separate argument at
parameters whose complete-information optimum is a boundary action. The first
is local. An incorrect slope can be offset by an
intercept that predicts the correct mean outcome at a particular action. Such
a mean-fitting perceived response, together with an action optimal for it,
forms a Berk--Nash equilibrium. If that action uniquely maximizes perceived
payoff and satisfies strict curvature and outcome-relevance conditions,
sufficiently concentrated priors near an associated high but finite precision
state remain near the incorrect equilibrium with probability arbitrarily
close to one, under every optimal policy and for every discount factor below
one. The second is global. For every fixed discount factor below one and
every fixed optimal policy, the probability of
incorrect learning on a Borel subset of regular parameters equals the prior
probability of that subset. Randomization among optimal actions and
history-dependent policies are allowed from the outset. Positive prior mass gives positive probability of incorrect
learning, and conull regularity gives probability one. The fixed-truth
implication holds for almost every regular parameter.

The examples show what these results mean economically. In advertising, the
interior Berk--Nash equilibria form a continuum along a quadratic curve of
perceived intercept--slope pairs, and the local theorem shows that a firm can
remain near an incorrect one even though it experiments optimally. This
connects to the mistaken self-assessment of \citet{hks2018}, whose decision
maker holds a fixed incorrect belief about ability; here both coefficients
are learned under a correctly specified observation model, and the persistent
error arises through the information generated by decisions. In pricing, the
closest economic comparison is optimal experimentation with an unknown demand
response \citep{rothschild1974} and with an unknown linear process
\citep{kiefernyarko1989optimal}. An optimally experimenting seller need not
learn the true demand curve. Our statements give primitive regularity
conditions for a common Gaussian regression model, distinguish local
persistence from a setwise global conclusion, and explicitly include optimal
randomization. In monetary policy,
\citet{wieland2000monetary,wieland2000learning} find that optimal
experimentation improves learning and stabilization relative to a passive
policy; our results show that these improvements need not culminate in
complete learning under a fixed discount factor.

Optimal decisions can therefore hinder learning even when future information
enters the decision maker's objective. Myopia is not the root cause---a
forward-looking policy may experiment and improve its expected payoff while
still failing to sustain the variation needed to learn the true slope. Readers
familiar with reinforcement learning may wonder how this fits with adaptive
strategies such as Thompson sampling, whose learning and payoff guarantees are
well known in suitable bandit environments. The distinction is between
eventually choosing an optimal action and optimally valuing the path toward it.
Under our conditions, a strategy that learns the true parameter and converges
to a truly optimal action attains the optimal average payoff when the decision
maker is infinitely patient. With finite patience, however, our results reveal
a dichotomy between optimality and learnability. For every fixed discount
factor below one, a strategy that learns the true parameter with positive
probability at regular parameters cannot be optimal. When regular
parameters have full prior probability, a rational decision maker chooses an
optimal strategy that it \emph{knows} will not reveal the true parameter.

The analysis is confined to a stable, nonstrategic linear environment with
conjugate Gaussian learning. Relevant information arrives through outcomes
generated by the chosen action; outside data, mandated experiments,
competitors' actions, or naturally occurring variation could change the
conclusion. Gaussian conjugacy makes the feedback tractable, while correct
specification and Bayesian updating sharpen the role of optimal action.

Section~\ref{sec:model} gives the model, optimal policies and the three
examples. Section~\ref{sec:approximation} presents the transfer theorem and
the smooth approximation. Section~\ref{sec:optimal} derives from them the
local and global results for optimal policies, with a separate argument at
boundary optima;
Section~\ref{sec:applications} applies
them to the examples and illustrates them by simulation, and
Section~\ref{sec:infinite} considers infinite patience. The
\hyperref[app:value]{Appendix} contains the proofs, and the
\hyperref[supplement]{Supplement} gives the construction of the smooth
approximation and the remaining proofs.

\section{Model}\label{sec:model}
We use the model of \citet{kky2026parti} and recall it briefly. At date
\(t=1,2,\dots\), a decision maker chooses an action
\(x_t\in X=[\underline x,\overline x]\), where \(\underline x<\overline x\),
and observes
\[
 y_t=\alpha+\beta x_t+\varepsilon_t,
 \qquad \varepsilon_t\stackrel{\mathrm{iid}}{\sim}N(0,1).
\]
The parameter \(\vartheta=(\alpha,\beta)\) is unknown and fixed, the prior
\(\mu=N_2((a_0,b_0)^\top,\Sigma_0)\) is nondegenerate Gaussian, and
\(\vartheta\) and \((\varepsilon_t)\) are mutually independent.

\subsection{Posterior beliefs}\label{sec:posterior}
The posterior after each history is Gaussian with mean \((a_t,b_t)\) and
covariance
\[
 \Sigma_t=
 \begin{pmatrix}
  A_t^{-1}+\bar x_t^2/V_t & -\bar x_t/V_t\\
  -\bar x_t/V_t & 1/V_t
 \end{pmatrix},
 \qquad
 A_t=A_0+t,\quad
 \bar x_t=\frac{A_0\bar x_0+\sum_{j=1}^t x_j}{A_0+t}.
\]
Here \(\bar x_t\) is the prior-augmented average action and \(V_t\) is
\textit{slope precision}, \(V_t^{-1}=\Var_{\mu_t}(\beta)\). We write
\(u_t=A_t^{-1}\) and \(w_t=V_t^{-1}\), and call
\(s_t=(a_t,b_t,u_t,\bar x_t,w_t)\) the \textit{posterior state}; a state
\(s\) is \textit{feasible} if \(u>0\) and \(w>0\). With the \textit{action
gap} \(D_t=x_{t+1}-\bar x_t\), \(\kappa_t=A_t/(A_t+1)\) and the prediction
error \(I_{t+1}=y_{t+1}-a_t-b_tx_{t+1}\),
\begin{equation}\label{eq:design}
 A_{t+1}=A_t+1,\qquad
 \bar x_{t+1}=\bar x_t+\frac{D_t}{A_t+1},\qquad
 V_{t+1}=V_t+\kappa_tD_t^2,
\end{equation}
\begin{equation}\label{eq:means}
 b_{t+1}=b_t+\frac{A_tD_t}{(A_t+1)V_{t+1}}I_{t+1},\qquad
 a_{t+1}=a_t+\frac{I_{t+1}}{A_t+1}-\bar x_{t+1}(b_{t+1}-b_t).
\end{equation}
An action away from the average action adds slope precision; an action at the
average action adds none. We write \(r_t=V_t/A_t\) for the \textit{precision
ratio} and \(V_\infty=\lim_tV_t\). By \PartILemComplete, stated for
randomized policies in \PartILemPosterior(c) and for history-dependent ones in
\PartIRemPredictable, complete learning
occurs exactly when \(V_t\to\infty\), up to an event of probability zero, and
\(b_\infty\ne\beta\) almost surely on \(\{V_\infty<\infty\}\).

\subsection{Payoffs and optimal policies}\label{sec:policies}
The decision maker receives a payoff \(U(x,y)\) each period. Define
\[
 G(x,m)=E[U(x,m+Z)],\qquad
 g^{\mathrm{full}}(x;\vartheta)=G(x,\alpha+\beta x),\qquad Z\sim N(0,1),
\]
and let \(g_t(x)\) be the expected payoff of \(x\) under the posterior at date
\(t\). So \(G(x,m)\) is the expected payoff from an action \(x\) when the mean
outcome is \(m\), and \(g^{\mathrm{full}}(x;\vartheta)\) is the expected payoff
when the parameter \(\vartheta\) is known. We maintain the following
assumption.

\begin{assumption}\label{ass:fl-payoff}
For every \(y\in\R\), \(x\mapsto U(x,y)\) is four times continuously
differentiable on \(X\), with one-sided derivatives at the endpoints, and for
every \(x\in X\), \(y\mapsto U(x,y)\) is Borel. There are finite constants
\(C\) and \(c_U\ge0\) such that
\(\lvert\partial_x^jU(x,y)\rvert\le Ce^{c_U\lvert y\rvert}\) for every
\(x\in X\), \(y\in\R\) and \(j\le4\).\footnote{This assumption ensures that the
expected payoffs defined here are finite and smooth.}
\end{assumption}

The payoff needs to be smooth in the action only. Because the outcome is
normal given the belief, the expected payoffs are smooth in the predicted mean
and variance even when the payoff is not smooth in the outcome. So the
assumption covers every polynomial payoff, including those of the examples
below, payoffs that grow exponentially in the outcome, such as the
exponential utility \(-e^{-\gamma y}-c(x)\) with \(\gamma>0\), and payoffs that
are kinked or discontinuous in the outcome. Examples
are an asymmetric (lin-lin) loss around a target \(y^*\) net of an action
cost, \(-\rho\max\{y-y^*,0\}-(1-\rho)\max\{y^*-y,0\}-c(x)\) with
\(\rho\in(0,1)\), a bonus for reaching the target net of the cost,
\(\1\{y\ge y^*\}-c(x)\), the profit \((x-k)\max\{y,0\}\) of a seller whose
demand is truncated at zero, and the capacity-constrained payoff
\(\min\{y,\overline y\}-c(x)\) with capacity \(\overline y\). Here the cost
\(c\) is four times continuously differentiable.

The decision maker discounts future payoffs by \(\delta\in[0,1)\). Given a prior
\(\mu\) and a policy \(\pi\), the discounted expected payoff is
\[
 U_\delta^\pi=(1-\delta)E^{\mu,\pi}
 \left[\sum_{t\ge0}\delta^tU(x_{t+1},y_{t+1})\right].
\]
The optimal action need not be unique, and the decision maker may randomize
among optimal actions. We therefore allow a policy to be \textit{randomized}.
At each date it draws the action from a distribution that may depend on the
whole past history. A policy is \textit{Markov} if this distribution depends
on the past history only through the posterior, that is, if every two
histories that induce the same posterior lead to the same distribution of
actions. For \(\delta>0\), a policy is \textit{optimal} if it maximizes the
discounted payoff \(U_\delta^\pi\); for \(\delta=0\), it is optimal if it is
\textit{myopically optimal}, that is, if it maximizes \(g_t\) in each period.
We write \(\pi_\delta\) for an optimal policy. Under
Assumption~\ref{ass:fl-payoff} the optimal value is attained and satisfies a
Bellman equation, optimal Markov policies exist, and a policy is optimal
exactly when it chooses, almost surely at every date, an action that
maximizes the right side of this equation at the current posterior
(Proposition~\ref{prop:value}, Definition~\ref{A-def:optimal} and
Lemma~\ref{lem:support} in Appendix~\ref{app:value}). The results below hold
for every optimal policy, Markov or not. Given a prior \(\mu\) and a policy \(\pi\),
\(P^{\mu,\pi}\) denotes the prior-predictive law and \(P^\pi_\vartheta\) the
fixed-truth law.

\subsection{Berk--Nash equilibria and examples}\label{sec:equilibria}
We use the terminology of \citet{kky2026parti}. Given the true parameter
\(\vartheta=(\alpha,\beta)\), a tuple \((a,b,x,w)\) with \(w\ge0\), consisting
of posterior means, an action and a slope variance, is a
\textit{(pure-strategy) Berk--Nash equilibrium} \citep{ep2016} if
\[
 x\in\arg\max_{z\in X}E\left[U\left(z,a+bz+\sqrt{1+w(z-x)^2}\,Z\right)\right]
 \quad\text{and}\quad
 a+bx=\alpha+\beta x.
\]
The first condition is \textit{incentive compatibility}, that is, \(x\) is
optimal given the Gaussian belief with mean \((a,b)\) and covariance
\(w\,nn^\top\), \(n=(-x,1)^\top\), which is supported on the line of
parameters that predict the mean outcome at \(x\) correctly. The second is the
\textit{no-surprise condition}, that is, at the equilibrium action, the
decision maker correctly anticipates the average outcome. This belief is the
\textit{Berk--Nash equilibrium belief}. Our results concern the case \(w=0\),
in which the belief is concentrated on \((a,b)\) and the two conditions read
\begin{align}
 & x\in\arg\max_{z\in X}G(z,a+bz), \label{eq:bne-ic}\\
 & a+bx=\alpha+\beta x. \label{eq:bne-ns}
\end{align}
We call such an equilibrium a \textit{complete-information Berk--Nash
equilibrium} and write it as the triple \((a,b,x)\), and we identify its
belief with the pair \((a,b)\). An equilibrium written as a triple, and an
equilibrium belief written as a pair, are always complete-information ones.
The true parameter is a Berk--Nash
equilibrium belief, but the examples below show that there are typically many
others. Let \(\Pi_X(z)=\min\{\overline x,\max\{\underline x,z\}\}\) denote the
projection onto the action interval \(X\).

\begin{example}[name=Advertising]\label{ex:fl-advertising}
A firm chooses a (signed) campaign reallocation \(x\), observes sales
performance \(y=\alpha+\beta x+\varepsilon\), and receives the payoff
\(U(x,y)=y-x^2/2\), where \(x^2/2\) is the cost of reallocation. Reallocation
can move effort in either direction around a neutral allocation. Simple
algebra shows that the myopically optimal action is the projection of the
posterior slope onto the action interval, \(\Pi_X(b_t)\). A
forward-looking firm has an incentive to experiment to learn about
\((\alpha,\beta)\), so its optimal action need not coincide with the myopic one.
Every \((a,b)\) with \(a=\alpha+\beta b-b^2\) and \(b\in\operatorname{int}X\) is
a Berk--Nash equilibrium belief, with action \(x=b\);
\PartIFigBeliefsAd{} shows them for \(\alpha=\beta=10\). These are the
equilibrium beliefs of the overconfidence model of \citet{hks2018}, in which
the decision maker holds a dogmatic incorrect belief about the intercept.
\end{example}

\begin{example}[name=Pricing]\label{ex:fl-pricing}
A seller posts a price \(x\in X\Subset(0,\infty)\),\footnote{We write
\(A\Subset B\) when the closure of \(A\) is compact and contained in the
interior of \(B\).} observes a noisy demand index
\(y=\alpha+\beta x+\varepsilon\), and receives \(U(x,y)=(x-k)y\), where \(k\ge0\)
is a constant marginal cost. Since the prior is nondegenerate Gaussian, the
true slope \(\beta\) may be positive, although the economically relevant case
is \(\beta<0\); since the noise is Gaussian, the demand index \(y\) may also
be negative. When \(b_t<0\), simple algebra shows that the myopically optimal price is
\(\Pi_X(k/2-a_t/(2b_t))\); a forward-looking seller may deviate from it to
experiment, as in the optimal-experimentation questions of
\citet{rothschild1974} and \citet{kiefernyarko1989optimal}. With true parameter
\((1,-1)\), every \((a,b)\) with \(b<0\),
\(a=b[2-k(1+b)]/(b-1)\) and \(k/2-a/(2b)\in\operatorname{int}X\) is a
Berk--Nash equilibrium belief (\PartIFigBeliefsPricing{} for \(k=0.1\)).
\end{example}

\begin{example}[name=Inflation]\label{ex:fl-inflation}
A central bank chooses a policy instrument \(x\in X\),
observes inflation \(y=\alpha+\beta x+\varepsilon\), and receives
\(U(x,y)=-[(y-y^*)^2+\omega x^2]\), where \(y^*\) is the target inflation rate
and \(\omega>0\) measures the cost of using the instrument. At a degenerate
belief \((a,b)\), the expected loss \((a+bx-y^*)^2+1+\omega x^2\) is strictly
convex in \(x\). For \(\beta b+\omega\ne0\), the Berk--Nash equilibrium beliefs
with an interior action are
\[
 a=\frac{(\alpha-y^*)b^2+\beta y^*b+\alpha\omega}{\beta b+\omega},
 \qquad
 \frac{b(y^*-a)}{b^2+\omega}\in\operatorname{int}X.
\]
\end{example}

\begin{remark}
As in \citet{kky2026parti}, attention can be restricted to pure-strategy
Berk--Nash equilibria \citep{epy2021}; in addition, the complete-information
optimum is unique at the regular parameters that our results concern.
\end{remark}

\section{Transfer and smooth approximation}\label{sec:approximation}
The results of \citet{kky2026parti} concern decision rules that are smooth
functions of the posterior. A forward-looking optimal policy need not be one.
As discussed in the introduction, the Bellman problem of this paper is not
concave, and there can be multiple optimal actions.

We first show that a policy inherits the conclusions for smooth rules whenever
it is close to a smooth map where it matters, and then that the optimal policy
is close to such a map near regular Berk--Nash equilibria and near complete
learning. For a state \(s=(a,b,u,\bar x,w)\), write \(V=1/w\) for its slope
precision. For a target \((a^e,b^e,x^e)\) and constants \(r_*>0\) and
\(u_*\in(0,1)\), the \textit{collar} at precision \(V^e\) is
\[
 \mathcal C(V^e)=\bigl\{(a,b,u,\bar x,w):
 \lvert a-a^e\rvert+\lvert b-b^e\rvert+\lvert\bar x-x^e\rvert<r_*,\ \lvert u\rvert<u_*,\ V^e/2\le1/w\le2V^e\bigr\},
\]
the set of states near the target belief and average action with slope
precision comparable to \(V^e\). On collars we use \(V\) in place of \(w\) as
a coordinate, so smoothness and norms of a map \(\varphi\) of the state refer to
the coordinates \((a,b,u,\bar x,V)\). Given such a map, write
\(D^\varphi=\varphi-\bar x\) for its action gap. A policy \(\pi\)
\textit{tracks} \(\varphi\) on a set \(\mathcal C\) with constant \(C_*\) for a
prior \(\mu\) if, \(P^{\mu,\pi}\)-almost surely, at every date with
\(s_t\in\mathcal C\),
\begin{equation}\label{eq:tracking}
 \lvert x_{t+1}-\varphi(s_t)\rvert\le C_*\bigl(u_t+\lvert D^\varphi(s_t)\rvert\bigr)^2.
\end{equation}
That is, the action differs from the smooth map by at most the square of the
current uncertainty about the mean outcome and the current action gap. A
smooth decision rule of \citet{kky2026parti} tracks itself with no error.

The smooth maps that matter satisfy the hypotheses of \PartILemTrapping, in
the \(C^{2,1}\) form stated after it. Fix a target
\((a^e,b^e,x^e)\) and constants \(r_*>0\), \(u_*\in(0,1)\), \(V_*<\infty\),
\(C_*<\infty\) and \(\lambda^0_*>0\), and write
\(\mathcal C_*=\bigcup_{V^e\ge V_*}\mathcal C(V^e)\). A map \(\varphi\) of the
state is a \textit{smooth target map} for \((a^e,b^e,x^e)\) with these
constants if, on every collar \(\mathcal C(V^e)\) with \(V^e\ge V_*\),
\(\varphi\) is \(C^{2,1}\), with the \(C^{2,1}\) norm defined after
\PartILemTrapping, and
\begin{equation}\label{eq:target-map}
 \lVert\varphi\rVert_{C^{2,1}}\le C_*,
 \qquad
 \varphi(a^e,b^e,0,x^e,1/V^e)=x^e,
 \qquad
 \lvert\varphi_b-\bar x\varphi_a\rvert\ge\lambda^0_*.
\end{equation}
The map is smooth, it chooses the target action when the mean outcome at that
action is known, and it satisfies transversality.

For the global result we use the \textit{coherent exclusion package} of
\PartIDefPackage. In words, a prior and a policy admit such a package on a set
\(\Theta_0\) of parameters if \(\Theta_0\) is covered by countably many Borel
sets, on each of which there are a smooth map of the state and a corridor of
states near complete learning with the following properties. The map is
smooth, satisfies transversality, and is flat in the covariance, that is, its
derivatives in the average action and in the logarithm of the slope precision
are at most proportional to the remaining uncertainty \(u+w\). On the corridor, the policy
tracks the map. And whenever learning is complete at a parameter of the set,
the state eventually enters the corridor and remains there. A smooth decision
rule admits such a package on its interior regular parameters, with (an
extension of) the rule itself as the map \citep{kky2026parti}. Let
\(\mathcal W=\{V_\infty<\infty,\ b_t\to b_\infty\ne\beta\}\) be the event of
an incorrect limiting slope. The following theorem shows that every policy
that tracks a smooth target map, or that admits a coherent exclusion package,
inherits the conclusions of \citet{kky2026parti}. The proof is in
Appendix~\ref{app:transfer}.

\begin{theorem}[Transfer]\label{thm:transfer}
\begin{enumerate}[label=\textup{(\roman*)},ref=(\roman*)]
\item \emph{Local.} Fix the true parameter \(\vartheta\), a target\label{thm:transfer:i}
\((a^e,b^e,x^e)\) with \(a^e+b^ex^e=\alpha+\beta x^e\), and a smooth target
map \(\varphi\) for it with constants \(r_*,u_*,V_*,C_*,\lambda^0_*\). Then
for every \(r\in(0,r_*)\), there
is \(\overline V_r\ge V_*\) such that, for every \(V^e\ge\overline V_r\) and
every \(\varepsilon\in(0,1)\), there are \(\overline A<\infty\) and \(r_0>0\)
with the following property. For every nondegenerate Gaussian prior \(\mu\)
with \(A_0>\overline A\) and
\(\lvert a_0-a^e\rvert+\lvert b_0-b^e\rvert+\lvert\bar x_0-x^e\rvert+\lvert V_0-V^e\rvert<r_0\), and every policy \(\pi\)
that tracks \(\varphi\) on \(\mathcal C_*\) with constant \(C_*\) for \(\mu\),
the following hold simultaneously with \(P^\pi_\vartheta\)-probability at
least \(1-\varepsilon\):
\begin{enumerate}[label=\textup{(\alph*)}]
\item \(\lvert a_t-a^e\rvert+\lvert b_t-b^e\rvert+\lvert\bar x_t-x^e\rvert<r\) and \(V^e/2\le V_t\le2V^e\) for
every \(t\ge0\);
\item \(V_\infty<\infty\), \(b_t\to b_\infty\), \(\bar x_t\to\bar x_\infty\),
\(x_{t+1}\to\bar x_\infty\) and
\(a_t\to\alpha+\beta\bar x_\infty-b_\infty\bar x_\infty\);
\item \(b_\infty\ne\beta\) when \(b^e\ne\beta\).
\end{enumerate}
These priors form a nonempty open set, and the thresholds do not depend on the
prior or the policy.
\item \emph{Global.} Fix a nondegenerate Gaussian prior \(\mu\) and a policy\label{thm:transfer:ii}
\(\pi\) that admit a coherent exclusion package on a Borel set \(\Theta_0\).
Then \(P^{\mu,\pi}(V_t\to\infty,\ \vartheta\in\Theta_0)=0\);
\(P^{\mu,\pi}(\mathcal W,\ \vartheta\in B)=\mu(B)\) for every Borel
\(B\subseteq\Theta_0\); and \(P^\pi_\vartheta(\mathcal W)=1\) for
Lebesgue-almost every \(\vartheta\in\Theta_0\).
\end{enumerate}
\end{theorem}

In the special case in which the policy is itself the smooth map,
Theorem~\ref{thm:transfer} gives the global result for smooth rules of
\citet{kky2026parti} and, for rules with \(C^{2,1}\) bounds that choose \(x^e\)
at every high slope precision, its local result (\PartIThmLocal) with priors
centered at \((a^e,b^e,x^e,V^e)\). The quadratic error allowed by the tracking condition is what the
contraction argument of that paper can absorb. An error of the order of the
square of the action gap does not change the multiplicative dynamics of the
gap to first order.

The next theorem shows that every optimal policy tracks such a map near
regular Berk--Nash equilibria and near complete learning. For its global part,
let \(\Theta^I\) be the set of parameters \(\vartheta\) at which
\(g^{\mathrm{full}}(\cdot;\vartheta)\) has a unique maximizer
\(x^\star(\vartheta)\in\operatorname{int}X\) with
\(\partial_{xx}g^{\mathrm{full}}(x^\star(\vartheta);\vartheta)<0\) and
\(G_m(x^\star(\vartheta),\alpha+\beta x^\star(\vartheta))\ne0\). The proof is
in Appendix~\ref{app:approximation}.

\begin{theorem}[Smooth approximation of optimal policies]\label{thm:approximation}
Fix \(\delta\in[0,1)\).
\begin{enumerate}[label=\textup{(\roman*)},ref=(\roman*)]
\item \emph{Local.} Fix \((\hat a,\hat b,\hat x)\) with\label{thm:approximation:i}
\(\hat x\in\operatorname{int}X\) the unique maximizer of
\(z\mapsto G(z,\hat a+\hat bz)\) over \(X\), with strictly negative second
derivative there and \(G_m(\hat x,\hat a+\hat b\hat x)\ne0\). There is a Borel
measurable smooth target map \(\varphi\) for \((\hat a,\hat b,\hat x)\), with
constants \(r_*,u_*,V_*,C_*,\lambda^0_*\), the map and the constants not
depending on a policy, a prior or the true parameter, such that:
\begin{enumerate}[label=\textup{(\alph*)}]
\item every action \(x\) in the Bellman first-action argmax at every feasible
state \(s\) of \(\mathcal C_*\) satisfies
\(\lvert x-\varphi(s)\rvert\le C_*(u+\lvert D^\varphi(s)\rvert)^2\);
\item every optimal policy for every prior tracks \(\varphi\) on
\(\mathcal C_*\) with constant \(C_*\) for that prior.
\end{enumerate}
\item \emph{Global.} \(\Theta^I\) is open, and for every nondegenerate Gaussian\label{thm:approximation:ii}
prior \(\mu\) and every optimal policy \(\pi_\delta\), \(\mu\) and
\(\pi_\delta\) admit a coherent exclusion package on \(\Theta^I\).
\end{enumerate}
\end{theorem}

The optimal policy need not equal the smooth map, but it is within a quadratic
error of it. Part~\ref{thm:approximation:ii} is the global counterpart. Near complete learning at
every regular parameter, the optimal policy is close to a smooth map in the
same sense, and these approximations form a coherent exclusion package.

The reason for the quadratic bound is a comparison of two values. Near a
concentrated belief, the smooth map is the maximizer of an auxiliary Bellman
problem that coincides with the true one as long as the posterior stays near
the target, and whose objective is strictly concave in the action. An action
at distance \(d\) from the map therefore loses at least a multiple of \(d^2\)
in the auxiliary problem. The true and auxiliary values differ only through
the event that the posterior leaves this neighborhood, and the difference is
at most a multiple of \((u+\lvert D^\varphi\rvert)^4\). An optimal action cannot give up
more than this difference, so \(d\) is at most a multiple of
\((u+\lvert D^\varphi\rvert)^2\). In economic terms, as in the introduction, both the
payoff cost of a small deviation and the value of the information it generates
are second order near a concentrated belief, so optimal experimentation may
continue, but only within a quadratic distance of the smooth map. The argument
does not require the optimal policy to be smooth.

\section{Implications for optimal policies}\label{sec:optimal}
\subsection{Local persistence of incorrect beliefs}\label{sec:local}
We have seen in the examples that there are many Berk--Nash equilibria at
which the posterior belief is incorrect. Their existence does not imply that
they arise as long-run outcomes of Bayesian learning, especially when the
decision maker deliberately experiments.

The result concerns equilibria with the following regularity. A
complete-information Berk--Nash equilibrium \((a,b,x)\) is \textit{interior regular} if \condtarget{cond:R1}{(R1)}
\(x\in\operatorname{int}X\) is the unique maximizer of \(z\mapsto G(z,a+bz)\)
over \(X\), \condtarget{cond:R2}{(R2)} this function has a strictly negative second derivative at
\(z=x\), and \condtarget{cond:R3}{(R3)} \(G_m(x,a+bx)\ne0\). Conditions \ref{cond:R1} and \ref{cond:R2} require the
action to be interior, uniquely optimal, and to satisfy the second-order
condition. Condition \ref{cond:R3} requires the mean outcome to be locally
payoff-relevant. In the language of \citet{kky2026parti}, it is the
transversality condition for the myopic policy, that is, a revision of beliefs that
leaves the predicted outcome at \(x\) unchanged moves the myopically optimal
action to first order. Differentiating the normal density in \(m\) gives
\(G_m(x,m)=E[U(x,m+Z)Z]\), which equals \(E[U_y(x,m+Z)]\) when
\(U(x,\cdot)\) is continuously differentiable with an exponentially bounded
derivative. Conditions
\ref{cond:R1}--\ref{cond:R3} are the hypotheses of
Theorem~\ref{thm:approximation}\ref{thm:approximation:i}. The following theorem shows that an
interior regular Berk--Nash equilibrium nevertheless persists under every
optimal policy. It is the first implication of
Theorems~\ref{thm:approximation}\ref{thm:approximation:i} and~\ref{thm:transfer}\ref{thm:transfer:i}; the proof is
in Appendix~\ref{app:optimal}.

\begin{theorem}[Local persistence]\label{thm:fl-local}
Fix the true parameter
\(\vartheta=(\alpha,\beta)\), a discount factor \(\delta\in[0,1)\), and an
interior regular Berk--Nash equilibrium \((a^e,b^e,x^e)\). There are
\(\overline V<\infty\) and \(\bar r>0\) such that, for every \(r\in(0,\bar r)\),
there is \(\overline V_r\ge\overline V\) with the following property. For every
\(V^e\ge\overline V_r\) and every \(\varepsilon\in(0,1)\), there exist
\(\overline A<\infty\) and \(r_0>0\) such that, for every nondegenerate Gaussian
prior \(\mu\) with
\[
 A_0>\overline A,
 \qquad
 \lvert a_0-a^e\rvert+\lvert b_0-b^e\rvert+\lvert\bar x_0-x^e\rvert+\lvert V_0-V^e\rvert<r_0,
\]
and every optimal policy \(\pi_\delta\) for \(\mu\), the following hold
simultaneously with \(P_\vartheta^{\pi_\delta}\)-probability at least
\(1-\varepsilon\):
\begin{enumerate}[label=\textup{(\roman*)},ref=(\roman*)]
\item For every \(t\ge0\),\label{thm:fl-local:i}
\[
 \lvert a_t-a^e\rvert+\lvert b_t-b^e\rvert+\lvert\bar x_t-x^e\rvert<r,
 \qquad
 \frac{V^e}{2}\le V_t\le2V^e.
\]
\item The posterior means converge to finite limits \((a_\infty,b_\infty)\),\label{thm:fl-local:ii}
\(V_\infty<\infty\), \(\bar x_t\to\bar x_\infty\) and
\(x_{t+1}\to\bar x_\infty\) for some finite \(\bar x_\infty\), so that
\(D_t\to0\), and
\(a_\infty=\alpha+\beta\bar x_\infty-b_\infty\bar x_\infty\).
\end{enumerate}
These priors form a nonempty open set. If \(b^e\ne\beta\), then
\(b_\infty\ne\beta\) on the event of \ref{thm:fl-local:i} and \ref{thm:fl-local:ii}.
\end{theorem}

The theorem shows that, starting from an open set of sufficiently concentrated,
nondegenerate priors near an equilibrium belief and for every fixed discount
factor, the posterior of a decision maker who optimizes remains near that belief
with probability arbitrarily close to one. When the equilibrium belief is
incorrect, the limiting slope is incorrect, even when \(r\) exceeds
\(\lvert b^e-\beta\rvert\); by
\ref{thm:fl-local:i}, it also lies within \(r\) of \(b^e\). At \(\delta=0\), the theorem is the
myopic case, which \citet{kky2026parti} obtains under weaker smoothness
conditions on the payoff, with three action derivatives rather than four.

This gives an interpretation of the advertising example. In the model of
\citet{hks2018}, the decision maker is dogmatically overconfident about the
intercept and consequently learns an incorrect slope. In our model, neither
parameter is fixed, and the decision maker values the information that a
different campaign would generate. Suppose that its initial prior is
concentrated near the same Berk--Nash equilibrium belief. The theorem shows
that, with probability arbitrarily close to one, the posterior means remain
nearby and converge. Thus overconfidence about the intercept can persist
endogenously even though the agent assigns positive probability to other
values, updates both parameters according to Bayes' rule, and experiments
optimally.

The local persistence problem differs from the one studied by
\citet{fls2021}, who consider a strict equilibrium in a finite-action
environment, so that the same action is chosen as long as beliefs stay near
the equilibrium. In our continuous-action model, a small change in beliefs
changes the optimal action, and a forward-looking decision maker may change it
further to experiment. Observations generated by two distinct actions identify
both coefficients, so this variation continually generates information that
points toward the truth. The question is whether its total amount is large
enough to overcome an initially concentrated incorrect belief. The theorem
shows that it need not be. The optimal policy is close to a smooth rule, and
under a smooth rule the action moves only in proportion to the belief
revision, which is itself of the order of the current action gap, so the
variation in actions shrinks and is too small to identify the slope.
\citet{kky2026parti} explains this mechanism in detail.

\subsection{Generic failure of complete learning}\label{sec:global}
The previous theorem concerns persistence near a \textit{selected} incorrect
equilibrium. We next ask whether optimal behavior can lead to complete
learning at all. Let \(\Theta^I\) be as in Section~\ref{sec:approximation}. In
words, it is the set of parameters \(\vartheta\) for which
\((\alpha,\beta,x^\star(\vartheta))\) is an interior regular Berk--Nash
equilibrium when the true parameter is \(\vartheta\). Let \(\Theta^B\) be the
set of parameters at which \(g^{\mathrm{full}}(\cdot;\vartheta)\) has a unique
maximizer at the upper endpoint \(\overline x\) with
\(g^{\mathrm{full}}_x(\overline x;\vartheta)>0\), or at the lower endpoint
\(\underline x\) with \(g^{\mathrm{full}}_x(\underline x;\vartheta)<0\), and let
\(\Theta^R=\Theta^I\cup\Theta^B\). The excluded parameters are those at which
the complete-information optimum is not unique or fails one of these local
regularity conditions. The following theorem shows that the event
\(\mathcal W=\{V_\infty<\infty,\ b_t\to b_\infty\ne\beta\}\) of
Section~\ref{sec:approximation}, an incorrect limiting slope, occurs with
probability one for almost all regular parameters. On \(\Theta^I\) it is the
second implication of Theorems~\ref{thm:approximation}\ref{thm:approximation:ii}
and~\ref{thm:transfer}\ref{thm:transfer:ii}; on \(\Theta^B\) it uses a separate argument at the
boundary optimum. The proof is in Appendix~\ref{app:optimal}.

\begin{theorem}[Generic incorrect learning under optimal decisions]\label{thm:fl-global}
Fix a nondegenerate Gaussian
prior \(\mu\), \(\delta\in[0,1)\), and an optimal policy \(\pi_\delta\). Then for
every Borel \(B\subseteq\Theta^R\),
\[
 P^{\mu,\pi_\delta}(\mathcal W,\ \vartheta\in B)=\mu(B).
\]
For Lebesgue-almost every \(\vartheta\in\Theta^R\),
\(P_\vartheta^{\pi_\delta}(\mathcal W)=1\).
\end{theorem}

This theorem implies that
\[
 \mu(\Theta^R)>0\ \Longrightarrow\ P^{\mu,\pi_\delta}(\mathcal W)>0,
 \qquad
 \mu(\Theta^R)=1\ \Longrightarrow\ P^{\mu,\pi_\delta}(\mathcal W)=1.
\]
So when regularity holds almost everywhere under the prior, the optimal policy
leads to incorrect learning almost surely. The following is an immediate
corollary, which describes a sharp tension between complete learning and
optimality. The proof is in Appendix~\ref{app:optimal}.

\begin{corollary}[Complete-learning policies are suboptimal]\label{cor:suboptimal}
Fix a nondegenerate Gaussian prior \(\mu\) and a policy \(\pi\) under which
complete learning at regular parameters has positive probability,
\[
 P^{\mu,\pi}(V_t\to\infty,\ \vartheta\in\Theta^R)>0.
\]
Then, for every discount factor \(\delta\in[0,1)\), \(\pi\) is not optimal
for \(\mu\).
\end{corollary}

In particular, a policy that attains complete learning on a Borel set
\(R\subseteq\Theta^R\) with \(\mu(R)>0\), in the sense that
\(P^{\mu,\pi}(V_t\to\infty,\ \vartheta\in R)=\mu(R)\), is not optimal.

To see the intuition for Theorem~\ref{thm:fl-global}, consider a parameter in
\(\Theta^I\). Suppose, contrary to the theorem, that complete learning occurs.
Then the posterior converges to the true parameter and the optimal action
converges to the complete-information optimum. Near complete information the
value of further experimentation vanishes, and the optimal policy is close to a
smooth map that responds to slope revisions. The noise keeps moving the
posterior means and hence the action, but by the mechanism of the local
theorem the resulting variation is too small to sustain \(V_t\to\infty\), a
contradiction. At a parameter in \(\Theta^B\), the optimal action eventually
stays at an endpoint, so slope precision stops growing.

The theorem has a stark implication when \(\mu(\Theta^R)=1\). At the beginning,
the decision maker knows the prior, the correctly specified model, and the
optimal policy. Under these beliefs, it assigns probability one to eventually
holding an incorrect belief about the slope. Thus the failure of complete
learning is not an unforeseen consequence of optimization---the decision maker
knows ex ante that the optimal policy will not reveal the true parameter.

As in \citet{kky2026parti}, the last assertion of
Theorem~\ref{thm:fl-global} and Fubini's theorem show that incorrect learning
does not require the true parameter to be drawn from the prior. If it is
drawn from a distribution with a density, a policy
\(\pi_\delta\) that is optimal for \(\mu\) leads to incorrect learning on every
Borel set \(B\subseteq\Theta^R\) with exactly the probability that this
distribution assigns to \(B\).

Theorems~\ref{thm:fl-local} and~\ref{thm:fl-global} describe different aspects
of the same problem. Theorem~\ref{thm:fl-local} shows that a particular
incorrect Berk--Nash equilibrium can persist from nearby concentrated priors,
with probability arbitrarily close to one. Theorem~\ref{thm:fl-global} does
not identify the limiting belief; instead, it rules out complete learning on
the regular parameter set. Neither result implies the other. Together, they
show both how a specific incorrect belief can be sustained and why optimal
Bayesian learning need not reach the truth.

\section{Applications}\label{sec:applications}
We now apply Theorems~\ref{thm:fl-local} and~\ref{thm:fl-global} to the
examples. Supplement~\ref{supp:applications} shows that their assumptions are
satisfied.

\begin{example}[name=Advertising,continues=ex:fl-advertising]
Every \(b^e\in\operatorname{int}X\) gives an interior regular Berk--Nash
equilibrium, so Theorem~\ref{thm:fl-local} applies to every such equilibrium.
The complete-information payoff is \(\alpha+\beta x-x^2/2\), so the
complete-information optimum is \(\Pi_X(\beta)\), and every parameter is
regular except when \(\beta\) equals one of the two endpoints of \(X\). Hence
\(\Theta^R\) has prior probability one under every nondegenerate Gaussian prior,
and Theorem~\ref{thm:fl-global} implies that incorrect learning occurs with
prior-predictive probability one under every optimal policy. The result applies
to forward-looking firms as well as myopic ones. Optimal experimentation need
not generate enough variation to learn the return to advertising
completely.\footnote{The managerial-myopia literature
\citep{mizikjacobson2007,mizik2010} gives a different motivation for short
horizons, but the forward-looking result does not depend on myopia.}
\end{example}

\begin{example}[name=Pricing,continues=ex:fl-pricing]
For negative true slopes, the complete-information profit is strictly concave.
For positive slopes it is strictly convex, so its unique maximizer is an
endpoint of \(X\) at which its derivative points strictly outward, unless the
two endpoint profits tie. Hence every parameter is regular except in the
knife-edge cases in which (i) \(\beta<0\), the optimal price is an endpoint of
\(X\), and the derivative of the complete-information profit vanishes there,
(ii) the optimal price is interior and equals marginal cost, so that
\(G_m=x-k=0\), (iii) \(\beta>0\) and the two endpoint profits tie, or (iv)
\(\beta=0\). These cases lie on five lines, so \(\Theta^R\) has prior probability
one, and incorrect learning occurs with prior-predictive probability one. For
a true parameter with \(\beta<0\) and \(\alpha+\beta k>0\), so that demand slopes
down and true expected demand at marginal cost is positive, the local theorem
applies to every Berk--Nash equilibrium \((a^e,b^e,x^e)\) with \(b^e<0\) and
\(x^e\in\operatorname{int}X\); for the true parameter \((1,-1)\) of
Example~\ref{ex:fl-pricing}, these are the equilibrium beliefs of
\PartIFigBeliefsPricing{} when \(k<1\). So an optimally experimenting seller need not learn
the true demand curve.
\end{example}

\begin{example}[name=Inflation,continues=ex:fl-inflation]
The complete-information payoff is strictly concave. At an interior optimum,
\(G_m=0\) together with the first-order condition would imply \(\omega x=0\),
hence \(x=0\) and \(\alpha=y^*\). So the irregular parameters lie among those
with \(\alpha=y^*\), at which inflation is on target at a zero instrument, and
those at which the optimum is an endpoint of \(X\) and the derivative of the
complete-information payoff vanishes there; these form a null set. Theorem~\ref{thm:fl-global} therefore gives prior-predictive
probability-one incorrect learning. A forward-looking bank can improve learning
and stabilization through experimentation, as in
\citet{wieland2000monetary,wieland2000learning}, while still failing to learn
completely.
\end{example}

\subsection{Simulations}\label{sec:simulations}
We illustrate the three examples using numerical Bellman policies with zero
terminal value, horizon \(28\), and discount factor \(0.6\), over \(T=1{,}000\)
periods with 256 paths per case. The true parameter is fixed across paths and
only the shocks vary. Each policy is computed by value iteration on a grid,
with Gaussian quadrature for the expectations.

\begin{figure}[htbp]
\centering
\begin{subfigure}[t]{0.31\textwidth}
\centering
\includegraphics[page=1]{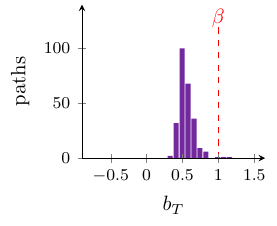}
\caption{Advertising, left.}
\label{figure:fl-distributions:ad-left}
\end{subfigure}
\
\begin{subfigure}[t]{0.31\textwidth}
\centering
\includegraphics[page=2]{tikz/fig_advertising_article_distributions.pdf}
\caption{Advertising, right.}
\label{figure:fl-distributions:ad-right}
\end{subfigure}
\
\begin{subfigure}[t]{0.31\textwidth}
\centering
\includegraphics[page=3]{tikz/fig_advertising_article_distributions.pdf}
\caption{Advertising, bimodal.}
\label{figure:fl-distributions:ad-bimodal}
\end{subfigure}

\medskip
\begin{subfigure}[t]{0.31\textwidth}
\centering
\includegraphics[page=1]{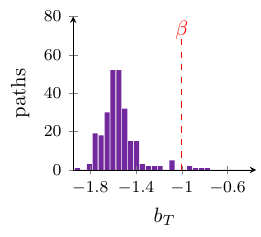}
\caption{Pricing, left.}
\label{figure:fl-distributions:pricing-left}
\end{subfigure}
\
\begin{subfigure}[t]{0.31\textwidth}
\centering
\includegraphics[page=2]{tikz/fig_pricing_article_distributions.pdf}
\caption{Pricing, right.}
\label{figure:fl-distributions:pricing-right}
\end{subfigure}

\medskip
\begin{subfigure}[t]{0.31\textwidth}
\centering
\includegraphics[page=1]{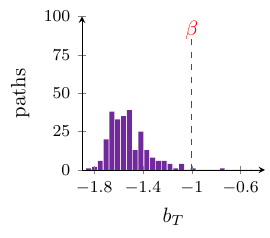}
\caption{Inflation, left.}
\label{figure:fl-distributions:inflation-left}
\end{subfigure}
\
\begin{subfigure}[t]{0.31\textwidth}
\centering
\includegraphics[page=2]{tikz/fig_inflation_article_distributions.pdf}
\caption{Inflation, right.}
\label{figure:fl-distributions:inflation-right}
\end{subfigure}
  \caption{Terminal slope distributions under numerical Bellman policies
  (horizon 28, discount factor 0.6), \(T=1{,}000\), 256 paths per case. In each example, the cases follow the order of the initial states in the text.}
\label{figure:fl-distributions}
\end{figure}

Figure~\ref{figure:fl-distributions} shows the terminal slope estimates. We
write initial states as \((a_0,b_0,\bar x_0,A_0,V_0)\). In the advertising
example, with true parameter \((1,1)\) and \(X=[-1,2]\), so that the
complete-information optimum \(\beta=1\) is interior (\PartIFigBeliefsAd{}
uses \(\alpha=\beta=10\) only to display the shape of the equilibrium set), the
concentrated priors \((1,0.5,0,5,4)\) and \((0.5,1.5,1,5,4)\) lead to terminal
slopes concentrated below and above the true slope, with median terminal slopes
\(0.53\) and \(1.36\). Along the equilibrium curve
\(a=\alpha+\beta b-b^2\), a positive slope below the true slope goes with an
intercept above the true one, the overconfidence of \citet{hks2018}, and a
slope above it with an intercept below the true one, underconfidence. A third prior, \((-0.3125,-0.75,-0.75,5,2.5)\), produces
paths that freeze near an incorrect slope and paths that keep learning. In the
pricing example, with true parameter \((1,-1)\) and \(k=0\), the priors
\((1,-1.5,0.5,5,4)\) and \((0.5,-0.5,1,5,8)\) lead to terminal slopes
concentrated below \(-1\), overstating the price effect, and above \(-1\),
understating it, with medians \(-1.57\) and \(-0.66\). In the inflation
example, with true parameter \((1,-1)\), \(y^*=0\) and \(\omega=1\), the priors
\((1,-1.5,0.5,5,4)\) and \((1,-0.75,0.5,5,8)\) give medians \(-1.55\) and
\(-0.82\). In the pricing and inflation examples, slope precision grows little
in the late periods of the simulation.

\section{Infinite patience}\label{sec:infinite}
The preceding results consider a fixed discount factor \(\delta<1\). We now
consider the limiting case of an infinitely patient decision maker. This
changes the conclusion: complete learning is compatible with optimal long-run
performance.

Given a policy \(\pi\), let \(U_{\delta,\vartheta}^\pi\) denote the normalized
discounted payoff \(U_\delta^\pi\) for \(\delta<1\) conditional on the true
parameter \(\vartheta\). In the limit as \(\delta\to1\), we evaluate this payoff by
\[
 U_{1,\vartheta}^\pi=\liminf_{\delta\to1}U_{\delta,\vartheta}^\pi,
\]
which is the usual Abel limit when the limit exists. We also consider the
long-run average payoff
\[
 U_{\mathrm{av},\vartheta}^\pi=\liminf_{T\to\infty}\frac1TE_\vartheta^\pi
 \left[\sum_{t=0}^{T-1}U(x_{t+1},y_{t+1})\right].
\]
The connection between learning and these criteria is transparent in terms of
regret. For a fixed \(\vartheta\), define
\[
 R_T^\pi(\vartheta)=\sum_{t<T}E_\vartheta^\pi\left[\max_{x\in X}
 g^{\mathrm{full}}(x;\vartheta)-g^{\mathrm{full}}(x_{t+1};\vartheta)\right].
\]
In words, \(R_T^\pi(\vartheta)\) is the payoff loss relative to a decision maker
who knows the true parameter. A policy \(\pi\) admits \textit{consistent
parameter estimates} if there are random parameters \(\widehat\vartheta_t\)
with \(\widehat\vartheta_t\to\vartheta\) in
\(P^\pi_\vartheta\)-probability, and it has \textit{vanishing maximization
error in probability} if
\(\max_{x\in X}g^{\mathrm{full}}(x;\widehat\vartheta_t)
-g^{\mathrm{full}}(x_{t+1};\widehat\vartheta_t)\to0\) in
\(P^\pi_\vartheta\)-probability, that is, if its action is asymptotically
optimal for the estimated parameter. The following result links both criteria
to this cumulative expected regret. It uses only that \(X\) is nonempty and
compact, that \(U\) is Borel on \(X\times\R\) with
\(\lvert U(x,y)\rvert\le Ce^{c_U\lvert y\rvert}\) for constants \(C\) and
\(c_U\ge0\), that \(G\) is
continuous on \(X\times\R\), and that
\(x\mapsto E\lvert U(x,\alpha+\beta x+Z)\rvert\) is continuous on \(X\) for
every \((\alpha,\beta)\).
Assumption~\ref{ass:fl-payoff} implies these conditions, and so does every
\(U\) that is continuous on \(X\times\R\) with this bound, so the result also
holds when \(X\) is a finite set of actions. The proof is in
Supplement~\ref{supp:infinite}.

\begin{theorem}[Optimal learning under infinite patience]\label{thm:infinite}
Fix \(\vartheta\) and a policy \(\pi\). The following are equivalent:
\begin{enumerate}[label=\textup{(\roman*)},ref=(\roman*)]
\item \(R_T^\pi(\vartheta)=o(T)\);\label{thm:infinite:i}
\item \(U_{1,\vartheta}^\pi=\max_{x\in X}g^{\mathrm{full}}(x;\vartheta)\);\label{thm:infinite:ii}
\item \(U_{\mathrm{av},\vartheta}^\pi=\max_{x\in X}g^{\mathrm{full}}(x;\vartheta)\).\label{thm:infinite:iii}
\end{enumerate}
Every policy that admits consistent parameter estimates together with
vanishing maximization error in probability gives \(R_T^\pi(\vartheta)=o(T)\).
In particular, a policy whose actions converge in
\(P^\pi_\vartheta\)-probability to a (possibly random) maximizer of
\(g^{\mathrm{full}}(\cdot;\vartheta)\), as when complete learning is
accompanied by convergence to an optimal action, attains the optimal
infinite-patience and long-run average payoffs.
\end{theorem}

The theorem contrasts with our results for a fixed discount factor. When
\(\delta<1\), experimentation is costly, and Corollary~\ref{cor:suboptimal}
shows that, on the regular set, a policy that induces enough experimentation for
complete learning cannot be optimal. When the decision maker is infinitely
patient, experimentation may continue long enough to identify the true
parameter while imposing only \(o(T)\) cumulative payoff loss, which has no
effect on the limiting average payoff.

This contrast clarifies the relation to stochastic approximation, bandits, and
undiscounted reinforcement learning, where successful learning is often
evaluated by convergence or by sublinear cumulative loss or regret
\citep{lairobbins1982,lr1985,kaufmannkordamunos2012,jakschortnerauer2010}. The
theorem explains how those criteria correspond to infinite patience. Every
policy with sublinear regret attains the complete-information payoff when
\(\delta=1\). They do not characterize optimality at a fixed \(\delta<1\), when
the decision maker maximizes present discounted payoff and therefore cares when
experimentation costs occur. Within our model, for every fixed discount factor
below one, a strategy that achieves complete learning with positive
probability at regular parameters cannot be optimal. Thus a learning rule can be optimal under the long-run
criterion used in adjacent fields and yet be rejected by every finitely
discounted decision maker.

\section{Conclusion}\label{sec:conclusion}
When decisions generate the observations from which beliefs are updated,
optimizing the economic objective can prevent complete learning, even when the
decision maker values information. A decision maker can continue to predict
outcomes accurately near its chosen action while remaining incorrect about
responses elsewhere. The local theorem explains persistence near a selected
incorrect Berk--Nash equilibrium. The global theorem shows that incorrect
learning occupies the full prior mass of regular parameters under each fixed
optimal policy. Both rest on a smooth approximation of the optimal policy,
through which the mechanism identified for smooth decision rules in
\citet{kky2026parti} applies to forward-looking decision makers; at boundary
optima the global theorem also uses a separate argument.

The results survive forward-looking incentives, randomization among optimal
actions, and history-dependent policies. Infinite patience gives a different comparison. Experimentation with
sublinear cumulative payoff cost can learn and attain the complete-information
average payoff. The distinction is between the economic value of information
under a fixed discounted objective and the long-run statistical accumulation of
information. Supplement~\ref{supp:technologies} considers finite action sets and
fixed action-change costs, under which optimal actions eventually become
constant.

\appendix
\numberwithin{equation}{section}
\numberwithin{theorem}{section}
\renewcommand{\theproposition}{\thetheorem}
\renewcommand{\thelemma}{\thetheorem}
\renewcommand{\thecorollary}{\thetheorem}
\renewcommand{\theclaim}{\thetheorem}
\renewcommand{\thedefinition}{\thetheorem}
\renewcommand{\theassumption}{\thetheorem}
\renewcommand{\theauxproposition}{\thetheorem}
\renewcommand{\theauxlemma}{\thetheorem}
\renewcommand{\theremark}{\thetheorem}

\section{Value foundation and optimal support}\label{app:value}

Throughout the appendices, a \emph{policy} is a randomized, possibly
history-dependent policy as in Section~\ref{sec:policies}; we write \(\Pi\) for
it where the main text writes \(\pi\). Formally, it is a sequence of Borel maps
\(\psi_t\), \(t\ge0\), with values in \(X\), and iid Uniform\((0,1)\) seeds
\(\zeta_1,\zeta_2,\ldots\) such that \(\vartheta\), the noise sequence
\((\varepsilon_t)_{t\ge1}\) and the seed sequence \((\zeta_t)_{t\ge1}\) are
mutually independent:
\[
 \mathcal H_t=\sigma(\zeta_1,y_1,\ldots,\zeta_t,y_t),\qquad
 \mathcal F_t=\mathcal H_t\vee\sigma(\zeta_{t+1}),
\]
\[
 x_{t+1}=\psi_t(s_0,\zeta_1,y_1,\ldots,\zeta_t,y_t,\zeta_{t+1}).
\]
This is a Borel predictable policy in the sense of \PartIRemPredictable. The
state \(s_t\) is \(\mathcal H_t\)-measurable, and the action \(x_{t+1}\) is
\(\mathcal F_t\)-measurable. A \emph{Markov} policy is a Borel randomized
posterior-state Markov kernel \(\Pi(s,dx)\) from the standard Borel space
\(\mathcal S\) of feasible states to \(X\), realized by a jointly Borel map
\(\psi:\mathcal S\times[0,1]\to X\) as \(x_{t+1}=\psi(s_t,\zeta_{t+1})\); a
deterministic Borel policy \(x_{t+1}=\phi(s_t)\) is the Dirac kernel,
\(\Pi(s,\cdot)\) the point mass at \(\phi(s)\). Given the initial
feasible state \(s_0\), with prior \(\mu=\mu_{s_0}\), \(P^{\mu,\Pi}\) is the
augmented prior-predictive law and \(P^\Pi_\vartheta\) the fixed-truth law;
\(P^\Pi_{s}\) and \(E^\Pi_{s}\) denote the prior-predictive law and its
expectation when the policy and the recursions start from the feasible state
\(s\). By \PartILemPosterior{} and \PartIRemPredictable, under \(P^{\mu,\Pi}\)
the posterior of \(\vartheta\) given \(\mathcal F_t\) equals the posterior
given \(\mathcal H_t\) and is the Gaussian posterior \(\mu_{s_t}\), and the
recursions \eqref{eq:design} and \eqref{eq:means} hold with the realized
action.

\paragraph{Part I results for history-dependent policies.} Part~I states
its posterior, trapping and exclusion results for posterior-state Markov kernels. They
hold for every policy above, with \(P^{\mu,\Pi}\) and \(P^\Pi_\vartheta\) its
laws: \PartILemPosterior{} and \PartILemDisintegration{} by
\PartIRemPredictable; \PartIDefTracking, \PartILemFiniteHistory{} and
\PartILemTrapping{} by \PartIRemTrapping; and \PartILemAnchored,
\PartIDefPackage{} and \PartIThmExclusion{} by \PartIRemSeeded.
\PartILemEventualConstant{} is pathwise.

For a feasible state \(s\) and an action \(x\in X\), write
\(g_s(x)=E_s[U(x,\tilde\alpha+\tilde\beta x+Z)]\), where
\((\tilde\alpha,\tilde\beta)\sim\mu_s\) and \(Z\sim N(0,1)\) are independent, so
that \(g_t=g_{s_t}\). Write \(\mathcal T(s,x,Z)\) for the successor state given
by \eqref{eq:design} and \eqref{eq:means} after action \(x\) at \(s\) when the
standardized prediction error \(I/[1+u+w(x-\bar x)^2]^{1/2}\) equals
\(Z\sim N(0,1)\), and \(s^+=\mathcal T(s,x,Z)\).

Let \(R_X=\max_{x\in X}\lvert x\rvert\) and
\(\lambda=c_U\max\{1,R_X\}\). In the appendices and in
Supplement~\ref{supp:construction} we write, for \(\nu\ge0\),
\[
 \mathsf W_\nu(s)=E_s\bigl[e^{\nu(\lvert\tilde\alpha\rvert+\lvert\tilde\beta\rvert)}\bigr],
\]
\[
 \mathcal C_\nu=\{f\in C(\mathcal S):\lVert f\rVert_\nu<\infty\},
 \qquad
 \lVert f\rVert_\nu\coloneqq\sup_{s\in\mathcal S}\frac{\lvert f(s)\rvert}{\mathsf W_\nu(s)}.
\]
The weight \(\mathsf W_\nu\) is finite, continuous and at least one, and
\(\mathsf W_\nu\le\mathsf W_{\nu'}\) for \(\nu\le\nu'\), so
\(\mathcal C_\lambda\subseteq\mathcal C_\nu\) for \(\nu\ge\lambda\)
(Supplement~\ref{S1-section-1}).
The value is written without the normalization \(1-\delta\) of
Section~\ref{sec:policies}. For a policy \(\pi\) and a feasible state \(s\),
\(J^\pi(s)=E^\pi_s\sum_{t\ge0}\delta^tU(x_{t+1},y_{t+1})\), so that
\(U^\pi_\delta=(1-\delta)J^\pi(s_0)\). The factor \(1-\delta>0\) changes
neither the maximizers nor the Bellman argmax below.

\begin{proposition}[Value foundation]\label{prop:value}
This proposition requires only that \(X\) is nonempty and compact, that \(U\)
is Borel on \(X\times\R\) with the growth bound
\(\lvert U(x,y)\rvert\le Ce^{c_U\lvert y\rvert}\), with constants \(C\) and
\(c_U\ge0\), for \(x\in X\),
\(y\in\R\), and that \(G(x,m)=E[U(x,m+Z)]\) is continuous on \(X\times\R\). A policy is \emph{admissible} if it is Borel
predictable, that is, its date-\(t\) action is a Borel function of the initial
state, the seeds realized through date \(t\) and the outcomes observed before
date \(t\); it may be history dependent, and the seeds are fresh and
independent of the parameter and the noise; its value from \(s\) is
\(J^\pi(s)=E_s^\pi\sum_{t\ge0}\delta^tU(x_{t+1},y_{t+1})\). Write
\(\bar v(\vartheta)=\sup_{x\in X}E_Z\lvert U(x,\alpha+\beta x+Z)\rvert\) for
\(\vartheta=(\alpha,\beta)\in\R^2\), \(Z\sim N(0,1)\). Fix \(\delta\in[0,1)\).

There is a constant \(C\), independent of the policy, such that
\(\lvert g^{\mathrm{full}}(x;\vartheta)\rvert\le\bar v(\vartheta)
\le Ce^{\lambda(\lvert\alpha\rvert+\lvert\beta\rvert)}\) for every
\(\vartheta\in\R^2\) and \(x\in X\),
\(\lvert g_s(x)\rvert\le C\mathsf W_\lambda(s)\),
\(E[\mathsf W_\lambda(s^+)\mid s,x]=\mathsf W_\lambda(s)\), and, for every
admissible policy and every \(N\ge0\),
\begin{equation}
 E_s\sum_{t\ge N}\delta^t\lvert U(x_{t+1},y_{t+1})\rvert
 \le\frac{C\delta^N}{1-\delta}\mathsf W_\lambda(s).
 \label{A-eq:tail}
\end{equation}
The Bellman operator
\(\mathcal B_\delta f(s)=\max_{x\in X}\{g_s(x)+\delta E_Z[f(\mathcal T(s,x,Z))]\}\)
is a contraction on \(\mathcal C_\lambda\). Its fixed point
\(J_\delta\) is the optimal value over all admissible policies,
\(J_\delta(s)=\sup_\pi J^\pi(s)\). \(J_\delta\) is continuous,
\(Q_\delta(s,x)=g_s(x)+\delta E_Z[J_\delta(\mathcal T(s,x,Z))]\) is jointly
continuous, maximizers exist, and the smallest maximizer
\(\xi(s)=\min\arg\max_{x\in X}Q_\delta(s,x)\) is a Borel selection on all of
\(\mathcal S\). Moreover, for every \(s\in\mathcal S\) and \(x\in X\),
\[
 Q_\delta(s,x)=\sup\{J^\pi(s):\pi\text{ admissible with first action }x\},
\]
and the supremum is attained by every admissible policy with first action
\(x\) whose action \(x_{t+1}\) lies in \(\arg\max_{x'\in X}Q_\delta(s_t,x')\)
for every \(t\ge1\) almost surely, for instance by playing \(x\) and then
\(x_{t+1}=\xi(s_t)\); in particular the stationary policy
\(x_{t+1}=\xi(s_t)\), \(t\ge0\), attains \(J_\delta(s)\).
\end{proposition}

The proof is in Supplement~\ref{supp:construction}
(Section~\ref{S1-value-proof}); it applies \PartILemPosterior{} to admissible
policies, which is justified by \PartIRemPredictable.
The action set \(X=[\underline x,\overline x]\) is nonempty and compact, and
Assumption~\ref{ass:fl-payoff} implies the other
hypotheses of Proposition~\ref{prop:value}. Being continuous in \(x\) and
Borel in \(y\), \(U\) is a Carath\'eodory function, hence Borel on
\(X\times\R\) \citep[Lemma~4.51]{ab2006}. Its bound with \(j=0\) is the growth
bound. And \(G\) is continuous on \(X\times\R\). Indeed,
\[
 G(x,m)=\frac1{\sqrt{2\pi}}\int_\R U(x,y)e^{-(y-m)^2/2}\,dy,
\]
the integrand is continuous in \((x,m)\in X\times\R\) for every \(y\), because
\(U(\cdot,y)\) is continuous on \(X\), and, since
\((y-m)^2\ge y^2/2-m^2\), it is bounded for \(\lvert m\rvert\le M\) by the
integrable function \(Ce^{c_U\lvert y\rvert}e^{M^2/2-y^2/4}/\sqrt{2\pi}\);
dominated convergence gives the continuity.

Put
\[
 \Gamma_\delta(s)=\arg\max_{x\in X}Q_\delta(s,x),
\]
the \emph{Bellman first-action argmax}. At \(\delta=0\),
\(J_0(s)=\max_{x\in X}g_s(x)\) and \(\Gamma_0(s)=\arg\max_{x\in X}g_s(x)\) is the
myopic argmax. By Proposition~\ref{prop:value}, \(J_\delta(s)=\max_{x\in
X}Q_\delta(s,x)\), so \(\Gamma_\delta(s)\) is nonempty and compact; its graph
\(\{(s,x):Q_\delta(s,x)=J_\delta(s)\}\) is closed, hence Borel; and
\(s\mapsto\Pi(s,\Gamma_\delta(s))\) is Borel for every Markov policy \(\Pi\).
Every policy as above is admissible in the sense of
Proposition~\ref{prop:value}. Its action is a Borel function of \(s_0\), the
seeds and the past outcomes by definition, and for a Markov policy because
\(s_t\) is a Borel function of \(s_0\), \(\zeta_1,\ldots,\zeta_t\) and
\(y_1,\ldots,y_t\).

\begin{definition}[Optimal policies]\label{A-def:optimal}
Fix \(\delta\in[0,1)\) and an initial feasible state \(s_0\), with prior
\(\mu=\mu_{s_0}\). A policy \(\Pi\) is \emph{optimal} for \(\mu\) if
\begin{equation}
 P^{\mu,\Pi}(x_{t+1}\in\Gamma_\delta(s_t)\ \text{for every }t\ge0)=1,
 \label{A-eq:on-path-support}
\end{equation}
that is, its action lies in the Bellman first-action argmax at the current
state at every date, almost surely. The policy may be randomized and history
dependent. We write \(\pi_\delta\) for an optimal policy. At
\(\delta=0\), \eqref{A-eq:on-path-support} is myopic optimality at every date. A
Borel selection \(\phi\) of \(\Gamma_\delta\), such as \(\xi\) of
Proposition~\ref{prop:value}, is optimal as the Dirac kernel at
\(\phi(s)\) for every prior, because its support property holds at
every state.
\end{definition}

For \(0<\delta<1\), the next lemma shows that a policy is optimal for \(\mu\) in
this sense if and only if it maximizes the discounted payoff
\(U^\pi_\delta\) of Section~\ref{sec:policies}, that is, if and only if
\(J^\Pi(s_0)=J_\delta(s_0)\); such policies exist by
Proposition~\ref{prop:value}. It also transfers the support property to every
fixed truth.

\begin{lemma}[Optimal support]\label{lem:support}
Assume the hypotheses of Proposition~\ref{prop:value}, and fix an initial
feasible state \(s_0\), with prior \(\mu=\mu_{s_0}\), and a policy \(\Pi\).
\begin{enumerate}[label=\textup{(\roman*)},ref=(\roman*)]
\item Let \(0<\delta<1\), and let \(J^\Pi_\delta=J^\Pi\) be the value of\label{lem:support:i}
\(\Pi\). Then
\begin{equation}
 J_\delta(s_0)-J_\delta^\Pi(s_0)
 =E_{s_0}^\Pi\sum_{t\ge0}\delta^t
 (J_\delta(s_t)-Q_\delta(s_t,x_{t+1})),
 \label{A-eq:residual}
\end{equation}
and every summand is nonnegative. Hence
\(J_\delta^\Pi(s_0)=J_\delta(s_0)\) if and only if \(\Pi\) is optimal for
\(\mu\), that is, \eqref{A-eq:on-path-support} holds. If \(\Pi\) is a Markov
policy and value equality holds from every feasible state, then
\(\Pi(s,\Gamma_\delta(s))=1\) for every feasible state; the converse also
holds.
\item Let \(0\le\delta<1\). If \(\Pi\) is optimal for \(\mu\), then for every\label{lem:support:ii}
\(\vartheta\in\R^2\),
\[
 P^\Pi_\vartheta\{x_{t+1}\in\Gamma_\delta(s_t)\ \text{for every }t\ge0\}=1.
\]
\end{enumerate}
\end{lemma}

\begin{proof}
\emph{Bounds at random states.} Let \(0<\delta<1\). By \eqref{A-eq:tail} with
\(N=0\), the value of every admissible policy, hence also \(J_\delta\), is
bounded in absolute value by \(C\mathsf W_\lambda/(1-\delta)\), and then
\(\lvert Q_\delta(s,x)\rvert\le C\mathsf W_\lambda(s)+\delta
C\mathsf W_\lambda(s)/(1-\delta)=C\mathsf W_\lambda(s)/(1-\delta)\). By
\PartILemPosterior, applied from the Gaussian prior \(\mu_{s_0}\), the
posterior given \(\mathcal F_t\) is \(\mu_{s_t}\), so
\(\mathsf W_\lambda(s_t)=E_{s_0}^\Pi[e^{\lambda(\lvert\alpha\rvert+\lvert\beta\rvert)}\mid\mathcal F_t]\)
and
\begin{equation}
 E_{s_0}^\Pi\mathsf W_\lambda(s_T)=\mathsf W_\lambda(s_0)\qquad(T\ge0).
 \label{A-eq:weight-conservation}
\end{equation}
Write \(U_{t+1}=U(x_{t+1},y_{t+1})\). Then \eqref{A-eq:weight-conservation}
and \eqref{A-eq:tail} with \(N=T\) give
\begin{equation}
 \delta^TE_{s_0}^\Pi\lvert J_\delta(s_T)\rvert
 +E_{s_0}^\Pi\sum_{t\ge T}\delta^t\lvert U_{t+1}\rvert
 \le\frac{2C\delta^T}{1-\delta}\mathsf W_\lambda(s_0)\to0.
 \label{A-eq:tail-random}
\end{equation}
These bounds concern the realized path from \(s_0\) only; they do not
restart the policy at \(s_T\), so they need no Markov property.

\emph{Part \ref{lem:support:i}.} The Bellman equation \(J_\delta=\max_{x\in X}Q_\delta(\cdot,x)\)
of Proposition~\ref{prop:value} gives the envelope \(J_\delta\ge Q_\delta\). By
\eqref{A-eq:weight-conservation} and the bounds above, all expectations below
are finite. For a finite horizon \(T\), insert the Bellman equation at each
date and telescope conditional expectations, using
\(E[U_{t+1}\mid\mathcal F_t]=g_{s_t}(x_{t+1})\) and
\(E[J_\delta(s_{t+1})\mid\mathcal F_t]=E_Z[J_\delta(\mathcal T(s_t,x_{t+1},Z))]\),
which follow from the conditional Gaussian law of the prediction error given
\(\mathcal F_t\) in \PartILemPosterior:
\[
 J_\delta(s_0)-E_{s_0}^\Pi\sum_{t=0}^{T-1}\delta^tU_{t+1}
 =E_{s_0}^\Pi\sum_{t=0}^{T-1}\delta^t
 (J_\delta(s_t)-Q_\delta(s_t,x_{t+1}))
 +\delta^TE_{s_0}^\Pi J_\delta(s_T).
\]
The Bellman envelope makes every residual nonnegative. As \(T\to\infty\), the
left side tends to \(J_\delta(s_0)-J_\delta^\Pi(s_0)\) and the last term to
zero, both by \eqref{A-eq:tail-random}, and monotone convergence of the
nonnegative residual sum gives \eqref{A-eq:residual}. A residual is zero if and only if the action lies in
the full Bellman argmax, including at tied states. If the left side of
\eqref{A-eq:residual} is zero, the expectation of the nonnegative sum is zero,
so each residual vanishes almost surely; intersecting the countably many
datewise probability-one events gives \eqref{A-eq:on-path-support}.
Conversely, \eqref{A-eq:on-path-support} makes every summand of
\eqref{A-eq:residual} zero almost surely, so the right side is zero. For a
Markov policy, the date-\(0\) residual from a state \(s\) vanishes almost
surely exactly when \(\Pi(s,\Gamma_\delta(s))=1\), so starting the same
argument from each feasible state gives statewise support.
Conversely, statewise support makes every residual zero, and the tail argument
gives value equality.

\emph{Part \ref{lem:support:ii}.} For each \(t\), the event
\(\{x_{t+1}\in\Gamma_\delta(s_t)\}\) is the event that
\((s_t,x_{t+1})\) lies in the Borel graph of \(\Gamma_\delta\). It is
determined by \(s_0\), the seeds \(\zeta_1,\ldots,\zeta_{t+1}\) and the
outcomes \(y_1,\ldots,y_t\), and the graph does not depend on the truth. Thus
\eqref{A-eq:on-path-support} is a countable collection of truth-independent
observable finite-date support properties that holds \(P^{\mu,\Pi}\)-almost
surely, and \PartILemFiniteHistory\ transfers it to \(P^\Pi_\vartheta\) for
every \(\vartheta\).
\end{proof}

\section{Proof of the transfer theorem}\label{app:transfer}

Throughout this appendix the nondegenerate Gaussian prior \(\mu\) and the
policy \(\Pi\), randomized and possibly history dependent as in
Appendix~\ref{app:value}, are fixed. All statements, including
their exceptional null sets, are for this prior and this policy.

By \eqref{eq:design}, \(V_t\) is nondecreasing, so
\(V_\infty=\lim_tV_t\in(0,\infty]\) exists and \(\{V_t\to\infty\}=\{V_\infty=\infty\}\).
On the Borel event that \(b_t\) converges, write \(b_\infty\) for its limit, and
set \(b_\infty=0\) on the complement. The incorrect-learning event
\[
 \mathcal W=\{V_\infty<\infty,\ b_t\to b_\infty\ne\beta\}
\]
is a joint Borel event in the parameter and the path, since every posterior
quantity is a Borel function of the path and \(\mathcal W\) also involves
\(\beta\). For a joint Borel event \(E\), \(P^\Pi_\vartheta(E)\) denotes the
fixed-truth probability of its \(\vartheta\)-section, in which \(\beta\) is the
slope of \(\vartheta\).

\begin{lemma}[From exclusion to incorrect learning]\label{C-lem:setwise}
Let \(R\subseteq\R^2\) be Borel, and suppose that
\[
 P^{\mu,\Pi}(V_\infty=\infty,\ \vartheta\in R)=0.
\]
Then:
\begin{enumerate}[label=\textup{(\alph*)},ref=(\alph*)]
\item \(P^{\mu,\Pi}(\mathcal W,\ \vartheta\in B)=\mu(B)\) for every Borel\label{C-lem:setwise:a}
\(B\subseteq R\);
\item \(P^\Pi_\vartheta(\mathcal W)=1\) for \(\mu\)-almost every, equivalently\label{C-lem:setwise:b}
Lebesgue-almost every, \(\vartheta\in R\), with an exceptional null set
determined by the fixed \(\mu\) and \(\Pi\).
\end{enumerate}
\end{lemma}

\begin{proof}
\ref{C-lem:setwise:a} Let \(B\subseteq R\) be Borel. The hypothesis gives
\(P^{\mu,\Pi}(V_\infty=\infty,\ \vartheta\in B)=0\). The marginal law of
\(\vartheta\) under \(P^{\mu,\Pi}\) is \(\mu\), so
\begin{equation}
 P^{\mu,\Pi}(V_\infty<\infty,\ \vartheta\in B)
 =\mu(B)-P^{\mu,\Pi}(V_\infty=\infty,\ \vartheta\in B)=\mu(B).
 \label{C-eq:setwise-identity}
\end{equation}
The event \(\mathcal W\) is contained in \(\{V_\infty<\infty\}\), and
\(\{V_\infty<\infty\}\setminus\mathcal W\) is contained in the union of the
event that \(b_t\) does not converge and the event
\(\{V_\infty<\infty,\ b_\infty=\beta\}\). By \PartILemPosterior, the posterior
means converge \(P^{\mu,\Pi}\)-almost surely and
\(P^{\mu,\Pi}(V_\infty<\infty,\ b_\infty=\beta)=0\). Hence \(\mathcal W\)
differs from \(\{V_\infty<\infty\}\) by a \(P^{\mu,\Pi}\)-null set, and
\eqref{C-eq:setwise-identity} gives \ref{C-lem:setwise:a}.

\parstartref{C-lem:setwise:b} Take \(B=R\) in \ref{C-lem:setwise:a}. The joint Borel event
\(E=\{\vartheta\in R\}\setminus\mathcal W\) satisfies
\(P^{\mu,\Pi}(E)=\mu(R)-P^{\mu,\Pi}(\mathcal W,\ \vartheta\in R)=0\). By
\PartILemDisintegration, \(\vartheta\mapsto P^\Pi_\vartheta(E_\vartheta)\) is
Borel and \(P^\Pi_\vartheta(E_\vartheta)=0\) for \(\mu\)-almost every
\(\vartheta\). For \(\vartheta\in R\), the section \(E_\vartheta\) is the
complement of the section of \(\mathcal W\), so \(P^\Pi_\vartheta(\mathcal W)=1\)
for \(\mu\)-almost every \(\vartheta\in R\). A nondegenerate Gaussian prior has
a strictly positive Lebesgue density, so the \(\mu\)-null Borel sets are the
Lebesgue-null Borel sets (\PartILemDisintegration), and ``\(\mu\)-almost
every'' may be replaced by ``Lebesgue-almost every''.
\end{proof}

\begin{proof}[Proof of Theorem~\ref{thm:transfer}]
\emph{Part \ref{thm:transfer:i}.} Part~\ref{thm:transfer:i} is \PartILemTrapping, which Part~I states for every
Borel randomized posterior-state Markov kernel that tracks the smooth map,
optimal or not, for a Borel map \(\varphi\), and, by \PartIRemTrapping, for
every policy of Appendix~\ref{app:value} that tracks \(\varphi\).  Its hypotheses on the truth,
the target and the constants are those of part~\ref{thm:transfer:i}.  On every collar with
\(V^e>0\), the \(C^{2,1}\) bound of part~\ref{thm:transfer:i} implies its hypothesis (a) on
\(\varphi\), by the sentence after that lemma, and the other hypotheses on
\(\varphi\) are those of part~\ref{thm:transfer:i}, except that \(\varphi\) is Borel there.
This is no restriction. Replace \(V_*\)
by \(V_*'=\max\{V_*,1\}\) and \(\varphi\) by the map \(\varphi'\) equal to
\(\varphi\) on \(\mathcal C_*'=\bigcup_{V^e\ge V_*'}\mathcal C(V^e)\) and to
\(0\) elsewhere.  Since \(V_*'>0\), each point of the Borel set
\(\mathcal C_*'\) has a relative neighborhood in a single collar, so
\(\varphi\) is continuous on \(\mathcal C_*'\) and \(\varphi'\) is Borel; it satisfies the collar hypotheses
for every \(V^e\ge V_*'\); every policy that tracks \(\varphi\) on
\(\mathcal C_*\) tracks \(\varphi'\) on \(\mathcal C_*'\subseteq\mathcal C_*\)
with the same constant; and the threshold \(\overline V_r\) obtained for
\(\varphi'\) is at least \(V_*'\ge V_*\).  The policy hypothesis of
\PartILemTrapping,
\PartIDefTracking, is the tracking condition \eqref{eq:tracking} imposed
\(P^{\mu,\Pi}\)-almost surely; within \PartILemTrapping\ it is transferred to
the fixed-truth law by \PartILemFiniteHistory.  Its conclusions for
\(\varphi'\), including the openness of the set of priors and the independence of the thresholds
from the prior and the policy, are those of part~\ref{thm:transfer:i}, at
every radius \(r<\lvert b^e-\beta\rvert/2\) when \(b^e\ne\beta\), with
\(\lvert b_\infty-\beta\rvert\ge\lvert b^e-\beta\rvert-r>0\) in place of (c).  For
a general \(r\in(0,r_*)\), put \(r'\coloneqq r\) if \(b^e=\beta\) and
\(r'\coloneqq\min\{r,\lvert b^e-\beta\rvert/4\}\) otherwise, apply the lemma at
\(r'\), and put \(\overline V_r\coloneqq\overline V_{r'}\).  Since \(r'\le r\),
the event it gives satisfies (a) and (b) at radius \(r\), and
\(\lvert b_\infty-\beta\rvert\ge\lvert b^e-\beta\rvert-r'>0\) gives (c).

\emph{Part \ref{thm:transfer:ii}.} Suppose that \(\mu\) and \(\Pi\) admit a coherent exclusion
package on the Borel set \(\Theta_0\) (\PartIDefPackage). \PartIThmExclusion,
stated in Part~I for every Borel randomized posterior-state Markov kernel and,
by \PartIRemSeeded, for every Borel predictable policy with iid uniform seeds,
gives
\[
 P^{\mu,\Pi}(V_t\to\infty,\ \vartheta\in\Theta_0)=0,
\]
which is the first conclusion of part~\ref{thm:transfer:ii}. Since
\(\{V_t\to\infty\}=\{V_\infty=\infty\}\), Lemma~\ref{C-lem:setwise} applies
with \(R=\Theta_0\), so its item~\ref{C-lem:setwise:a} gives
\(P^{\mu,\Pi}(\mathcal W,\ \vartheta\in B)=\mu(B)\) for every Borel
\(B\subseteq\Theta_0\), and its item~\ref{C-lem:setwise:b} gives
\(P^\Pi_\vartheta(\mathcal W)=1\) for Lebesgue-almost every
\(\vartheta\in\Theta_0\), with an exceptional set determined by the fixed prior
and policy. If \(\mu(\Theta_0)=0\), then \(\Theta_0\) is Lebesgue-null and
the last two conclusions hold trivially.
\end{proof}

\section{Smooth approximation}\label{app:approximation}

This appendix proves Theorem~\ref{thm:approximation}.
Section~\ref{B-sec:scope} fixes the primitives and coordinates,
Section~\ref{B-sec:virtual} states the virtual-policy lemma, whose proof is
in Supplement Section~\ref{S1-section-4},
Section~\ref{B-sec:matching} derives exact matching and transversality,
Section~\ref{B-sec:tails} records the limits of complete learning under
optimal decisions and the localization data, and
Section~\ref{B-sec:proof} proves the two parts of the theorem.  The trapping
argument is \PartILemTrapping{} and the abstract exclusion theorem is
\PartIThmExclusion; neither is repeated here.

In this appendix \(\varphi\) denotes the \emph{virtual policy}, the smooth
map constructed in Lemma~\ref{B-lem:virtual}, an auxiliary object; the
decision maker uses a policy \(\Pi\).  Policies \(\Pi\) and their laws
\(P^{\mu,\Pi}\) and \(P^\Pi_\vartheta\) are as in
Appendix~\ref{app:value}.  The posterior state is
\(s=(a,b,u,\bar x,w)\), as in Section~\ref{sec:model}, and we use
\(A=u^{-1}\) and \(V=w^{-1}\) freely; the chart \eqref{B-eq:coordinates}
below replaces \(w\) by \(\ell_V=\log w\).

\subsection{Scope, primitives and coordinates}\label{B-sec:scope}

Throughout, assume Assumption~\ref{ass:fl-payoff}, that is,
\(U(\cdot,y)\) is \(C^4\) on \(X\) for every \(y\), \(U(x,\cdot)\) is Borel
for every \(x\), and
\(\lvert\partial_x^jU(x,y)\rvert\le Ce^{c_U\lvert y\rvert}\)
for \(x\in X\), \(y\in\R\) and \(j\le4\).  The construction
uses \(U\) only through the growth bound and its Gaussian expectations
\[
 G_{\mathrm{var}}(x,m,\tau_{\mathrm{var}})
 =E_Z\bigl[U\bigl(x,m+\sqrt{1+\tau_{\mathrm{var}}}\,Z\bigr)\bigr],
 \qquad x\in X,\ m\in\R,\ \tau_{\mathrm{var}}>-1,
\]
so that \(G(x,m)=G_{\mathrm{var}}(x,m,0)\).  By Supplement
Lemma~\ref{S1-lem:smoothing}, \(G_{\mathrm{var}}\) is \(C^4\) on
\(\operatorname{int}X\times\R\times(-1,\infty)\), with derivatives bounded
on every set on which \(m\) is bounded and \(\tau_{\mathrm{var}}\) lies in a
compact subset of \((-1,\infty)\); it satisfies
\(G_{\mathrm{var},\tau_{\mathrm{var}}}=\tfrac12G_{\mathrm{var},mm}\); and
\(\lvert G_{\mathrm{var}}(x,m,\tau_{\mathrm{var}})-G(x,m)\rvert
\le B\tau_{\mathrm{var}}\) for \(x\in X\), \(\lvert m\rvert\le R\) and
\(0\le\tau_{\mathrm{var}}\le3\), with \(B\) depending on \(R\).  By
Appendix~\ref{app:value}, \(G\) is continuous on \(X\times\R\).  Nothing
below differentiates \(U\) in the outcome or more than four times in the
action.  The target uniqueness, curvature and
outcome-relevance conditions are conditions~\ref{B-local:uniqueness}--\ref{B-local:relevance}
of Proposition~\ref{B-prop:local}; the truth localization of
Proposition~\ref{B-prop:global} imposes them on the truth through the set
\(\Theta^I\) of regular mean pairs defined below.
Each discount factor \(0\le\delta<1\) is fixed; at zero discount
an optimal policy is myopically optimal.

Write \(g^{\mathrm{full}}(x;(a,b))=G(x,a+bx)\).  At a regular interior
mean pair, let \(x^\star(a,b)\) be its unique maximizer and put
\(m_{\mathrm{out}}(a,b)=G_m(x^\star(a,b),a+bx^\star(a,b))\).
A mean pair \((a,b)\) is \emph{regular} if \(g^{\mathrm{full}}(\cdot;(a,b))\)
has a unique maximizer \(x^\star(a,b)\in\operatorname{int}X\) with
\(\partial_{xx}g^{\mathrm{full}}(x^\star(a,b);(a,b))<0\) and
\(m_{\mathrm{out}}(a,b)\ne0\); the set of regular mean pairs is the set
\(\Theta^I\subset\R^2\) of the main text.  The map
\(((a,b),x)\mapsto g^{\mathrm{full}}(x;(a,b))\) is continuous on
\(\R^2\times X\) and \(C^4\) on \(\R^2\times\operatorname{int}X\).  The set
of mean pairs with a
unique interior maximizer and strictly negative curvature is open, and
\(x^\star\) is \(C^3\) on it. Strict curvature persists near the maximizer,
uniqueness and compactness give a positive payoff gap away from it that
persists under small parameter changes, and the implicit-function theorem
gives the maximizer nearby.  Since \(m_{\mathrm{out}}\) is continuous there,
\(\Theta^I\) is open as well.  Outcome relevance, \(m_{\mathrm{out}}\ne0\),
enters first in Section~\ref{B-sec:matching} (transversality).
Fix a nonempty compact family \(K^{\mathrm{mean}}\) of mean pairs, each
with a unique interior maximizer and strictly negative curvature; a compact
regular mean family is one such family, and for a target
\((a^e,b^e,x^e)\) with reference slope precision \(V^e\), as in
Section~\ref{B-sec:proof}, take \(K^{\mathrm{mean}}=\{(a^e,b^e)\}\).

Put \(g_s(x)=E_s[U(x,\tilde\alpha+\tilde\beta x+Z)]\), with
independent standard Gaussian \(Z\), and let \(J_\delta\) be the optimal
discounted value over admissible policies.
Proposition~\ref{prop:value} supplies the value foundation.  The exact
updates of the posterior state are those of
\eqref{eq:design} and \eqref{eq:means}; for a randomized policy they hold
with the realized seed-measurable action, by \PartILemPosterior.
For a continuation \(F\) on posterior states, write
\(Q_\delta[F](s,x)=g_s(x)+\delta E_Z[F(\mathcal T(s,x,Z))]\),
where \(\mathcal T(s,x,Z)\) is the exact successor posterior state of
these recursions after action \(x\), written as a function of
the standardized prediction error \(Z\sim N(0,1)\), and
\(Q_\delta=Q_\delta[J_\delta]\).  The Bellman first-action argmax is
\(\Gamma_\delta(s)=\arg\max_{x\in X}Q_\delta(s,x)\).

Use independent coordinates \(z_\ell=(a,b,u,\bar x,\ell_V)\), with
\begin{equation}\label{B-eq:coordinates}
 \iota_\ell(s)=(a,b,A^{-1},\bar x,-\log V),\quad
 \iota_\epsilon=S_\epsilon\circ\iota_\ell,\quad
 S_\epsilon(a,b,u,\bar x,\ell_V)
 =(a,b,u/\epsilon_{\mathrm{sc}},\bar x,\ell_V).
\end{equation}
Thus \(\ell_V=-\log V=\log w\).
The scaled coordinates are \(z_\epsilon=(a,b,u_\epsilon,\bar x,\ell_V)\),
where \(u_\epsilon=u/\epsilon_{\mathrm{sc}}\).
On the current-state domain \(H=\R^2\times(-1,\infty)\times\R\times\R\),
put
\[
\begin{gathered}
 D=x-\bar x,\qquad \sigma_I^2=1+u+e^{\ell_V}D^2,\qquad
 \boldsymbol e_a=(1,0),\\
 \boldsymbol n(\bar x)=(-\bar x,1),\qquad
 \boldsymbol k_I=\frac{u\boldsymbol e_a+e^{\ell_V}D\boldsymbol n(\bar x)}{\sigma_I}.
\end{gathered}
\]
For \(x\in X\), the variance and denominators are positive.  With
\(\boldsymbol m=(a,b)\), the true-form transition is
\begin{equation}\label{B-eq:transition}
 \mathcal T^{[\ell]}(z_\ell,x,Z)
 =\left(\boldsymbol m+\boldsymbol k_I Z,\frac{u}{1+u},
 \ \
 \frac{\bar x+ux}{1+u},
 \ \
 \ell_V-\log\left(1+\frac{e^{\ell_V}D^2}{1+u}\right)\right).
\end{equation}
Its codomain is \(\R^5\); successors need not remain in \(H\) or a
matching neighborhood.  On feasible states it equals
\(\iota_\ell\circ\mathcal T\).  Define
\(\mathcal T^{[\epsilon]}=S_\epsilon\circ\mathcal T^{[\ell]}\circ S_\epsilon^{-1}\)
on its corresponding domain.  For globally bounded Borel \(F:\R^5\to\R\),
set
\[
\begin{aligned}
 g_{z_\ell}^{[\ell]}(x)&=E_Z[U(x,a+bx+\sigma_I Z)],\\
 Q_\delta^{[\ell]}[F](z_\ell,x)
 &=g_{z_\ell}^{[\ell]}(x)+\delta E_Z[F(\mathcal T^{[\ell]}(z_\ell,x,Z))].
\end{aligned}
\]
Here \(Z\) is standard normal.  On feasible states these expressions are
the posterior expected reward and continuation; off the feasible domain they
are defined by the same formulas.  Since \(\sigma_I^2=1+u+e^{\ell_V}D^2\),
\(g_{z_\ell}^{[\ell]}(x)=G_{\mathrm{var}}(x,a+bx,u+e^{\ell_V}D^2)\), which is
\(C^4\) in \((z_\ell,x)\) for \(z_\ell\in H\) and \(x\in\operatorname{int}X\).  The chart covers only \(e^{\ell_V}>0\).
The lemma's global regularity norms use \(z_\epsilon\).  Derivatives of
the true-form reward, transition and hybrid objective, and the calculations
in Section~\ref{B-sec:matching}, use the independent unscaled chart with
physical \(x\) fixed.  The abbreviation \(g_{z_\ell}=g_{z_\ell}^{[\ell]}\)
applies only in Sections~\ref{B-sec:virtual} and~\ref{B-sec:matching} and
in Supplement Section~\ref{S1-covariance-proof}; elsewhere, local and
physical-time conventions remain as stated.

\subsection{Virtual-policy lemma}\label{B-sec:virtual}

The following lemma collects the construction needed by both parts of
Theorem~\ref{thm:approximation}.  It supplies a smooth virtual policy
without assuming smoothness of an actual optimal policy.

\begin{auxlemma}[Virtual policy near regular interior means]
\label{B-lem:virtual}
Under Assumption~\ref{ass:fl-payoff}, for fixed \(\delta\in[0,1)\)
and a nonempty compact family \(K^{\mathrm{mean}}\) of mean pairs, each
with a unique maximizer \(x^\star(a,b)\in\operatorname{int}X\) and
\(\partial_{xx}g^{\mathrm{full}}(x^\star(a,b);(a,b))<0\), there are bounded open mean neighborhoods
\(U_{\mathrm{in}},U_{\mathrm{out}}\) and compact intervals
\(J_{\mathrm{in}},J_{\mathrm{out}}\) such that
\[
 K^{\mathrm{mean}}\subset U_{\mathrm{in}},\qquad
 \overline U_{\mathrm{in}}\subset U_{\mathrm{out}},\qquad
 X\subset J_{\mathrm{in}}\subset\operatorname{int}J_{\mathrm{out}},
\]
every mean pair in \(\overline U_{\mathrm{out}}\) has a unique interior
maximizer with strictly negative curvature, and
\(\overline U_{\mathrm{out}}\subset\Theta^I\) when
\(K^{\mathrm{mean}}\subset\Theta^I\).
There is a compact action interval
\(\mathcal X(a,b)\Subset\operatorname{int}X\) for each
\((a,b)\in U_{\mathrm{out}}\), containing \(x^\star(a,b)\) in its
interior.  These choices are independent of the scale.  For some
\(\epsilon_*>0\), all sufficiently small
\(0<\epsilon_{\mathrm{sc}}\le\epsilon_*\) admit the following objects
and conclusions.  In coordinate order \(((a,b),u,\bar x,\ell_V)\), put
\[
\begin{aligned}
 \mathcal M_\epsilon
 &=U_{\mathrm{out}}\times(-2\epsilon_{\mathrm{sc}},2\epsilon_{\mathrm{sc}})
   \times\operatorname{int}J_{\mathrm{out}}
   \times(-\infty,\log\epsilon_{\mathrm{sc}}],\\
 \mathcal I_\epsilon
 &=U_{\mathrm{in}}\times(0,\epsilon_{\mathrm{sc}}]
   \times J_{\mathrm{in}}\times(-\infty,\log\epsilon_{\mathrm{sc}}].
\end{aligned}
\]
The threshold satisfies \(\epsilon_*\le1/4\), so that
\(\mathcal I_\epsilon\subset\mathcal M_\epsilon\subset H\) and
\(u>-1/2\) on \(\mathcal M_\epsilon\).
The matching domain \(\mathcal M_\epsilon\) is two-sided in \(u\);
\(\mathcal I_\epsilon\) consists of feasible states.

\begin{enumerate}[label=\textup{(\roman*)},ref=\textup{(\roman*)}]
\item\label{B-virtual:regularity}
There are globally bounded maps
\(J_{\delta,\epsilon}^{\mathrm{aux}}:\R^5\to\R\) and
\(\varphi_\epsilon:\R^5\to X\), respectively \(C^{3,1}\) and
\(C^{2,1}\) in the scaled coordinates, with norms uniform over the
small-scale family.  Define
\[
 J^{\mathrm{aux},[\ell]}=J_{\delta,\epsilon}^{\mathrm{aux}}\circ S_\epsilon,
 \quad \varphi^{[\ell]}=\varphi_\epsilon\circ S_\epsilon,
 \quad J_\delta^{\mathrm{aux}}=J^{\mathrm{aux},[\ell]}\circ\iota_\ell,
 \quad \varphi=\varphi^{[\ell]}\circ\iota_\ell.
\]
Derivatives not involving \(u\) inherit the uniform bounds.  Conversion
to \(u\)-derivatives introduces fixed powers of
\(\epsilon_{\mathrm{sc}}^{-1}\); their bounds hold once the scale is
fixed, uniformly in a lower \(\ell_V\) cutoff.  The global
map is a fixed two-sided extension of its restriction to the matching
domain, with the same global regularity available on successor segments.

\item\label{B-virtual:hybrid}
Write \(Q^{[\ell]}=Q_\delta^{[\ell]}[J^{\mathrm{aux},[\ell]}]\).
For \(z_\ell\in\mathcal M_\epsilon\),
\[
 J^{\mathrm{aux},[\ell]}(z_\ell)
 =\max_{x\in\mathcal X(a,b)}Q^{[\ell]}(z_\ell,x),\quad
 (Q^{[\ell]})_{xx}(z_\ell,x)\le-c_Q<0
 \quad(x\in\mathcal X(a,b)),
\]
where \(c_Q\) is independent of the small scale and lower tail cutoff.
The unique maximizer is \(\varphi^{[\ell]}(z_\ell)\), interior to
\(\mathcal X(a,b)\).  The expectation evaluates the global auxiliary
continuation for every Gaussian realization, including successors outside
\(\mathcal M_\epsilon\).

\item\label{B-virtual:tracking}
After fixing the scale, for every feasible \(s\) with
\(\iota_\ell(s)\in\mathcal I_\epsilon\) and every
\(x\in\arg\max_{x'\in X}Q_\delta(s,x')\),
\begin{equation}\label{B-eq:virtual-tracking}
\left.
\begin{aligned}
 \lvert J_\delta(s)-J_\delta^{\mathrm{aux}}(s)\rvert
 &\le C\bigl(A^{-1}+\lvert\varphi(s)-\bar x\rvert\bigr)^4,\\
 \lvert x-\varphi(s)\rvert
 &\le C\bigl(A^{-1}+\lvert\varphi(s)-\bar x\rvert\bigr)^2.
\end{aligned}
\quad
\right\}
\end{equation}
The constant is determined by the fixed construction and is the same for
every lower \(\ell_V\) cutoff and every exact maximizer.  At \(\delta=0\),
the virtual policy maximizes the current expected payoff.
\end{enumerate}
\end{auxlemma}

The proof is given at the end of Supplement
Section~\ref{S1-section-4}, using the constructions and estimates
developed in Supplement Sections~\ref{S1-section-1}--\ref{S1-section-4}.
The bound controls every exact optimal
action.\footnote{\citet{struloviciszydlowski2015} approximate the optimum by
piecewise-constant controls; here the virtual policy is smooth, and the bound
controls every exact optimal action.}
Choose the small-scale family using the fixed-point, gap and tracking
estimates.  The matching and transversality arguments below may then select
a smaller member and narrower graph and precision neighborhoods.
All constants are for the fixed discount factor and the fixed family
\(K^{\mathrm{mean}}\).

\subsection{Exact matching and covariance derivative cancellations}
\label{B-sec:matching}

In this subsection use the independent unscaled
coordinates in \eqref{B-eq:coordinates} and
\(Q^{[\ell]}=Q_\delta^{[\ell]}[J^{\mathrm{aux},[\ell]}]\) supplied by
Lemma~\ref{B-lem:virtual}; abbreviate \(g_{z_\ell}=g_{z_\ell}^{[\ell]}\).
Derivatives hold physical action \(x\) fixed.

\subsubsection{Covariance derivative bounds}

Recall from Section~\ref{B-sec:scope} that
\(g_{z_\ell}(x)=G_{\mathrm{var}}(x,a+bx,u+e^{\ell_V}D^2)\) and
\(G_{\mathrm{var},\tau_{\mathrm{var}}}=\tfrac12G_{\mathrm{var},mm}\).

\begin{auxlemma}[Covariance derivative bounds]\label{B-lem:covariance-bounds}
Use the extended hybrid objective and the independent unscaled chart just
defined, on the two-sided matching domain \(\mathcal M_\epsilon\) of
Lemma~\ref{B-lem:virtual}, or a smaller bounded-coordinate
neighborhood and lower tail inside it, with local actions in
\(\mathcal X(a,b)\).  Fix the discount factor and construction scale.
Then the virtual policy satisfies
\[
\lvert\varphi^{[\ell]}_{\bar x}\rvert\le C(\lvert u\rvert+e^{\ell_V}),\qquad
\lvert\varphi^{[\ell]}_{\ell_V}\rvert
\le C(\lvert u\rvert+e^{\ell_V}\lvert D^\varphi\rvert),
\]
where \(D^\varphi=\varphi^{[\ell]}-\bar x\), with constants independent of
a lower \(\ell_V\)-cutoff.
\end{auxlemma}

The proof is in Supplement Section~\ref{S1-covariance-proof}.  The
zero-set matching and transversality arguments follow here.

\subsubsection{Zero-set matching}

Consider
\(u=0\), \(\bar x=x_0=x^\star(a,b)\), and \(e^{\ell_V}>0\)
inside \(\mathcal M_\epsilon\), with \(x_0\) interior to
\(\mathcal X(a,b)\).  At action \(x_0\), the successor is the current state
for every \(Z\), and
\(\mathcal T^{[\ell]}_x=e^{\ell_V}(-x_0,1,0,0,0)Z\).
Thus
\((Q^{[\ell]})_x(z_\ell,x_0)=\partial_xg_{z_\ell}(x_0)=0\).
Strict concavity gives
\begin{equation}\label{B-eq:matching}
\varphi^{[\ell]}(z_\ell)=x_0,\qquad
J^{\mathrm{aux},[\ell]}(z_\ell)=\frac{G(x_0,a+bx_0)}{1-\delta}.
\end{equation}
In particular, \(J^{\mathrm{aux},[\ell]}_{\ell_V}=0\) on this zero set.

Put
\[
 m_{\mathrm{out}}^e\coloneqq G_m(x^e,a^e+b^ex^e).
\]
At the target reference
\(z_\ell^e=(a^e,b^e,0,x^e,\log(1/V^e))\),
we therefore have
\[
(Q^{[\ell]})_{xb}-x^e(Q^{[\ell]})_{xa}=m_{\mathrm{out}}^e.
\]
Moreover,
\begin{equation}\label{B-eq:matching-curvature}
(Q^{[\ell]})_{xx}
=\partial_{xx}g^{\mathrm{full}}(x^e;(a^e,b^e))+e^{\ell_V}G_{mm}
+\delta e^{2\ell_V}(\partial_b-x^e\partial_a)^2J^{\mathrm{aux},[\ell]}.
\end{equation}
In deriving \eqref{B-eq:matching-curvature}, the term \(e^{\ell_V}G_{mm}\) is
\(2e^{\ell_V}G_{\mathrm{var},\tau_{\mathrm{var}}}\) at \(\tau_{\mathrm{var}}=0\),
and the \(\ell_V\)-component of
\(\mathcal T^{[\ell]}_{xx}\) is \(-2e^{\ell_V}\). Its contribution
\(\delta(-2e^{\ell_V})J^{\mathrm{aux},[\ell]}_{\ell_V}\) vanishes by the
zero-set identity.

By \eqref{B-eq:matching}, \(\varphi^{[\ell]}(z_\ell^e)=x^e\). We differentiate
the identity \((Q^{[\ell]})_x(z_\ell,\varphi^{[\ell]}(z_\ell))=0\) in \(a\)
and in \(b\) at \(z_\ell^e\), where \(e^{\ell_V}=1/V^e\). With the two
displays above, this gives the exact identity
\[
\varphi^{[\ell]}_b-x^e\varphi^{[\ell]}_a
=-\frac{m_{\mathrm{out}}^e}{\hat Q+\Delta},
\qquad
\Delta=\frac{G_{mm}}{V^e}
+\frac{\delta}{(V^e)^2}(\partial_b-x^e\partial_a)^2J^{\mathrm{aux},[\ell]},
\]
with \(\hat Q=\partial_{xx}g^{\mathrm{full}}(x^e;(a^e,b^e))\), and \(G_{mm}\) and
the derivative of \(J^{\mathrm{aux},[\ell]}\) evaluated at
\((x^e,a^e+b^ex^e)\) and \(z_\ell^e\). Let \(C_G\) bound \(\lvert G_{mm}\rvert\) at the
target and \(C_J\) bound \(\lvert(\partial_b-x^e\partial_a)^2J^{\mathrm{aux},[\ell]}\rvert\)
on the matching domain, a mean-coordinate second derivative covered by
item~\ref{B-virtual:regularity} of Lemma~\ref{B-lem:virtual}. For
\(V^e\ge\max\{1,2(C_G+C_J)/\lvert\hat Q\rvert\}\), \(\lvert\Delta\rvert\le(C_G+C_J)/V^e\le\lvert\hat Q\rvert/2\),
so
\[
\left\lvert\varphi^{[\ell]}_b-x^e\varphi^{[\ell]}_a
+\frac{m_{\mathrm{out}}^e}{\hat Q}\right\rvert
\le\frac{2\lvert m_{\mathrm{out}}^e\rvert(C_G+C_J)}{\hat Q^2}\cdot\frac1{V^e}.
\]
This is matching at the zero-\(u\) reference.

\subsubsection{Transversality and extension}

For a local action in \(\mathcal M_\epsilon\), with
\(\bar x^+=(\bar x+ux)/(1+u)\),
\[
\partial_{xb}g_{z_\ell}-\bar x^+\partial_{xa}g_{z_\ell}
=G_{\mathrm{var},m}+(x-\bar x^+)
  \left(G_{\mathrm{var},xm}+bG_{\mathrm{var},mm}
        +2e^{\ell_V}DG_{\mathrm{var},\tau_{\mathrm{var}}m}\right).
\]
Since
\(x-\bar x^+=D/(1+u)\) and the derivatives of \(G_{\mathrm{var}}\) are
continuous and locally bounded (Section~\ref{B-sec:scope}),
this expression converges uniformly to \(m_{\mathrm{out}}(a,b)\) on a
shrinking neighborhood of the regular truth graph.  The continuation
contribution is \(O(\lvert u\rvert+e^{\ell_V})\), and
at the virtual action, using
\(\lvert\varphi^{[\ell]}-x^\star\rvert\le C(\lvert u\rvert+e^{\ell_V})\),
for which the argument of Supplement Section~\ref{S1-section-4} applies
verbatim on the two-sided domain,
\((Q^{[\ell]})_{xx}\bigl(z_\ell,\varphi^{[\ell]}(z_\ell)\bigr)
=\partial_{xx}g^{\mathrm{full}}(x^\star(a,b);(a,b))
+O(\lvert u\rvert+e^{\ell_V})\).
Choose a fixed two-sided graph neighborhood inside \(\mathcal M_\epsilon\),
with \(\lvert\bar x-x^\star(a,b)\rvert<\nu\),
and then a sufficiently small precision threshold on a lower tail.
Compactness and
\(\inf_{K^{\mathrm{mean}}}\lvert m_{\mathrm{out}}/
\partial_{xx}g^{\mathrm{full}}(x^\star(a,b);(a,b))\rvert>0\) give
\[
\lvert\varphi^{[\ell]}_b-\bar x^{+,\varphi}\varphi^{[\ell]}_a\rvert
\ge\lambda_0>0,
\]
where \(\bar x^{+,\varphi}=(\bar x+u\varphi^{[\ell]})/(1+u)\) is the
successor average action at the virtual action.

Lemma~\ref{B-lem:virtual} supplies a global \(C^{2,1}\)
virtual policy under Assumption~\ref{ass:fl-payoff}.  Use that same global map as
the fixed two-sided extension of its restriction to the matching domain.
Its global bounds apply on every successor segment hull.
After the construction scale is fixed, let \(C_1^\varphi\) bound the first
derivatives and \(C_2^\varphi\) bound the second derivatives of the extended
virtual policy.  Splitting distances at one gives
\(\lvert\mathrm D\varphi^{[\ell]}(z)-\mathrm D\varphi^{[\ell]}(z')\rvert
\le\max\{C_2^\varphi,2C_1^\varphi\}\lvert z-z'\rvert^\eta\) for
\(0<\eta<1\).
Thus every fixed \(\eta\in(6/7,1)\) is available on the entire
lower-\(\ell_V\) tail, and first derivatives are uniformly Lipschitz in
\(\ell_V\).  The constants are determined by the fixed scale and are the
same for every lower tail cutoff.

\subsection{Coherent tails and global localization}\label{B-sec:tails}

Throughout this subsection, \(\mu\) denotes the fixed nondegenerate
Gaussian initial prior of the unknown coefficients
\(\vartheta=(\alpha,\beta)\). It is the Gaussian posterior \(\mu_{s_0}\) at
the fixed initial feasible posterior state \(s_0\) from which the posterior
recursions start, and under the prior-predictive laws used below
\(\vartheta\) has law \(\mu\).  For a truth \(\vartheta\), write
\(g^{\mathrm{full}}(x;\vartheta)=G(x,\alpha+\beta x)\).  A truth is
\emph{regular} if, as a mean pair, it lies in \(\Theta^I\); for
\(\vartheta\in\Theta^I\), \(x^\star(\vartheta)\in\operatorname{int}X\) is
the unique global maximizer of \(g^{\mathrm{full}}(\cdot;\vartheta)\).

Lemma~\ref{B-lem:virtual} supplies, for a compact regular mean family, one
virtual policy on \(\ell_V\le\log\epsilon_{\mathrm{sc}}\).  It never
switches policies as \(V\) grows.  By openness, every regular truth has a
compact neighborhood \(K^{\mathrm{mean}}\subset\Theta^I\) of mean pairs,
with the truth in its interior, that is a compact regular mean family in
the sense of Section~\ref{B-sec:scope}.

\subsubsection{Complete learning under optimal decisions}

\begin{auxlemma}[Limits of complete learning under optimal decisions]
\label{B-lem:optimal-collapse}
Fix \(\delta\in[0,1)\) and a policy \(\Pi\), randomized and possibly
history dependent as in Appendix~\ref{app:value}, and suppose that
\(x_{t+1}\in\Gamma_\delta(s_t)\) for every \(t\ge0\),
\(P^{\mu,\Pi}\)-almost surely.  Then, \(P^{\mu,\Pi}\)-almost surely on
\(\{V_t\to\infty,\ \vartheta\in\Theta^I\}\),
\begin{equation}\label{B-eq:collapse-limits}
 (a_t,b_t,A_t^{-1},\bar x_t,V_t^{-1})
 \to(\alpha,\beta,0,x^\star(\vartheta),0),
 \qquad
 x_{t+1}\to x^\star(\vartheta),
 \qquad
 V_t/A_t\to0.
\end{equation}
\end{auxlemma}

\begin{proof}
On \(\{V_t\to\infty\}\), \PartILemPosterior(c) gives
\((a_t,b_t,\Sigma_t)\to(\alpha,\beta,0)\) almost surely under
\(P^{\mu,\Pi}\).  Let
\(g_*^{\mathrm{full}}(\vartheta)=\max_{x\in X}g^{\mathrm{full}}(x;\vartheta)\)
and \(g_s^*=\max_{x\in X}g_s(x)\).
A constant action maximizing \(g_s(\cdot)\) gives
\(J_\delta(s)\ge g_s^*/(1-\delta)\).
Perfect information gives the upper bound
\(J_\delta(s)\le E_s[g_*^{\mathrm{full}}(\tilde\vartheta)]/(1-\delta)\).
Here \(\tilde\vartheta\sim\mu_s\) is a posterior draw.
Therefore every optimal first action satisfies
\begin{equation}\label{B-eq:optimal-gap}
g_s^*-g_s(x)
\le\frac{\delta}{1-\delta}
   \bigl(E_s[g_*^{\mathrm{full}}(\tilde\vartheta)]-g_s^*\bigr).
\end{equation}
At \(\delta=0\) the right side is zero, so every optimal first action is a
myopic maximizer.  On covariance collapse with means converging to the
truth, the joint continuity of \(g^{\mathrm{full}}\) on \(\R^2\times X\),
which follows from that of \(G\) (Appendix~\ref{app:value}), Gaussian
exponential-moment bounds and the exponential growth of
\(g^{\mathrm{full}}\) and \(\bar v\) in Proposition~\ref{prop:value} imply
uniform convergence of \(g_s(\cdot)\) to \(g^{\mathrm{full}}(\cdot;\vartheta)\)
on \(X\), and the right side of \eqref{B-eq:optimal-gap} tends to zero.
For a regular truth, the unique maximizer therefore gives
\(x_{t+1}\to x^\star(\vartheta)\) for every realized optimal action.
\PartILemAnchored{} then gives \(\bar x_t\to x^\star(\vartheta)\),
\(D_t\to0\), and \(V_t/A_t\to0\) on the same event.  Together these give
\eqref{B-eq:collapse-limits}.
\end{proof}

\subsubsection{Localization data}\label{B-sec:localization}

We now fix the deterministic data of the coherent exclusion package of
\PartIDefPackage, clauses (P1)--(P5), in its symbols
\(\mathcal B_m^{\rm in}\Subset\mathcal B_m^{\rm mid}\),
\(\ell_m^{\rm in}<\ell_m^{\rm out}\), \(\mathcal S_m^{\mathrm{tail}}\),
\(\mathcal K_m^0\), \(\mathcal K_m\), and \(\varphi_m\), with
\(z_b=(a,b,u,\bar x)\) and \(\ell_V=\log w\) as there.  Around each
regular truth, first obtain the virtual policy from
Lemma~\ref{B-lem:virtual} on an enlarged compact regular
mean neighborhood.  The matching and transversality neighborhoods of
Section~\ref{B-sec:matching}, inside \(\mathcal M_\epsilon\), are open in
the bounded coordinates \(z_b=(a,b,u,\bar x)\), uniformly on a sufficiently
low \(\ell_V\)-tail.  Shrink the truth neighborhood so that the graph
\(\vartheta\mapsto(\alpha,\beta,0,x^\star(\vartheta))\) fits inside one
bounded rectangle \(\mathcal B_m^{\rm in}\) of the form required by (P1), a
product of open intervals in \((a,b,\bar x)\) with an interval
\([0,u_m^{\rm in})\) in \(u\), whose closure lies in a second such
rectangle \(\mathcal B_m^{\rm mid}\), with
\(\overline{\mathcal B}_m^{\rm mid}\times(-\infty,\ell_m^{\rm out}]\) inside
those neighborhoods for some \(\ell_m^{\rm out}\), and with the feasible
part of \(\mathcal B_m^{\rm mid}\times(-\infty,\ell_m^{\rm out})\) inside
\(\mathcal I_\epsilon\).  This is possible by continuity of \(x^\star\).
Choose compact truth rectangles \(K_m^{\mathrm{truth}}\) in these
neighborhoods with interiors covering \(\Theta^I\), and take a
countable subcover.  Fix the associated data in advance:
\[
 \left\{(\alpha,\beta,0,x^\star(\vartheta)):\vartheta\in K_m^{\mathrm{truth}}\right\}
 \subset\mathcal B_m^{\rm in}\Subset\mathcal B_m^{\rm mid},
 \qquad\ell_m^{\rm in}<\ell_m^{\rm out},
\]
\(\mathcal S_m^{\mathrm{tail}}=\mathcal B_m^{\rm mid}\times(-\infty,\ell_m^{\rm out})\)
and
\(\mathcal K_m^0=\overline{\mathcal B}_m^{\rm in}\times(-\infty,\ell_m^{\rm in}]\);
all bounds below are uniform on the enlarged compact mean neighborhood.
Set \(\Theta_m=K_m^{\mathrm{truth}}\), a compact subset of \(\Theta^I\), so
that \(\Theta^I=\bigcup_m\Theta_m\), and let \(\varphi_m\) be the one fixed
virtual policy restricted to \(\mathcal S_m^{\mathrm{tail}}\).  These data
are fixed before the policy, the truth and the path.

\subsection{Proof of Theorem~\ref{thm:approximation}}\label{B-sec:proof}

\paragraph{Eligible policies.}
Fix a discount factor \(\delta\in[0,1)\).  Given an initial prior, a policy
is \emph{eligible} for it, that is, optimal for it in the sense of
Definition~\ref{A-def:optimal}, if its action lies in the infinite-horizon
Bellman argmax \(\Gamma_\delta(s_t)\) at every date almost surely under the
prior-predictive law generated by that initial prior; a Borel optimal
selection, as a Dirac kernel, is eligible for every initial prior.  If
every \(x\) in the Bellman argmax at every feasible \(s\in\mathcal C\)
satisfies
\(\lvert x-\varphi(s)\rvert\le C_*(A^{-1}+\lvert D^\varphi(s)\rvert)^2\),
then every policy eligible for an initial prior tracks \(\varphi\) on
\(\mathcal C\) with constant \(C_*\) for that prior in the sense of
\PartIDefTracking, because the support property and the bound together give
the tracking inequality at every date with \(s_t\in\mathcal C\).

\subsubsection{Part (i): local approximation at a regular target}

Part~\ref{thm:approximation:i} of Theorem~\ref{thm:approximation} is the following proposition.
It concerns optimal policies and the target alone; the truth enters only
through the condition \(\hat a+\hat b\hat x=\alpha+\beta\hat x\), a
hypothesis of \PartILemTrapping{} that is verified where that lemma is
applied.

\begin{proposition}[Smooth approximation of optimal policies at a regular
target]\label{B-prop:local}
Fix \(\delta\in[0,1)\) and a target \((\hat a,\hat b,\hat x)\) with
\(\hat x\in\operatorname{int}X\).  Put \(\hat m\coloneqq\hat a+\hat b\hat x\), and
suppose that \(U\) satisfies Assumption~\ref{ass:fl-payoff} and that
\begin{enumerate}[label=\arabic*.,ref=\arabic*]
\item \emph{Unique global optimality:} the\label{B-local:uniqueness}
perceived-line maximization problem
\[
 \max_{x\in X}G\bigl(x,\hat m+\hat b(x-\hat x)\bigr)
\]
has the unique solution \(\hat x\);
\item \emph{Strict local curvature:}\label{B-local:curvature}
\(\hat Q\coloneqq G_{xx}(\hat x,\hat m)+2\hat b\,G_{xm}(\hat x,\hat m)
+\hat b^2G_{mm}(\hat x,\hat m)<0\);
\item \emph{Outcome relevance:}\label{B-local:relevance}
\(\hat M\coloneqq G_m(\hat x,\hat m)\ne0\).
\end{enumerate}
Then there are \(r_*>0\), \(u_*\in(0,1)\), \(V_*<\infty\), a finite \(C_*\),
a positive \(\lambda^0_*\) and a Borel map \(\varphi\) of the coordinates
\((a,b,u,\bar x,V)\), none of which depends on a selection, a policy, an
initial prior or the truth \((\alpha,\beta)\), such that:
\begin{enumerate}
\item[\condtarget{B-prop:local:i}{(i)}] with the target \((a^e,b^e,x^e)=(\hat a,\hat b,\hat x)\), the
constants \(r_*,u_*,V_*,C_*,\lambda^0_*\) and the map \(\varphi\) satisfy
hypotheses~(a)--(c) of \PartILemTrapping{} on every collar
\(\mathcal C(V^e)\), \(V^e\ge V_*\);
\item[\condtarget{B-prop:local:ii}{(ii)}] at every feasible state \(s\) of
\(\mathcal C_*=\bigcup_{V^e\ge V_*}\mathcal C(V^e)\), every \(x\) in the
infinite-horizon Bellman first-action argmax at \(s\) satisfies
\[
 \lvert x-\varphi(s)\rvert\le C_*\bigl(A^{-1}+\lvert\varphi(s)-\bar x\rvert\bigr)^2;
\]
\item[\condtarget{B-prop:local:iii}{(iii)}] consequently, for every initial prior, every policy eligible
for that prior, in particular every Borel optimal selection, tracks
\(\varphi\) on \(\mathcal C_*\) with constant \(C_*\) for that prior, in the
sense of \PartIDefTracking.
\end{enumerate}
\end{proposition}

\begin{proof}
\emph{The target is a regular mean pair.}
Put \((a^e,b^e,x^e)=(\hat a,\hat b,\hat x)\).  Since
\(g^{\mathrm{full}}(z;(\hat a,\hat b))=G(z,\hat a+\hat bz)
=G(z,\hat m+\hat b(z-\hat x))\), condition~\ref{B-local:uniqueness} says
that \(\hat x\) is the unique maximizer of
\(g^{\mathrm{full}}(\cdot;(\hat a,\hat b))\) on \(X\), and
\(\hat x\in\operatorname{int}X\) by hypothesis; differentiating twice,
\(\partial_{xx}g^{\mathrm{full}}(\hat x;(\hat a,\hat b))=\hat Q<0\) by
condition~\ref{B-local:curvature}; and
\(m_{\mathrm{out}}^e=G_m(\hat x,\hat m)=\hat M\ne0\) by
condition~\ref{B-local:relevance}.  Thus \((a^e,b^e)\) is a regular mean
pair with \(x^\star(a^e,b^e)=x^e\), and \(K^{\mathrm{mean}}=\{(a^e,b^e)\}\)
is a compact regular mean family.  Assumption~\ref{ass:fl-payoff} gives the
standing assumptions of Section~\ref{B-sec:scope}.

\emph{The scale, the threshold and the virtual policy.}  Apply
Lemma~\ref{B-lem:virtual} with the fixed \(\delta\) and
\(K^{\mathrm{mean}}=\{(a^e,b^e)\}\), and fix the scale
\(\epsilon_{\mathrm{sc}}\) so small that the zero-set matching and
transversality statements of Section~\ref{B-sec:matching} hold on a fixed
two-sided graph neighborhood of the target graph inside
\(\mathcal M_\epsilon\) and on a lower tail \(\ell_V\le\ell_\ast\) with
\(\ell_\ast\le\log\epsilon_{\mathrm{sc}}\).  Put
\[
 \bar V\coloneqq2e^{-\ell_\ast}.
\]
After the scale is fixed, \(\varphi^{[\ell]}\) is a globally bounded
\(C^{2,1}\) map of the chart coordinates \((a,b,u,\bar x,\ell_V)\).  Define
\[
 \bar\varphi(a,b,u,\bar x,V)\coloneqq\varphi^{[\ell]}(a,b,u,\bar x,-\log V)
 \qquad(V>0),
\]
which is \(C^{2,1}\) on \(\{V>0\}\) because \(V\mapsto-\log V\) is smooth
there, and which agrees with \(\varphi^{[\ell]}\circ\iota_\ell\) at
feasible states.  Let \(\bar r>0\) be so small that the closed
radius-\(\bar r\) neighborhood of \((a^e,b^e,x^e)\) in the coordinates
\((a,b,\bar x)\) lies in the graph neighborhood just fixed and in
\(U_{\mathrm{in}}\times\operatorname{int}J_{\mathrm{in}}\).  The threshold
\(\bar r\) depends only on the target, \(\delta\) and the fixed scale.

\emph{The constants.}  Put \(\varphi\coloneqq\bar\varphi\), \(V_*\coloneqq\bar V\) and
\(r_*\coloneqq\bar r\).  Choose \(u_*\in(0,\epsilon_{\mathrm{sc}}]\)
so small that the transversality margin of Section~\ref{B-sec:matching}
gives \(\lvert\varphi_b-\bar x\varphi_a\rvert\ge\lambda^0_*>0\) on every
collar.  The choice is uniform in \(V^e\). The margin bounds
\(\lvert\varphi_b-\bar x^{+,\varphi}\varphi_a\rvert\) below by \(\lambda_0\)
on the lower tail, the two quantities differ by
\(uD^\varphi\varphi_a/(1+u)\), and on every collar
\(\lvert D^\varphi\rvert\le B_D\coloneqq\sup_{x\in X}\lvert x\rvert+\lvert\hat
x\rvert+r_*\), because \(\varphi\) takes values in \(X\), and
\(\lvert\varphi_a\rvert\le B_a\), the first-derivative
bound of item~\ref{B-virtual:regularity} of Lemma~\ref{B-lem:virtual},
which does not depend on the lower tail; so \(u_*\le1/2\) with
\(2u_*B_DB_a\le\lambda_0/2\) gives \(\lambda^0_*=\lambda_0/2\).

\emph{Verification.}  Hypothesis~(a) of \PartILemTrapping{} follows, by the
sentence after that lemma and since \(V^e\ge V_*>0\), from the
\(C^{2,1}\) bound of item~\ref{B-virtual:regularity} of
Lemma~\ref{B-lem:virtual}, which does not depend on the
lower tail and hence not on \(V^e\), composed with \(V\mapsto-\log V\),
whose derivatives of order at most three are bounded on \(V\ge V_*/2\);
hypothesis~(b) is the zero-set matching \eqref{B-eq:matching} at
\((\hat a,\hat b,0,\hat x,V^e)\), which lies on the lower tail for every
\(V^e\ge V_*\); and hypothesis~(c) is the margin just chosen.  This proves
\ref{B-prop:local:i}.  For \ref{B-prop:local:ii}, the chart image of every feasible state of \(\mathcal C_*\)
lies in \(\mathcal I_\epsilon\), so item~\ref{B-virtual:tracking} of
Lemma~\ref{B-lem:virtual} gives the displayed bound for every maximizer,
with a constant independent of the state and of the maximizer.  Take
\(C_*\) at least this constant and the \(C^{2,1}\) bound just used.
Part~\ref{B-prop:local:iii} is the paragraph on eligible policies above. The support
property and the bound in \ref{B-prop:local:ii} give the tracking inequality of
\PartIDefTracking{} at every date with \(s_t\in\mathcal C_*\).  At
\(\delta=0\) the same holds with myopic maximization, as recorded in
Lemma~\ref{B-lem:virtual}.
\end{proof}

\subsubsection{Part (ii): global approximation on \texorpdfstring{\(\Theta^I\)}{Theta I}}

Part~\ref{thm:approximation:ii} of Theorem~\ref{thm:approximation} is the following proposition,
applied to every nondegenerate Gaussian initial prior \(\mu\) and every
optimal policy.

\begin{proposition}[Regular-truth localization]\label{B-prop:global}
Assume \(0\le\delta<1\), that \(X\) is a compact interval, and that \(U\)
satisfies Assumption~\ref{ass:fl-payoff}.  Fix \(\eta\in(6/7,1)\), the
nondegenerate Gaussian initial prior \(\mu\), and one policy \(\Pi\),
randomized and possibly history dependent as in Appendix~\ref{app:value}, and
suppose that under \(P^{\mu,\Pi}\) from the fixed initial posterior
\[
 x_{t+1}\in\Gamma_\delta(s_t)
 \quad\text{for every }t\ge0\quad\text{almost surely},
\]
where \(\Gamma_\delta(s)\) is the infinite-horizon Bellman first-action
argmax; a fixed Borel measurable infinite-horizon optimal first-action
selection is the Dirac kernel, for which the support property holds at
every state.  Then the following hold.
\begin{enumerate}
\item[\condtarget{B-prop:global:i}{(i)}] The set \(\Theta^I\) is open, and the optimizer map
\(x^\star\) is locally \(C^1\).
\item[\condtarget{B-prop:global:ii}{(ii)}] There is a countable Borel cover
\(\Theta^I=\bigcup_{m\ge1}\Theta_m\) with the following property.
For each \(m\), one virtual policy on a semi-infinite precision tail and
one deterministic constant/localization package, chosen before the truth
and path are realized, are fixed, and the family \((\Theta_m)\) with these
data is a coherent exclusion package for \(\mu\) and \(\Pi\) on \(\Theta^I\)
in the sense of \PartIDefPackage{} (clauses (P1)--(P5), with exponent \(2\)
in the approximation clause (P4)); hence every hypothesis of
\PartIThmExclusion{} holds on all of \(\Theta^I\).
\end{enumerate}
These conclusions are for the fixed prior \(\mu\) and the fixed policy
\(\Pi\).
\end{proposition}

\begin{proof}
The posterior recursions \eqref{eq:design}--\eqref{eq:means} hold with the
realized seed-measurable action, by \PartILemPosterior; for the Dirac kernel
of a Borel selection the support property holds at every state.  Write
\(\bar\varphi\) for the virtual policy of Lemma~\ref{B-lem:virtual} when it
appears without index, to distinguish it from the actual selection.

\emph{Openness and the optimizer map.}  Section~\ref{B-sec:scope} proves
that \(\Theta^I\) is open and that \(x^\star\) is \(C^3\), in
particular \(C^1\), on a neighborhood of every nonempty compact regular
mean family, hence of every regular truth; this is \ref{B-prop:global:i}.  By openness,
every regular truth has a compact neighborhood
\(K^{\mathrm{mean}}\subset\Theta^I\) of mean pairs, with the truth
in its interior, that is a compact regular mean family, and
Lemma~\ref{B-lem:virtual} applied to such a family supplies one virtual
policy \(\bar\varphi\) on \(\ell_V\le\log\epsilon_{\mathrm{sc}}\).

\emph{Limits on complete learning.}  By Lemma~\ref{B-lem:optimal-collapse},
\(P^{\mu,\Pi}\)-almost surely on \(\{V_t\to\infty,\ \vartheta\in\Theta^I\}\),
the limits \eqref{B-eq:collapse-limits} hold.

\emph{The coherent exclusion package.}  Take the cover
\(\Theta^I=\bigcup_m\Theta_m\), the rectangles, the tail levels and the maps
\(\varphi_m\) fixed in Section~\ref{B-sec:localization}.

Clause (P1) holds by construction.  For (P2), the extension argument in
Section~\ref{B-sec:matching} gives \(\varphi_m\) a finite \(C^{1,\eta}\)
norm on \(\mathcal S_m^{\mathrm{tail}}\) in the coordinates \((z_b,\ell_V)\)
for the fixed \(\eta\in(6/7,1)\), with first derivatives uniformly Lipschitz
in \(\ell_V\), so that on \(\mathcal K_m^0\) the derivatives
\(\varphi_{m,a}\) and \(\varphi_{m,b}\) have a common Lipschitz constant in
\(\ell_V\); on \(\mathcal K_m^0\), transversality gives
\(\lvert\varphi_{m,b}-\bar x_m^{+,\varphi}\varphi_{m,a}\rvert\ge\lambda_0\eqqcolon\lambda_m\),
and Lemma~\ref{B-lem:covariance-bounds} gives
\(\lvert\varphi_{m,\bar x}\rvert\le C(u+e^{\ell_V})\) and
\(\lvert\varphi_{m,\ell_V}\rvert\le C(u+e^{\ell_V}\lvert D^{\varphi_m}\rvert)
\le C_m(u+e^{\ell_V})\), because \(u\ge0\) there and \(D^{\varphi_m}\) is
bounded on the bounded rectangle; enlarging \(C_m\) bounds the sum.  For
(P3), fix the four positive thresholds so that they satisfy the finitely
many inequalities that clause (P3) of \PartIDefPackage{} requires of them;
they depend only on \(\eta\), \(X\), \(\mu\), the data just fixed and the
tracking constant \(C_{2,m}\) below, and they are fixed before any path is
observed.  For (P4), quadratic on-path tracking,
item~\ref{B-virtual:tracking} of Lemma~\ref{B-lem:virtual},
holds for every action in the Bellman argmax at every feasible state of
\(\mathcal K_m\), with \(u+\lvert D^{\varphi_m}\rvert=A^{-1}+\lvert\varphi_m-\bar x\rvert\); since
the policy chooses an action in that argmax at every date almost
surely under \(P^{\mu,\Pi}\), this gives, \(P^{\mu,\Pi}\)-almost surely, at
every date \(t\) with \(s_t\in\mathcal K_m\),
\(\lvert x_{t+1}-\varphi_m(s_t)\rvert\le C_{2,m}(u_t+\lvert D^{\varphi_m}(s_t)\rvert)^2\) with
\(C_{2,m}\) the tracking constant.  For (P5), matching in
\eqref{B-eq:matching} and the uniform first-derivative bound in the bounded
coordinates give \(\varphi_m(s)\to x^\star(\vartheta)\) as the bounded
coordinates of \(s\) approach the graph point
\((\alpha,\beta,0,x^\star(\vartheta))\) along the tail.  By the limits
established above, \(P^{\mu,\Pi}\)-almost surely on
\(\{V_t\to\infty,\ \vartheta\in\Theta_m\}\): \(V_t/A_t\to0\); the bounded
coordinates \((a_t,b_t,A_t^{-1},\bar x_t)\) converge to the graph point,
which lies in \(\mathcal B_m^{\rm in}\), a set open relative to
\(\{u\ge0\}\), so they eventually stay in \(\overline{\mathcal B}_m^{\rm in}\);
\(\ell_{V_t}\to-\infty\), \(0<A_t^{-1}\to0\), \(V_t^{-1}\to0\) and
\(u_te^{-\ell_{V_t}}=V_t/A_t\to0\); and
\(\lvert D^{\varphi_m}(s_t)\rvert=\lvert\varphi_m(s_t)-\bar x_t\rvert\to0\).  Hence the state
eventually remains in the corridor \(\mathcal K_m\), which verifies (P5).
Every virtual policy, rectangle and threshold is fixed before the path is
observed, for the one fixed policy.  This proves \ref{B-prop:global:ii}.
\end{proof}

\begin{proof}[Proof of Theorem~\ref{thm:approximation}]
Part~\ref{thm:approximation:i} is Proposition~\ref{B-prop:local}.  Part~\ref{thm:approximation:ii} is
Proposition~\ref{B-prop:global}, applied to the given nondegenerate
Gaussian prior and optimal policy.
\end{proof}

\section{Proofs for optimal policies}\label{app:optimal}

Throughout this appendix Assumption~\ref{ass:fl-payoff} holds, and
\(\delta\in[0,1)\) is fixed. Policies, optimal policies and the Bellman objects
\(J_\delta\), \(Q_\delta\) and \(\Gamma_\delta\) are those of
Appendix~\ref{app:value}; as there, \(\Pi\) denotes a policy and
\(\pi_\delta\) an optimal policy, the latter as in the main-text statements
proved here; \(\mathcal T\) is the successor map defined there,
and \(R_X=\max\{\lvert\underline x\rvert,\lvert\overline x\rvert\}\).

\subsection{Regular parameter sets}

\begin{lemma}[Regular parameter sets]\label{D-lem:regular}
The map \((\vartheta,x)\mapsto g^{\mathrm{full}}(x;\vartheta)\) is continuous on
\(\R^2\times X\). For every \(\vartheta\), \(g^{\mathrm{full}}(\cdot;\vartheta)\)
is continuously differentiable on \(X\), one-sidedly at the endpoints, with
\begin{equation}
 g^{\mathrm{full}}_x(x;\vartheta)
 =E\bigl[U_x(x,\alpha+\beta x+Z)\bigr]+\beta E\bigl[Z\,U(x,\alpha+\beta x+Z)\bigr],
 \qquad Z\sim N(0,1),
 \label{D-eq:gfull-derivative}
\end{equation}
where \(U_x\) is one-sided at the endpoints, and
\((\vartheta,x)\mapsto g^{\mathrm{full}}_x(x;\vartheta)\) is continuous on
\(\R^2\times X\). The set \(\Theta^I\) is open, the set \(\Theta^B\) is Borel,
\(\Theta^I\cap\Theta^B=\emptyset\), and \(\Theta^R=\Theta^I\cup\Theta^B\) is
Borel.
\end{lemma}

\begin{proof}
Since \(g^{\mathrm{full}}(x;\vartheta)=G(x,\alpha+\beta x)\) and \(G\) is
continuous on \(X\times\R\) (Appendix~\ref{app:value}),
\(g^{\mathrm{full}}\) is continuous on \(\R^2\times X\). For the derivative,
write \(g^{\mathrm{full}}(x;\vartheta)=\int_\R F(x,y;\vartheta)\,dy\) with
\[
 F(x,y;\vartheta)=\frac1{\sqrt{2\pi}}\,U(x,y)\,e^{-(y-\alpha-\beta x)^2/2}.
\]
For every \(y\), \(F(\cdot,y;\vartheta)\) is continuously differentiable on
\(X\), one-sidedly at the endpoints, with
\[
 F_x(x,y;\vartheta)=\frac1{\sqrt{2\pi}}
 \bigl[U_x(x,y)+\beta(y-\alpha-\beta x)U(x,y)\bigr]e^{-(y-\alpha-\beta x)^2/2},
\]
and \(y\mapsto U_x(x,y)\) is Borel, as a pointwise limit of one-sided
difference quotients. Let \(\lvert\alpha\rvert,\lvert\beta\rvert\le K\) and
\(M=K(1+R_X)\), so that \(\lvert\alpha+\beta x\rvert\le M\) on \(X\).
Assumption~\ref{ass:fl-payoff} with \(j\le1\) and
\((y-\alpha-\beta x)^2\ge y^2/2-M^2\) give, for \(x\in X\),
\[
 \lvert F(x,y;\vartheta)\rvert+\lvert F_x(x,y;\vartheta)\rvert
 \le\frac{C(2+K)(1+M)}{\sqrt{2\pi}}(1+\lvert y\rvert)e^{c_U\lvert y\rvert}e^{M^2/2-y^2/4},
\]
an integrable function of \(y\). By the mean value theorem it dominates the
difference quotients of \(F(\cdot,y;\vartheta)\) on \(X\), so dominated
convergence gives the derivative of \(g^{\mathrm{full}}(\cdot;\vartheta)\) at
every \(x\in X\), one-sidedly at the endpoints, as
\(\int_\R F_x(x,y;\vartheta)\,dy\); the substitution \(y=\alpha+\beta x+Z\)
turns this into \eqref{D-eq:gfull-derivative}. For every \(y\),
\(F_x(x,y;\vartheta)\) is continuous in \((\vartheta,x)\in\R^2\times X\),
because \(U(\cdot,y)\) and \(U_x(\cdot,y)\) are continuous on \(X\), and
dominated convergence with the same bound gives the joint continuity of
\((\vartheta,x)\mapsto g^{\mathrm{full}}_x(x;\vartheta)\). No derivative of
\(U\) in the outcome is used.

\(\Theta^I\) is open by Theorem~\ref{thm:approximation}\ref{thm:approximation:ii}. The upper endpoint
\(\overline x\) is the unique maximizer of \(g^{\mathrm{full}}(\cdot;\vartheta)\)
if and only if
\(\min_{x\in[\underline x,\overline x-1/n]}\{g^{\mathrm{full}}(\overline x;\vartheta)-g^{\mathrm{full}}(x;\vartheta)\}>0\)
for every integer \(n>1/(\overline x-\underline x)\). Each such minimum is
continuous in \(\vartheta\), by joint continuity and compactness, and so is
\(g^{\mathrm{full}}_x(\overline x;\vartheta)\). So the upper-endpoint part of
\(\Theta^B\) is a countable intersection of open sets; likewise the
lower-endpoint part. Hence \(\Theta^B\) is Borel. A parameter in \(\Theta^I\)
has its unique maximizer in \(\operatorname{int}X\), and a parameter in
\(\Theta^B\) at an endpoint, so the two sets are disjoint, and \(\Theta^R\) is
Borel.
\end{proof}

\subsection{Optimal support and boundary exclusion}

For a feasible state \(s\) and \(x\in X\), put
\[
 W_s(x)=E_Z\bigl[J_\delta(\mathcal T(s,x,Z))\bigr],
 \qquad\text{so that}\qquad
 Q_\delta(s,x)=g_s(x)+\delta W_s(x).
\]
\(W_s(x)\) is the supremum, over all Borel predictable continuation rules after
the first experiment \(x\), including history-dependent rules that need not
compute the Bayesian posterior, of the expected discounted continuation reward
\(E\sum_{j\ge0}\delta^jU(X_j,y_j)\), where \(X_0,X_1,\ldots\) are the
continuation actions and \(y_j\) their outcomes. Such rules are Borel
predictable policies in the sense of \PartIRemPredictable, so
\PartILemPosterior{} applies to them. For \(0<\delta<1\), this is the
last part of Proposition~\ref{prop:value}. A policy with first action \(x\) has
value \(g_s(x)+\delta E\sum_{j\ge0}\delta^jU(X_j,y_j)\), so
\(\delta W_s(x)=Q_\delta(s,x)-g_s(x)\) is \(\delta\) times that supremum. For
\(\delta=0\), \(W_s(x)=E_Z[\max_{x'}g_{s^+}(x')]\) with \(s^+=\mathcal T(s,x,Z)\).
By \PartILemPosterior(a)--(b), conditional on the first outcome and the seeds, the next
outcome has the posterior-predictive law at \(s^+\), so a continuation action
\(X_0\) earns at most \(g_{s^+}(X_0)\le\max_{x'}g_{s^+}(x')\) in conditional
expectation, with equality for \(X_0=\xi(s^+)\) of
Proposition~\ref{prop:value}.

\begin{lemma}[Stability of the first experiment]\label{D-lem:experiment}
With \(C\) and \(c_U\) from Assumption~\ref{ass:fl-payoff}, put
\[
 B(s)=\frac{\sqrt2\,Ce^{c_U^2}}{1-\delta}
 \Bigl(E_s\bigl[e^{2c_U(\lvert\tilde\alpha\rvert+R_X\lvert\tilde\beta\rvert)}\bigr]\Bigr)^{1/2},
\]
where \((\tilde\alpha,\tilde\beta)\sim\mu_s\).
\begin{enumerate}[label=\textup{(\roman*)},ref=(\roman*)]
\item If feasible states \(s_n\) have posterior means converging to\label{D-lem:experiment:i}
\(\vartheta\) and posterior covariances converging to zero, then
\[
 \sup_{x\in X}\bigl\lvert g_{s_n}'(x)-g^{\mathrm{full}}_x(x;\vartheta)\bigr\rvert\longrightarrow0,
\]
with one-sided derivatives at the endpoints.
\item For every feasible state \(s\) and every \(x,x'\in X\),\label{D-lem:experiment:ii}
\[
 \lvert W_s(x)-W_s(x')\rvert\le B(s)\,w^{1/2}\lvert x-x'\rvert,
\]
where \(w=V^{-1}\) is the slope variance at \(s\).
\item The formula for \(B\) defines \(B\) for every mean in \(\R^2\) and every\label{D-lem:experiment:iii}
positive-semidefinite covariance, and \(B\) is finite and continuous there,
zero covariance included.
\end{enumerate}
\end{lemma}

\begin{proof}
\parstartref{D-lem:experiment:i} By Fubini's theorem,
\(g_s(x)=E_s[g^{\mathrm{full}}(x;\tilde\vartheta)]\) with
\(\tilde\vartheta=(\tilde\alpha,\tilde\beta)\sim\mu_s\). By
\eqref{D-eq:gfull-derivative} and Assumption~\ref{ass:fl-payoff} with
\(j\le1\), for \(x\in X\),
\[
 \lvert g^{\mathrm{full}}_x(x;\vartheta)\rvert
 \le C(1+\lvert\beta\rvert)e^{c_U(\lvert\alpha\rvert+R_X\lvert\beta\rvert)}
 E\bigl[(1+\lvert Z\rvert)e^{c_U\lvert Z\rvert}\bigr],
\]
an exponential envelope in \(\vartheta\), integrable under every Gaussian law.
By the mean value theorem it dominates the difference quotients of
\(g^{\mathrm{full}}(\cdot;\tilde\vartheta)\) on \(X\), so dominated
convergence gives
\(g_s'(x)=E_s[g^{\mathrm{full}}_x(x;\tilde\vartheta)]\) at every \(x\in X\),
one-sidedly at the endpoints. Write the posterior draw
as \((a,b)+\Sigma^{1/2}\zeta\) with \(\zeta\sim N(0,I_2)\). Along \(s_n\),
the square root \(\Sigma_n^{1/2}\) tends to zero, by its
continuity on the positive-semidefinite cone, so for every \(\zeta\) the
draws \(\tilde\vartheta_n\) converge to \(\vartheta\), and uniform continuity
of \(g^{\mathrm{full}}_x\) on compact subsets of \(\R^2\times X\)
(Lemma~\ref{D-lem:regular}) gives
\(\sup_{x\in X}\lvert g^{\mathrm{full}}_x(x;\tilde\vartheta_n)-g^{\mathrm{full}}_x(x;\vartheta)\rvert\to0\)
(a measurable supremum, over the rational points and endpoints of \(X\)).
The exponential envelope, evaluated at the draws, dominates these suprema
uniformly in \(n\), since the means and \(\Sigma_n^{1/2}\) are bounded along
\(s_n\), and dominated convergence gives \ref{D-lem:experiment:i}.

\parstartref{D-lem:experiment:ii} Recenter the first observation as \(T_x=y^+-(a+bx)\), and let \(P^0_x\)
be the joint law of the parameter and \(T_x\) under the posterior at \(s\).
Given the parameter, \(T_x\) is unit normal, and its means under \(x\) and
\(x'\) differ by \((\beta-b)(x-x')\). For the unnormalized Hellinger distance
\(H^2(P,Q)=\int(\sqrt p-\sqrt q)^2\), the unit-normal affinity and
\(1-e^{-z}\le z\) give
\begin{equation}
 H^2(P^0_x,P^0_{x'})
 =2E_s\left[1-\exp\left(-\frac{(\beta-b)^2(x-x')^2}{8}\right)\right]
 \le\frac14w(x-x')^2.
 \label{D-eq:hellinger}
\end{equation}
Every Borel predictable continuation rule after \(x\) can be transported to
one after \(x'\). After observing \(y_1'\), feed the rule the pseudo-history
\((x,y_1'+b(x-x'))\), which has the same recentered signal, and then the
actual subsequent outcomes. Given the parameter and the recentered signal,
both systems use the same future kernel, so their expected continuation
rewards are \(E_{P^0_x}F\) and \(E_{P^0_{x'}}F\) for one Borel function \(F\).
Since \(\lvert y_j\rvert\le\lvert\alpha\rvert+R_X\lvert\beta\rvert+\lvert\varepsilon_j\rvert\),
the bound on \(\lvert U\rvert\), the independence of \(\varepsilon_j\) and the
parameter, and \(E e^{2c_U\lvert\varepsilon_j\rvert}\le2e^{2c_U^2}\) give
\(\lVert U(X_j,y_j)\rVert_2\le\sqrt2\,Ce^{c_U^2}
\bigl(E_s[e^{2c_U(\lvert\alpha\rvert+R_X\lvert\beta\rvert)}]\bigr)^{1/2}\)
for every predictable continuation. Minkowski's inequality over \(j\), with
\(\sum_j\delta^j=(1-\delta)^{-1}\), and Jensen's inequality for the
conditional expectation \(F\) then give
\(\lVert F\rVert_{L^2(P^0_x)},\lVert F\rVert_{L^2(P^0_{x'})}\le B(s)\). The
Cauchy--Schwarz inequality gives
\(\lvert E_PF-E_QF\rvert\le H(P,Q)(\lVert F\rVert_{L^2(P)}+\lVert F\rVert_{L^2(Q)})\),
so by \eqref{D-eq:hellinger} the transported rule earns at least the reward
of the original minus \(B(s)w^{1/2}\lvert x-x'\rvert\). Applying this to
near-optimal continuations in both directions and using the supremum
representation of \(W_s\) gives \ref{D-lem:experiment:ii}.

\parstartref{D-lem:experiment:iii} In the representation of (i), the formula reads
\[
 B(s)=\frac{\sqrt2\,Ce^{c_U^2}}{1-\delta}
 \Bigl(E\bigl[e^{2c_U(\lvert a+(\Sigma^{1/2}\zeta)_1\rvert
 +R_X\lvert b+(\Sigma^{1/2}\zeta)_2\rvert)}\bigr]\Bigr)^{1/2},
\]
which makes sense for every \((a,b)\in\R^2\) and every positive-semidefinite
\(\Sigma\). The integrand is continuous in \((a,b,\Sigma)\) for every \(\zeta\),
and on every bounded set of \((a,b,\Sigma)\) it is dominated by a function
\(e^{K(1+\lvert\zeta\rvert)}\), which is integrable. Dominated convergence makes \(B\)
finite and continuous there, zero covariance included.
\end{proof}

For \(e\in\{\underline x,\overline x\}\), let \(\mathcal E_e\) be the set of
parameters \(\vartheta\) for which there is \(c_\vartheta>0\) with
\begin{equation}
 g^{\mathrm{full}}(e;\vartheta)-g^{\mathrm{full}}(x;\vartheta)
 \ge c_\vartheta\lvert x-e\rvert
 \qquad(x\in X),
 \label{D-eq:cone}
\end{equation}
and put \(\mathcal E=\mathcal E_{\underline x}\cup\mathcal E_{\overline x}\). The
condition is stronger than uniqueness of an endpoint maximizer. It requires a
uniform linear margin away from the endpoint. The set \(\mathcal E\) is Borel.
By Lemma~\ref{D-lem:regular},
\[
 \mathcal E_e=\bigcup_{n\ge1}\bigcap_{q\in X\cap\mathbb Q}
 \left\{\vartheta:g^{\mathrm{full}}(e;\vartheta)-g^{\mathrm{full}}(q;\vartheta)\ge\frac{\lvert q-e\rvert}{n}\right\}
\]
is a countable union of closed sets, continuity in \(x\) extending the rational
inequalities to every \(x\in X\).

\begin{lemma}[Strict active boundary]\label{D-lem:boundary}
\begin{enumerate}[label=\textup{(\roman*)},ref=(\roman*)]
\item Let \(\vartheta\in\Theta^B\), with unique maximizer\label{D-lem:boundary:i}
\(e\in\{\underline x,\overline x\}\) of \(g^{\mathrm{full}}(\cdot;\vartheta)\).
Then \eqref{D-eq:cone} holds for some \(c_\vartheta>0\). Hence
\(\Theta^B\subseteq\mathcal E\).
\item Fix a nondegenerate Gaussian prior \(\mu\) and a policy \(\Pi\) that is\label{D-lem:boundary:ii}
optimal for \(\mu\) (Definition~\ref{A-def:optimal}). Then
\[
 P^{\mu,\Pi}(V_\infty=\infty,\ \vartheta\in\mathcal E)=0,
 \qquad\text{in particular}\qquad
 P^{\mu,\Pi}(V_\infty=\infty,\ \vartheta\in\Theta^B)=0,
\]
and \(P^{\mu,\Pi}(\mathcal W,\ \vartheta\in B)=\mu(B)\) for every Borel
\(B\subseteq\mathcal E\).
\end{enumerate}
\end{lemma}

\begin{proof}
\ref{D-lem:boundary:i} Let \(e=\overline x\), so that
\(g^{\mathrm{full}}_x(\overline x;\vartheta)>0\). By
Lemma~\ref{D-lem:regular}, \(g^{\mathrm{full}}(\cdot;\vartheta)\) is
continuously differentiable on \(X\), so there are \(c_0>0\) and
\(h\in(0,\overline x-\underline x)\) with
\(g^{\mathrm{full}}_x(x;\vartheta)\ge c_0\) for \(x\in[\overline x-h,\overline x]\),
and the fundamental theorem of calculus gives
\(g^{\mathrm{full}}(\overline x;\vartheta)-g^{\mathrm{full}}(x;\vartheta)\ge c_0(\overline x-x)\)
on that interval. On the compact set \([\underline x,\overline x-h]\),
uniqueness of the maximizer and continuity give
\(d=\min_x\{g^{\mathrm{full}}(\overline x;\vartheta)-g^{\mathrm{full}}(x;\vartheta)\}>0\),
so \(g^{\mathrm{full}}(\overline x;\vartheta)-g^{\mathrm{full}}(x;\vartheta)\ge d(\overline x-x)/(\overline x-\underline x)\)
there. The constant \(c_\vartheta=\min\{c_0,d/(\overline x-\underline x)\}\)
gives \eqref{D-eq:cone}. The case \(e=\underline x\) follows by reversing the
inequalities.

\parstartref{D-lem:boundary:ii} By \PartILemPosterior, the posterior means converge almost surely, and
almost surely on \(\{V_\infty=\infty\}\) the means converge to \(\vartheta\) and
\(\Sigma_t\to0\). Let \(\Omega_0\) be the intersection of this probability-one
event with the event \eqref{A-eq:on-path-support}. Take a path of \(\Omega_0\)
with \(V_\infty=\infty\) and \(\vartheta\in\mathcal E_e\). By
Lemma~\ref{D-lem:experiment}\ref{D-lem:experiment:i}, eventually
\(\sup_x\lvert g_{s_t}'(x)-g^{\mathrm{full}}_x(x;\vartheta)\rvert\le c_\vartheta/2\),
and the fundamental theorem of calculus and \eqref{D-eq:cone} give
\(g_{s_t}(e)-g_{s_t}(x)\ge(c_\vartheta/2)\lvert x-e\rvert\) for \(x\in X\). The
coefficient \(B(s_t)\) is eventually bounded, because \(B\) is continuous at
mean \(\vartheta\) and zero covariance (Lemma~\ref{D-lem:experiment}\ref{D-lem:experiment:iii}), while
\(w_t\to0\). Hence, eventually, by Lemma~\ref{D-lem:experiment}\ref{D-lem:experiment:ii},
\[
 Q_\delta(s_t,e)-Q_\delta(s_t,x)
 \ge\left(\frac{c_\vartheta}{2}-\delta B(s_t)w_t^{1/2}\right)\lvert x-e\rvert>0
 \qquad(x\ne e).
\]
At \(\delta=0\) the flow margin alone gives this. Thus
\(\Gamma_\delta(s_t)=\{e\}\) from some finite date \(T\) onward, and the support
property \eqref{A-eq:on-path-support} gives \(x_{t+1}=e\) for every \(t\ge T\).
\PartILemEventualConstant\ gives \(\sum_{t\ge T}\kappa_tD_t^2<\infty\), so
\(V_\infty=V_T+\sum_{t\ge T}\kappa_tD_t^2<\infty\) by \eqref{eq:design},
contradicting \(V_\infty=\infty\). Hence
\(\{V_\infty=\infty,\ \vartheta\in\mathcal E\}\cap\Omega_0=\emptyset\), which
proves the first display; the second follows from \ref{D-lem:boundary:i}. The last assertion is
Lemma~\ref{C-lem:setwise}\ref{C-lem:setwise:a} with \(R=\mathcal E\).
\end{proof}

\subsection{Proofs of the main results}

\begin{proof}[Proof of Theorem~\ref{thm:fl-local}]
Fix the truth \(\vartheta=(\alpha,\beta)\), the discount factor
\(\delta\in[0,1)\) and the interior regular Berk--Nash equilibrium
\((a^e,b^e,x^e)\). The equilibrium gives \(a^e+b^ex^e=\alpha+\beta x^e\), and
its three regularity conditions say that \(x^e\in\operatorname{int}X\) is the
unique maximizer of \(z\mapsto G(z,a^e+b^ez)\) over \(X\), with strictly
negative second derivative there, and that \(G_m(x^e,a^e+b^ex^e)\ne0\). These
are the hypotheses of Theorem~\ref{thm:approximation}\ref{thm:approximation:i} with
\((\hat a,\hat b,\hat x)=(a^e,b^e,x^e)\). It supplies a Borel smooth target
map \(\varphi\) for \((a^e,b^e,x^e)\), with constants
\(r_*,u_*,V_*,C_*,\lambda^0_*\), the map and the constants not depending on a
policy, a prior or the truth, such that every optimal policy for every prior tracks \(\varphi\) on
\(\mathcal C_*\) with constant \(C_*\) for that prior. The tracking
requirement of Theorem~\ref{thm:transfer}\ref{thm:transfer:i} is therefore met by every policy
that is optimal for the initial prior in question.

Put \(\overline V=V_*\) and \(\bar r=r_*\). For \(r\in(0,\bar r)\),
Theorem~\ref{thm:transfer}\ref{thm:transfer:i} gives \(\overline V_r\ge V_*=\overline V\), and,
for every \(V^e\ge\overline V_r\) and \(\varepsilon\in(0,1)\), thresholds
\(\overline A<\infty\) and \(r_0>0\). Let \(\mu\) be a nondegenerate Gaussian
prior with \(A_0>\overline A\) and
\(\lvert a_0-a^e\rvert+\lvert b_0-b^e\rvert+\lvert\bar x_0-x^e\rvert+\lvert V_0-V^e\rvert<r_0\),
and let \(\pi_\delta\) be optimal for \(\mu\). Since \(\pi_\delta\) tracks
\(\varphi\) on \(\mathcal C_*\) with constant \(C_*\) for \(\mu\),
Theorem~\ref{thm:transfer}\ref{thm:transfer:i} gives, simultaneously with
\(P^{\pi_\delta}_\vartheta\)-probability at least \(1-\varepsilon\): the bounds
of conclusion~\ref{thm:fl-local:i} for every \(t\ge0\); \(V_\infty<\infty\),
\(b_t\to b_\infty\), \(\bar x_t\to\bar x_\infty\), \(x_{t+1}\to\bar x_\infty\)
and \(a_t\to\alpha+\beta\bar x_\infty-b_\infty\bar x_\infty\); and, when
\(b^e\ne\beta\), \(b_\infty\ne\beta\).
On this event the posterior means converge to the finite limits
\((a_\infty,b_\infty)\), with \(a_\infty=\alpha+\beta\bar x_\infty-b_\infty\bar x_\infty\),
\(\bar x_t\) and \(x_{t+1}\) converge to \(\bar x_\infty\), and
\(D_t=x_{t+1}-\bar x_t\to0\), which is conclusion~\ref{thm:fl-local:ii}. The priors form a
nonempty open set, and the thresholds \(\overline V\), \(\bar r\),
\(\overline V_r\), \(\overline A\) and \(r_0\) do not depend on the prior or
the optimal policy, by Theorems~\ref{thm:approximation}\ref{thm:approximation:i}
and~\ref{thm:transfer}\ref{thm:transfer:i}.
\end{proof}

\begin{proof}[Proof of Theorem~\ref{thm:fl-global}]
Fix the nondegenerate Gaussian prior \(\mu\), the discount factor
\(\delta\in[0,1)\) and one policy \(\pi_\delta\) that is optimal for \(\mu\). By
Lemma~\ref{D-lem:regular}, \(\Theta^I\) is open and \(\Theta^R\) is Borel.

\emph{Interior-regular parameters.} Theorem~\ref{thm:approximation}\ref{thm:approximation:ii} shows
that \(\mu\) and \(\pi_\delta\) admit a coherent exclusion package on
\(\Theta^I\). Theorem~\ref{thm:transfer}\ref{thm:transfer:ii} therefore gives
\(P^{\mu,\pi_\delta}(V_t\to\infty,\ \vartheta\in\Theta^I)=0\).

\emph{Strict active-boundary parameters.} Lemma~\ref{D-lem:boundary}\ref{D-lem:boundary:ii} gives
\(P^{\mu,\pi_\delta}(V_\infty=\infty,\ \vartheta\in\Theta^B)=0\).

\emph{Composition.} Since \(V_t\) is nondecreasing,
\(\{V_t\to\infty\}=\{V_\infty=\infty\}\), and the two cases give
\(P^{\mu,\pi_\delta}(V_\infty=\infty,\ \vartheta\in\Theta^R)=0\).
Lemma~\ref{C-lem:setwise} with \(R=\Theta^R\) gives
\(P^{\mu,\pi_\delta}(\mathcal W,\ \vartheta\in B)=\mu(B)\) for every Borel
\(B\subseteq\Theta^R\), and \(P^{\pi_\delta}_\vartheta(\mathcal W)=1\) for
Lebesgue-almost every \(\vartheta\in\Theta^R\). The conclusions, including
the null set, are for the fixed prior and optimal policy.
\end{proof}

\begin{proof}[Proof of Corollary~\ref{cor:suboptimal}]
Suppose that \(P^{\mu,\Pi}(V_t\to\infty,\ \vartheta\in\Theta^R)>0\), and that
\(\Pi\) is optimal for \(\mu\) for some \(\delta\in[0,1)\).
Theorem~\ref{thm:fl-global} with \(B=\Theta^R\) gives
\(P^{\mu,\Pi}(\mathcal W,\ \vartheta\in\Theta^R)=\mu(\Theta^R)\). Since
\(\mathcal W\subseteq\{V_\infty<\infty\}\) and \(\vartheta\) has marginal law
\(\mu\),
\[
\begin{aligned}
 P^{\mu,\Pi}(V_\infty=\infty,\ \vartheta\in\Theta^R)
 &=\mu(\Theta^R)-P^{\mu,\Pi}(V_\infty<\infty,\ \vartheta\in\Theta^R)\\
 &\le\mu(\Theta^R)-P^{\mu,\Pi}(\mathcal W,\ \vartheta\in\Theta^R)=0.
\end{aligned}
\]
As \(\{V_t\to\infty\}=\{V_\infty=\infty\}\), this contradicts the hypothesis.
Hence \(\Pi\) is not optimal for \(\mu\), that is,
\eqref{A-eq:on-path-support} fails, for every \(\delta\in[0,1)\). In
particular, for every \(0<\delta<1\), \(J^\Pi_\delta(s_0)\ne J_\delta(s_0)\) by
Lemma~\ref{lem:support}\ref{lem:support:i}, so \(\Pi\) does not maximize the discounted payoff
\(U^\pi_\delta\). The special case after the corollary follows from
\(P^{\mu,\Pi}(V_t\to\infty,\ \vartheta\in\Theta^R)\ge
P^{\mu,\Pi}(V_t\to\infty,\ \vartheta\in R)=\mu(R)>0\).
\end{proof}

\renewcommand{\bibsection}{\section*{\refname}\pdfbookmark[1]{References}{part2-article-references}}
\bibliographystyle{econ}
\bibliography{part2}

\clearpage
\pagenumbering{arabic}
\setcounter{section}{0}
\setcounter{subsection}{0}
\setcounter{theorem}{0}
\setcounter{equation}{0}
\setcounter{figure}{0}
\setcounter{table}{0}
\setcounter{footnote}{0}
\renewcommand{\thesection}{S\arabic{section}}
\renewcommand{\thefigure}{S\arabic{figure}}
\renewcommand{\thetable}{S\arabic{table}}
\pdfbookmark[0]{Supplement}{part2-supplement}
\phantomsection\label{supplement}
\thispagestyle{plain}
\begin{center}
\vspace*{1.4cm}
{\LARGE Supplement to ``\ArticleTitle''\par}
\vspace{1.2cm}
{\large \AuthorOne\textsuperscript{1}, \AuthorTwo\textsuperscript{2},
and \AuthorThree\textsuperscript{3}\par}
\bigskip
{\normalsize
\textsuperscript{1}\AffiliationOne\par
\textsuperscript{2}\AffiliationTwo\par
\textsuperscript{3}\AffiliationThree\par}
\end{center}
\begin{abstract}
This supplement gives the construction of the smooth approximation and
its supporting calculations, including the proof of
Proposition~\ref{prop:value}, the proof of Theorem~\ref{thm:infinite},
the verifications for the examples, and the analysis of finite action sets
and fixed action-change costs.
\end{abstract}

\section{Virtual-policy construction and supporting calculations}\label{supp:construction}
This section proves Proposition~\ref{prop:value} and
Lemma~\ref{B-lem:virtual}, the latter under Assumption~\ref{ass:fl-payoff}
through the Gaussian smoothing facts of Section~\ref{S1-section-smoothing}, in
the coordinates \eqref{B-eq:coordinates} and with the transition
\eqref{B-eq:transition}. The value foundation requires only a Borel payoff
with exponential growth and a continuous \(G\). The final covariance calculations prove
Lemma~\ref{B-lem:covariance-bounds} of Section~\ref{B-sec:matching}.

\subsection{Weighted value foundation}\label{S1-section-1}\label{S1-value-proof}
The rate \(\lambda\), the weights \(\mathsf W_\nu\) and the spaces
\(\mathcal C_\nu\) are those of Appendix~\ref{app:value}.

\begin{proof}[Proof of Proposition~\ref{prop:value}]
\emph{Weights.} For \(\nu\ge0\), write the posterior draw as
\((a,b)+\Sigma^{1/2}\zeta\) with \(\zeta\sim N(0,I_2)\). The integrand
\(e^{\nu(\lvert\tilde\alpha\rvert+\lvert\tilde\beta\rvert)}\) is continuous in
\((a,b,\Sigma)\) and, on bounded sets, dominated by \(e^{K(1+\lvert\zeta\rvert)}\),
so \(\mathsf W_\nu\) is finite and continuous. Jensen's inequality gives the
lower bound below, and \(2pq\le p^2+q^2\), the coordinate marginals
\(N(a,\Sigma_{11})\), \(N(b,\Sigma_{22})\) and
\(e^{t\lvert u\rvert}\le e^{tu}+e^{-tu}\) give the upper bound,
\begin{equation}
 e^{\nu(\lvert a\rvert+\lvert b\rvert)}\le\mathsf W_\nu(s)
 \le e^{2\nu\lvert a\rvert+2\nu^2\Sigma_{11}}+e^{2\nu\lvert b\rvert+2\nu^2\Sigma_{22}}.
 \label{S1-eq:weight-bounds}
\end{equation}

\emph{Bounds.} For \(x\in X\),
\(\lvert\alpha+\beta x+\epsilon\rvert\le\lvert\alpha\rvert+R_X\lvert\beta\rvert+\lvert\epsilon\rvert\),
so the growth bound gives the pointwise envelope
\(\lvert U(x,\alpha+\beta x+\epsilon)\rvert
 \le Ce^{c_U\lvert\epsilon\rvert}e^{\lambda(\lvert\alpha\rvert+\lvert\beta\rvert)}\).
Integrating over \(\epsilon\sim N(0,1)\), with
\(Ee^{c_U\lvert\epsilon\rvert}\le2e^{c_U^2/2}\), gives
\(\lvert g^{\mathrm{full}}(x;\vartheta)\rvert\le\bar v(\vartheta)
 \le Ce^{\lambda(\lvert\alpha\rvert+\lvert\beta\rvert)}\), the growth
clause. Since \(g_s(x)=E_s[U(x,\tilde\alpha+\tilde\beta x+Z)]\) with
\((\tilde\alpha,\tilde\beta)\sim\mu_s\) independent of \(Z\), the pointwise
envelope under the posterior gives the reward bound
\(\lvert g_s(x)\rvert\le C\mathsf W_\lambda(s)\).
By \PartILemPosterior(a)--(b),
which applies to admissible policies by \PartIRemPredictable, the posterior
of \(\vartheta\) given \(\mathcal F_t\) is \(\mu_{s_t}\), and that given
\(\mathcal H_{t+1}\supseteq\mathcal F_t\) is \(\mu_{s_{t+1}}\). The tower
property, with the current seed independent of the current posterior, gives
\(E[\mathsf W_\lambda(s_{t+1})\mid\mathcal F_t]=\mathsf W_\lambda(s_t)\), hence
\(E_s[\mathsf W_\lambda(s_t)]=\mathsf W_\lambda(s)\); with
\(E[\lvert U(x_{t+1},y_{t+1})\rvert\mid\mathcal F_t]\le C\mathsf W_\lambda(s_t)\),
from the same envelope, this proves \eqref{A-eq:tail}.

\emph{Contraction and continuity.} For
\(f,h\in\mathcal C_\lambda\),
\[
 \lvert\mathcal B_\delta f(s)-\mathcal B_\delta h(s)\rvert
 \le\delta\lVert f-h\rVert_\lambda E_s[\mathsf W_\lambda(s^+)]
 =\delta\mathsf W_\lambda(s)\lVert f-h\rVert_\lambda;
\]
the space is complete, since \(f\mapsto f/\mathsf W_\lambda\) identifies it with a closed
space of bounded continuous functions. On a compact current-state/action set,
successor means are affine in \(Z\) with bounded coefficients and successor
covariances are bounded, so
\(\mathsf W_\lambda(\mathcal T(s,x,Z))\le C_Ke^{c_K\lvert Z\rvert}\), and
dominated convergence makes \((s,x)\mapsto E_Z[f(\mathcal T(s,x,Z))]\) jointly
continuous for every \(f\in\mathcal C_\lambda\). So is
\(g_s(x)=E_Z[U(x,a+bx+\sigma_IZ)]\), with
\(\sigma_I^2=1+A^{-1}+V^{-1}(x-\bar x)^2\). Splitting the Gaussian
\(\sigma_IZ\) into independent parts with variances \(1\) and
\(\tau_{\mathrm{var}}=\sigma_I^2-1\ge0\), Fubini's theorem gives
\(g_s(x)=E_W[G(x,a+bx+\sqrt{\tau_{\mathrm{var}}}\,W)]\) with
\(W\sim N(0,1)\); here \(\tau_{\mathrm{var}}\) is continuous and bounded on
compact sets, \(G\) is continuous, and
\(\lvert G(x,m)\rvert\le Ce^{c_U\lvert m\rvert}\) gives the dominating
function
\(Ce^{c_U(\lvert a+bx\rvert+\sqrt{\tau_{\mathrm{var}}}\lvert W\rvert)}\).
Hence, by compactness
of \(X\), \(\mathcal B_\delta f\) is continuous with attained maximum, and at
the fixed point \(J_\delta\) is continuous and \(Q_\delta\) jointly continuous.
The smallest maximizer is Borel. It is less than \(c\) if and only if
\(\max_{x\in X}Q_\delta(s,x)=\max_{x\in X\cap(-\infty,q]}Q_\delta(s,x)\) for
some rational \(q<c\) with nonempty restricted set, and both maxima are
continuous in \(s\).

\emph{Verification.} Let \(\pi\) be admissible, write
\(U_{t+1}=U(x_{t+1},y_{t+1})\), let \(\mathcal F_t\) be generated by the
initial state, the seeds through date \(t+1\) and the outcomes through date
\(t\), and put \(\omega_t=J_\delta(s_t)-Q_\delta(s_t,x_{t+1})\), nonnegative by
the fixed-point identity \(J_\delta=\max_{x\in X}Q_\delta(\cdot,x)\). Given
\(\mathcal F_t\), the outcome \(y_{t+1}\) has the posterior-predictive law at
\((s_t,x_{t+1})\), by \PartILemPosterior(a)--(b) and \PartIRemPredictable;
hence
\(J_\delta(s_t)=\omega_t+E_s^\pi[U_{t+1}+\delta J_\delta(s_{t+1})\mid\mathcal F_t]\).
Every term is integrable, since
\(\lvert J_\delta(s_t)\rvert\le\lVert J_\delta\rVert_\lambda\mathsf W_\lambda(s_t)\),
with \(E_s^\pi\mathsf W_\lambda(s_t)=\mathsf W_\lambda(s)\) and \eqref{A-eq:tail}.
Multiplying by \(\delta^t\), taking expectations,
summing over \(m\le t<n\), and letting \(n\to\infty\), with remainder
\(\lvert\delta^nE_s^\pi J_\delta(s_n)\rvert
\le\delta^n\lVert J_\delta\rVert_\lambda\mathsf W_\lambda(s)\), gives
\[
 \delta^mE_s^\pi J_\delta(s_m)
 =E_s^\pi\sum_{t\ge m}\delta^tU_{t+1}+\sum_{t\ge m}\delta^tE_s^\pi\omega_t
 \qquad(m\ge0).
\]
At \(m=0\),
\(J_\delta(s)=J^\pi(s)+\sum_{t\ge0}\delta^tE_s^\pi\omega_t\ge J^\pi(s)\), with
equality when \(\omega_t=0\) almost surely for every \(t\), in particular for
the stationary policy \(x_{t+1}=\xi(s_t)\), admissible because the posterior
recursions are Borel; so \(J_\delta\) is the optimal value and this policy
attains it. If the first action of \(\pi\) is \(x\), then
\(E_s^\pi U_1=g_s(x)\) and \(E_s^\pi J_\delta(s_1)=E_Z[J_\delta(\mathcal T(s,x,Z))]\),
so \(m=1\) gives
\(Q_\delta(s,x)=J^\pi(s)+\sum_{t\ge1}\delta^tE_s^\pi\omega_t\ge J^\pi(s)\),
with equality when every later action lies in the argmax almost surely, as
for the admissible policy that plays \(x\) and then \(\xi(s_t)\). This proves
the first-action identity.
\end{proof}

The following comparison is used in Section~\ref{S1-section-4}.

\begin{auxlemma}[Comparison on an invariant set]\label{S1-lem:comparison}
Let \(\nu\ge\lambda\), let \(F\subseteq\mathcal S\) satisfy
\(\mathcal T(s,x,z)\in F\) for every \(s\in F\), \(x\in X\) and \(z\in\R\),
and let \(h\) be continuous on \(\mathcal S\) with
\(\lvert h\rvert\le A\mathsf W_\nu\) on \(F\) for a constant \(A\).
\begin{enumerate}[label=\textup{(\alph*)},ref=(\alph*)]
\item If \(g_s(x)+\delta E_Z[h(\mathcal T(s,x,Z))]\le h(s)\) for every
\(s\in F\) and \(x\in X\), then \(J_\delta\le h\) on \(F\).\label{S1-lem:comparison:a}
\item If for every \(s\in F\) some \(x\in X\) satisfies
\(h(s)\le g_s(x)+\delta E_Z[h(\mathcal T(s,x,Z))]\), then \(h\le J_\delta\)
on \(F\).\label{S1-lem:comparison:b}
\end{enumerate}
\end{auxlemma}

\begin{proof}
Since \(\mathsf W_\lambda\le\mathsf W_\nu\),
\(\lvert J_\delta\rvert\le\lVert J_\delta\rVert_\lambda\mathsf W_\nu\), and
\(E_Z[\mathsf W_\nu(\mathcal T(s,x,Z))]=\mathsf W_\nu(s)\) by the tower property,
as for \(\mathsf W_\lambda\). In \ref{S1-lem:comparison:a}, induction on \(n\)
with the Bellman equation \(J_\delta=\max_{x\in X}Q_\delta(\cdot,x)\) gives
\(J_\delta-h\le\delta^n(\lVert J_\delta\rVert_\lambda+A)\mathsf W_\nu\) on \(F\),
because every successor stays in \(F\); let \(n\to\infty\). Part
\ref{S1-lem:comparison:b} is symmetric, with
\(J_\delta\ge Q_\delta(\cdot,x)\) at the action supplied by the hypothesis.
\end{proof}

\subsection{Gaussian smoothing of the payoff}\label{S1-section-smoothing}
The construction uses the payoff only through the growth bound and
\(G_{\mathrm{var}}\) of Section~\ref{B-sec:scope}. For \(x\in X\), \(m\in\R\) and
\(\tau_{\mathrm{var}}>-1\),
\begin{equation}\label{S1-eq:density}
 G_{\mathrm{var}}(x,m,\tau_{\mathrm{var}})
 =\int_\R U(x,y)\,\mathsf n(y-m,\tau_{\mathrm{var}})\,dy,
\end{equation}
where
\(\mathsf n(d,\tau)\coloneqq[2\pi(1+\tau)]^{-1/2}\exp\bigl(-d^2/[2(1+\tau)]\bigr)\)
is the density of \(N(0,1+\tau)\). The function \(G=G_{\mathrm{var}}(\cdot,\cdot,0)\) is continuous on
\(X\times\R\) by Appendix~\ref{app:value}.

\begin{auxlemma}[Gaussian smoothing]\label{S1-lem:smoothing}
Suppose that \(U\) satisfies Assumption~\ref{ass:fl-payoff}.
\begin{enumerate}[label=\textup{(\alph*)},ref=(\alph*)]
\item \(G_{\mathrm{var}}\) is \(C^4\) on\label{S1-lem:smoothing:a}
\(\operatorname{int}X\times\R\times(-1,\infty)\), and its derivatives of
order at most four are obtained by differentiating under the integral in
\eqref{S1-eq:density}. For every \(R<\infty\) and \(\tau_0\in(-1,R]\), these
derivatives are bounded on
\(\operatorname{int}X\times[-R,R]\times[\tau_0,R]\).
\item \(G_{\mathrm{var},\tau_{\mathrm{var}}}=\tfrac12G_{\mathrm{var},mm}\)\label{S1-lem:smoothing:b}
on \(\operatorname{int}X\times\R\times(-1,\infty)\).
\item For every \(R<\infty\) there is \(B<\infty\) such that\label{S1-lem:smoothing:c}
\(\lvert G_{\mathrm{var}}(x,m,\tau_{\mathrm{var}})-G(x,m)\rvert\le B\tau_{\mathrm{var}}\)
for every \(x\in X\), \(\lvert m\rvert\le R\) and
\(\tau_{\mathrm{var}}\in[0,3]\).
\end{enumerate}
\end{auxlemma}

\begin{proof}
In \eqref{S1-eq:density}, \((m,\tau_{\mathrm{var}})\) enters only through
the Gaussian density \(\mathsf n\), which is smooth on
\(\R\times(-1,\infty)\), and \(x\) enters only through \(U\), which is
\(C^4\) in \(x\). So at an interior action every derivative of order
\(j\le4\) of the integrand in \((x,m,\tau_{\mathrm{var}})\) is a finite sum of
terms \(\partial_x^iU(x,y)\,P(y-m)\,\mathsf n(y-m,\tau_{\mathrm{var}})\) with
\(i\le j\) and \(P\) a polynomial of degree at most \(2j\) whose coefficients
are continuous in \(\tau_{\mathrm{var}}>-1\). The functions
\(\partial_x^iU(x,\cdot)\) are Borel, as pointwise limits of difference
quotients. For \(\lvert m\rvert\le R\) and
\(\tau_{\mathrm{var}}\in[\tau_0,R]\),
\(\mathsf n(y-m,\tau_{\mathrm{var}})\le[2\pi(1+\tau_0)]^{-1/2}
\exp\bigl(-(\lvert y\rvert-R)_+^2/[2(1+R)]\bigr)\), so, by
Assumption~\ref{ass:fl-payoff}, each term is dominated, uniformly in
\(x\in\operatorname{int}X\), by a multiple of
\((1+\lvert y\rvert)^8e^{c_U\lvert y\rvert}\exp\bigl(-(\lvert y\rvert-R)_+^2/[2(1+R)]\bigr)\),
which is integrable. Differentiation under the integral and continuity of the
derivatives follow from dominated convergence, and integrating the dominating
function bounds the derivatives; this gives \ref{S1-lem:smoothing:a}. No
derivative of \(U\) in the outcome is used. A direct calculation gives
\(\partial_\tau\mathsf n(d,\tau)=\tfrac12\partial_d^2\mathsf n(d,\tau)\), and
\(\partial_m^2\mathsf n(y-m,\tau)=\partial_d^2\mathsf n(y-m,\tau)\), so
differentiating under the integral gives \ref{S1-lem:smoothing:b}.

For \ref{S1-lem:smoothing:c}, fix \(x\in X\). Only the Borel measurability
of \(U(x,\cdot)\) and the growth bound are used, so the endpoints of \(X\) are
included. For \(\tau\in[0,3]\),
\[
 \partial_\tau\mathsf n(d,\tau)
 =\frac{1}{2(1+\tau)}\left(\frac{d^2}{1+\tau}-1\right)\mathsf n(d,\tau),
 \qquad
 \lvert\partial_\tau\mathsf n(d,\tau)\rvert
 \le\frac{1+d^2}{2\sqrt{2\pi}}e^{-d^2/8}\le c\,e^{-d^2/16}
\]
for a numerical constant \(c\), so
\(\lvert\mathsf n(d,\tau)-\mathsf n(d,0)\rvert\le c\tau e^{-d^2/16}\). Hence,
for \(\lvert m\rvert\le R\) and \(\tau_{\mathrm{var}}\in[0,3]\),
\[
\begin{aligned}
 \lvert G_{\mathrm{var}}(x,m,\tau_{\mathrm{var}})-G(x,m)\rvert
 &\le c\tau_{\mathrm{var}}\int_\R Ce^{c_U\lvert y\rvert}e^{-(y-m)^2/16}\,dy\\
 &\le c\tau_{\mathrm{var}}Ce^{c_UR}\int_\R e^{c_U\lvert v\rvert-v^2/16}\,dv,
\end{aligned}
\]
which is \ref{S1-lem:smoothing:c}.
\end{proof}

\subsection{A direct fixed-point lemma}\label{S1-section-2}\label{S1-fixed-point-proof}
In this subsection, \(J^{\mathrm{aux}}=J_{\delta,\epsilon}^{\mathrm{aux}}\)
and \(e_{\mathrm{ctr}}^\star\) suppress discount and scale indices; all
stated uniformity in \(\epsilon_{\mathrm{sc}}\) is retained.

\begin{auxlemma}[Direct fixed-point lemma]\label{S1-lem:fixed-point}
Let \(k_{\mathrm{reg}}=4\). Suppose
\(\mathcal T_\epsilon^{\mathrm{aux}}(z,e_{\mathrm{ctr}},Z)
=z+\epsilon_{\mathrm{sc}}F_\epsilon^{\mathrm{aux}}(z,e_{\mathrm{ctr}},Z)\),
\(e_{\mathrm{ctr}}\in[-h_x,h_x]\) for some \(h_x>0\), where
\(F_\epsilon^{\mathrm{aux}}(\cdot,\cdot,Z)\) is \(C^{k_{\mathrm{reg}}}\) on a
neighborhood of each point of the control strip
\(\{\lvert e_{\mathrm{ctr}}\rvert\le h_x\}\) for every \(Z\),
\(F_\epsilon^{\mathrm{aux}}(z,e_{\mathrm{ctr}},\cdot)\) is measurable for
every \((z,e_{\mathrm{ctr}})\) in the strip, and, in ordinary multilinear
operator norms,
\[
\sup_{z,e_{\mathrm{ctr}}}\sum_{j=1}^{k_{\mathrm{reg}}}\lvert\mathrm D^jF_\epsilon^{\mathrm{aux}}(z,e_{\mathrm{ctr}},Z)\rvert
\le K_{\mathrm{der}}(Z),\qquad E[(1+K_{\mathrm{der}})^{k_{\mathrm{reg}}}]<\infty.
\]
Suppose \(g_\epsilon^{\mathrm{aux}}\) is \(C^{k_{\mathrm{reg}}}\) on a
neighborhood of each point of the control strip, its derivatives through order
\(k_{\mathrm{reg}}\), including its value, are bounded by
\(C_{\mathrm{der}}\) on the strip, and
\[
{g_\epsilon^{\mathrm{aux}}}_{e_{\mathrm{ctr}}e_{\mathrm{ctr}}}\le-C_{\mathrm{curv}}^{\mathrm{fp}}<0,\quad
{g_\epsilon^{\mathrm{aux}}}_{e_{\mathrm{ctr}}}(z,-h_x)\ge C_{\mathrm{end}}>0,\quad
{g_\epsilon^{\mathrm{aux}}}_{e_{\mathrm{ctr}}}(z,h_x)\le-C_{\mathrm{end}}.
\]
For fixed \(\delta\in[0,1)\), sufficiently small \(\epsilon_{\mathrm{sc}}\)
gives a unique bounded continuous fixed point
\[
J^{\mathrm{aux}}(z)=\max_{\lvert e_{\mathrm{ctr}}\rvert\le h_x}
 \bigl(g_\epsilon^{\mathrm{aux}}(z,e_{\mathrm{ctr}})+\delta E[J^{\mathrm{aux}}(\mathcal T_\epsilon^{\mathrm{aux}}(z,e_{\mathrm{ctr}},Z))]\bigr).
\]
It satisfies \(J^{\mathrm{aux}}\in C^{k_{\mathrm{reg}}-1,1}\), and its unique
selector satisfies \(e_{\mathrm{ctr}}^\star\in C^{k_{\mathrm{reg}}-2,1}\).
Their stated norms are bounded uniformly for every sufficiently small
\(\epsilon_{\mathrm{sc}}\).
\end{auxlemma}

\begin{proof}[Proof of Lemma~\ref{S1-lem:fixed-point}]
Fix \(\delta\in[0,1)\) and reward and transition families satisfying the
hypotheses; the scale is chosen after the derivative radii below.

\emph{Invariant set.} For a symmetric scalar \(j\)-tensor \(\mathsf A\), put
\(\lvert\mathsf A\rvert_{\mathrm{diag}}=\sup_{\lvert\boldsymbol h\rvert=1}\lvert\mathsf A[\boldsymbol h,\ldots,\boldsymbol h]\rvert\).
Polarization gives
\(\lvert\mathsf A[\boldsymbol h_1,\ldots,\boldsymbol h_j]\rvert
 \le C_j^{\mathrm{pol}}\lvert\mathsf A\rvert_{\mathrm{diag}}\prod_i\lvert\boldsymbol h_i\rvert\)
with \(C_j^{\mathrm{pol}}\) depending only on \(j\). Put
\(C_j^{\mathrm{mom}}=E[(1+K_{\mathrm{der}})^j]\) and \(C_B=(1+\delta)/2\).
For radii \(C_0^{\mathrm{der}},\ldots,C_{k_{\mathrm{reg}}}^{\mathrm{der}}\),
let \(\mathcal K_{\mathrm{der}}\) consist of the bounded continuous functions
with
\begin{equation}
\lVert F\rVert_\infty\le C_0^{\mathrm{der}},\qquad
\lVert\mathrm D^jF\rVert_{\mathrm{diag}}\le C_j^{\mathrm{der}}\quad(1\le j\le k_{\mathrm{reg}}),
\label{S1-eq-2-5}
\end{equation}
the derivative inequalities being distributional. Testing each directional
inequality against compactly supported test functions shows that this set is
closed in the supremum metric. By polarization, its elements have bounded weak
partial derivatives, hence \(C^{k_{\mathrm{reg}}-1,1}\) representatives;
mollification preserves \eqref{S1-eq-2-5} and, the first weak derivative
being bounded, converges uniformly.

\emph{Objective bounds.} Fix a smooth trial \(F\in\mathcal K_{\mathrm{der}}\)
(at the fixed point, \(F=J^{\mathrm{aux}}\)) and write
\(Q^{\mathrm{aux}}(z,e_{\mathrm{ctr}})=g_\epsilon^{\mathrm{aux}}(z,e_{\mathrm{ctr}})
+\delta E[F(\mathcal T_\epsilon^{\mathrm{aux}}(z,e_{\mathrm{ctr}},Z))]\). Then
\begin{align}
Q^{\mathrm{aux}}_{e_{\mathrm{ctr}}e_{\mathrm{ctr}}}&\le-C_{\mathrm{curv}}^{\mathrm{fp}}+
 \delta\bigl(\epsilon_{\mathrm{sc}}^2C_2^{\mathrm{der}}E[K_{\mathrm{der}}^2]+\epsilon_{\mathrm{sc}}C_1^{\mathrm{der}}E[K_{\mathrm{der}}]\bigr),
\label{S1-eq-2-6}
\\
\lvert Q^{\mathrm{aux}}_{e_{\mathrm{ctr}}}-{g_\epsilon^{\mathrm{aux}}}_{e_{\mathrm{ctr}}}\rvert&\le\delta\epsilon_{\mathrm{sc}}C_1^{\mathrm{der}}E[K_{\mathrm{der}}],
\label{S1-eq-2-7}
\\
\lVert Q^{\mathrm{aux}}_{ze_{\mathrm{ctr}}}\rVert
&\le C_{\mathrm{der}}+\delta\epsilon_{\mathrm{sc}}
 \bigl(C_2^{\mathrm{pol}}C_2^{\mathrm{der}}E[K_{\mathrm{der}}(1+\epsilon_{\mathrm{sc}}K_{\mathrm{der}})]+C_1^{\mathrm{der}}E[K_{\mathrm{der}}]\bigr).
\label{S1-eq-2-8}
\end{align}
For \(\epsilon_{\mathrm{sc}}\) sufficiently small, the maximizer is unique
and interior,
\(Q^{\mathrm{aux}}_{e_{\mathrm{ctr}}e_{\mathrm{ctr}}}\le-C_{\mathrm{curv}}^{\mathrm{fp}}/2\)
and
\(\lVert\mathrm De_{\mathrm{ctr}}^\star\rVert\le C_{e^\star}^{\mathrm{der}}\coloneqq2(C_{\mathrm{der}}+1)/C_{\mathrm{curv}}^{\mathrm{fp}}\).
Set \(C_{\mathrm{chain}}=1+C_{e^\star}^{\mathrm{der}}\).

\emph{First order.} The envelope identity gives
\[
 \mathrm D(\mathcal B_{\delta,\epsilon}^{\mathrm{aux}}F)
 ={g_\epsilon^{\mathrm{aux}}}_{z}+\delta E[\mathrm DF(\mathcal T_\epsilon^{\mathrm{aux}}){\mathcal T_\epsilon^{\mathrm{aux}}}_{z}],
 \qquad
 \lVert\mathrm D\mathcal B_{\delta,\epsilon}^{\mathrm{aux}}F\rVert
 \le C_{\mathrm{der}}+\delta E[(1+\epsilon_{\mathrm{sc}}K_{\mathrm{der}})C_1^{\mathrm{der}}].
\]

\emph{Orders two through four.} Restrict to a unit state line, and put
\(\boldsymbol\gamma=(z,e_{\mathrm{ctr}}^\star(z))\) and
\(f_{\mathrm{line}}=Q^{\mathrm{aux}}\circ\boldsymbol\gamma\).
Differentiating the first-order condition gives
\[
f_{\mathrm{line}}''={Q^{\mathrm{aux}}}^{(2)}[\boldsymbol\gamma'^2],\ \
f_{\mathrm{line}}'''={Q^{\mathrm{aux}}}^{(3)}[\boldsymbol\gamma'^3],\ \
f_{\mathrm{line}}''''={Q^{\mathrm{aux}}}^{(4)}[\boldsymbol\gamma'^4]-3Q^{\mathrm{aux}}_{e_{\mathrm{ctr}}e_{\mathrm{ctr}}}({e_{\mathrm{ctr}}^\star}'')^2.
\]
Fix \(2\le j\le k_{\mathrm{reg}}\). The partition formula for
differentiating a composition and polarization give
\(\lVert\mathrm D^jQ^{\mathrm{aux}}\rVert_{\mathrm{op}}\le C_j^Q\) for
\(\epsilon_{\mathrm{sc}}\le1\), with \(C_j^Q<\infty\) depending only on \(j\),
\(\delta\), \(C_{\mathrm{der}}\), \(C_1^{\mathrm{der}},\ldots,C_j^{\mathrm{der}}\)
and \(C_1^{\mathrm{mom}},\ldots,C_j^{\mathrm{mom}}\). In that formula for
\(\mathrm D^j(F\circ\mathcal T_\epsilon^{\mathrm{aux}})[\boldsymbol\gamma',\ldots,\boldsymbol\gamma']\),
the partition into singletons gives the diagonal leading term, with
coefficient one rather than \(C_j^{\mathrm{pol}}\):
\[
\bigl\lvert
\delta E[\mathrm D^jF(\mathcal T_\epsilon^{\mathrm{aux}})
       [\mathrm D\mathcal T_\epsilon^{\mathrm{aux}}\boldsymbol\gamma',\ldots,\mathrm D\mathcal T_\epsilon^{\mathrm{aux}}\boldsymbol\gamma']]
\bigr\rvert
\le\delta E[(1+\epsilon_{\mathrm{sc}}K_{\mathrm{der}}C_{\mathrm{chain}})^jC_j^{\mathrm{der}}].
\]
Every other partition has fewer than \(j\) blocks, so it involves
\(\mathrm D^rF\) with \(r\le j-1\), bounded through \(C_r^{\mathrm{der}}\) and
polarization; each block of size at least two contributes a factor
\(\epsilon_{\mathrm{sc}}\mathrm D^{\lvert B\rvert}F_\epsilon^{\mathrm{aux}}\);
and \(\lvert\boldsymbol\gamma'\rvert\le C_{\mathrm{chain}}\). Hence some
\(C_j^{\mathrm{jet}}<\infty\), depending only on \(j\), \(\delta\),
\(C_{\mathrm{der}}\), \(C_{\mathrm{chain}}\),
\(C_1^{\mathrm{der}},\ldots,C_{j-1}^{\mathrm{der}}\) and
\(C_1^{\mathrm{mom}},\ldots,C_j^{\mathrm{mom}}\), bounds these lower-jet terms
together with the reward term
\(\mathrm D^jg_\epsilon^{\mathrm{aux}}[\boldsymbol\gamma',\ldots,\boldsymbol\gamma']\).
The differentiated first-order condition gives
\(Q^{\mathrm{aux}}_{e_{\mathrm{ctr}}e_{\mathrm{ctr}}}{e_{\mathrm{ctr}}^\star}''
=-{Q^{\mathrm{aux}}}^{(3)}[\boldsymbol e_{\mathrm{ctr}},\boldsymbol\gamma',\boldsymbol\gamma']\),
with \(\boldsymbol e_{\mathrm{ctr}}\) the unit action direction. Using
\(\lvert\boldsymbol\gamma'\rvert\le C_{\mathrm{chain}}\) and, once in each
bound,
\(\lvert Q^{\mathrm{aux}}_{e_{\mathrm{ctr}}e_{\mathrm{ctr}}}\rvert\ge C_{\mathrm{curv}}^{\mathrm{fp}}/2\),
the second after writing
\(-3Q^{\mathrm{aux}}_{e_{\mathrm{ctr}}e_{\mathrm{ctr}}}({e_{\mathrm{ctr}}^\star}'')^2
=3({Q^{\mathrm{aux}}}^{(3)}[\boldsymbol e_{\mathrm{ctr}},\boldsymbol\gamma',\boldsymbol\gamma'])^2/\lvert Q^{\mathrm{aux}}_{e_{\mathrm{ctr}}e_{\mathrm{ctr}}}\rvert\),
\[
 \lvert {e_{\mathrm{ctr}}^\star}''\rvert\le\frac2{C_{\mathrm{curv}}^{\mathrm{fp}}}C_3^QC_{\mathrm{chain}}^2,\qquad
 -3Q^{\mathrm{aux}}_{e_{\mathrm{ctr}}e_{\mathrm{ctr}}}({e_{\mathrm{ctr}}^\star}'')^2\le\frac6{C_{\mathrm{curv}}^{\mathrm{fp}}}{C_3^Q}^2C_{\mathrm{chain}}^4.
\]
Put \(C_1^{\mathrm{lower}}=C_{\mathrm{der}}\),
\(C_j^{\mathrm{lower}}=C_j^{\mathrm{jet}}\) for \(j=2,3\), and
\(C_4^{\mathrm{lower}}=C_4^{\mathrm{jet}}+6{C_3^Q}^2C_{\mathrm{chain}}^4/C_{\mathrm{curv}}^{\mathrm{fp}}\).
Then, for \(1\le j\le k_{\mathrm{reg}}\) and every smooth trial
\(F\in\mathcal K_{\mathrm{der}}\),
\[
\lVert\mathrm D^j(\mathcal B_{\delta,\epsilon}^{\mathrm{aux}}F)\rVert_{\mathrm{diag}}
\le C_j^{\mathrm{lower}}
+\delta E[(1+\epsilon_{\mathrm{sc}}K_{\mathrm{der}}C_{\mathrm{chain}})^j]\,C_j^{\mathrm{der}},
\]
where \(C_j^{\mathrm{lower}}\) depends, besides \(\delta\) and \(j\), only on
\(C_{\mathrm{der}}\), \(C_{\mathrm{curv}}^{\mathrm{fp}}\),
\(C_1^{\mathrm{der}},\ldots,C_{j-1}^{\mathrm{der}}\) and
\(C_1^{\mathrm{mom}},\ldots,C_j^{\mathrm{mom}}\), through
\(C_{\mathrm{chain}}\), \(C_j^{\mathrm{jet}}\) and \(C_3^Q\).

\emph{Order of choices.} Choose
\(C_0^{\mathrm{der}}>C_{\mathrm{der}}/(1-\delta)\) and, recursively,
\(C_j^{\mathrm{der}}>C_j^{\mathrm{lower}}/(1-C_B)\). Only after choosing every
radius, choose \(\epsilon_{\mathrm{sc}}\) so that, through
\eqref{S1-eq-2-6}--\eqref{S1-eq-2-8},
\(Q^{\mathrm{aux}}_{e_{\mathrm{ctr}}e_{\mathrm{ctr}}}\le-C_{\mathrm{curv}}^{\mathrm{fp}}/2\),
\(\delta\epsilon_{\mathrm{sc}}C_1^{\mathrm{der}}E[K_{\mathrm{der}}]<C_{\mathrm{end}}\)
(interiority of the maximizer), the \(\epsilon_{\mathrm{sc}}\)-term of
\eqref{S1-eq-2-8} is at most \(1\) (whence
\(\lVert\mathrm De_{\mathrm{ctr}}^\star\rVert\le C_{e^\star}^{\mathrm{der}}\)),
and
\(\delta E[(1+\epsilon_{\mathrm{sc}}K_{\mathrm{der}}C_{\mathrm{chain}})^{k_{\mathrm{reg}}}]\le C_B\).
These choices are acyclic and make the smooth part of
\(\mathcal K_{\mathrm{der}}\) invariant; mollification, closure and
supremum-norm continuity of the Bellman operator extend invariance to all of
\(\mathcal K_{\mathrm{der}}\). The operator contracts by \(\delta\) in the
supremum metric, so it has a fixed point in this complete set. Finally,
differentiating \(Q^{\mathrm{aux}}_{e_{\mathrm{ctr}}}(z,e_{\mathrm{ctr}}^\star(z))=0\)
and using
\(\lvert Q^{\mathrm{aux}}_{e_{\mathrm{ctr}}e_{\mathrm{ctr}}}\rvert\ge C_{\mathrm{curv}}^{\mathrm{fp}}/2\)
gives uniform \(C^{k_{\mathrm{reg}}-2,1}\) bounds for the selector; at the
highest order, Lipschitz continuity follows from the Lipschitz highest
derivatives of \(Q^{\mathrm{aux}}_{e_{\mathrm{ctr}}}\), the bounded lower
derivatives of \(e_{\mathrm{ctr}}^\star\), and the denominator bound.
\end{proof}

\subsection{An explicit global Gaussian artificial problem}\label{S1-section-3}
Use the compact family \(K^{\mathrm{mean}}\) of Lemma~\ref{B-lem:virtual}
(for Theorem~\ref{thm:fl-local}, \(\{(a^e,b^e)\}\)), its optimizer
\(x^\star(a,b)\), and the regular-optimizer preliminary, standing
assumptions and charts of Section~\ref{B-sec:scope}. The scale
\(\epsilon_{\mathrm{sc}}\) is chosen last, after the derivative radii of
Section~\ref{S1-section-2} and the tracking constants of
Section~\ref{S1-section-4}.

\paragraph{Collars and action center.}
Choose \(r_{\mathrm{mean}}>0\) so that the closed
\(6r_{\mathrm{mean}}\)-neighborhood \(K^{\mathrm{mean}}_{6r_{\mathrm{mean}}}\)
lies in the neighborhood of \(K^{\mathrm{mean}}\) on which
Section~\ref{B-sec:scope} gives a unique interior maximizer, strictly negative
curvature and \(x^\star\in C^3\), and in \(\Theta^I\) when
\(K^{\mathrm{mean}}\subset\Theta^I\). Choose
\(h_x,C_{\mathrm{curv}}^{\mathrm{stat}}>0\) such that, on
\(K^{\mathrm{mean}}_{6r_{\mathrm{mean}}}\),
\[
x^\star(a,b)\pm2h_x\in\operatorname{int}X,
\qquad
(g^{\mathrm{full}})_{xx}(x;a,b)\le-C_{\mathrm{curv}}^{\mathrm{stat}}
\quad(\lvert x-x^\star(a,b)\rvert\le2h_x).
\]
By compactness and uniqueness, there is also a uniform positive global gap
away from this local maximizing neighborhood. Choose a globally smooth
bounded \(x_{\mathrm{ctr}}(a,b)\) (a smooth-cutoff convex combination of
\(x^\star\) with a fixed interior action, mollified) with
\(\lvert x_{\mathrm{ctr}}-x^\star\rvert\le h_x/4\) on
\(K^{\mathrm{mean}}_{4r_{\mathrm{mean}}}\) and
\(x_{\mathrm{ctr}}(a,b)+[-h_x,h_x]\subset\operatorname{int}X\) globally, and
use the action coordinate \(x=x_{\mathrm{ctr}}(a,b)+e_{\mathrm{ctr}}\),
\(\lvert e_{\mathrm{ctr}}\rvert\le h_x\). With a nonnegative smooth
unit-integral mollifier \(\varrho^{\mathrm{sm}}\) supported in the unit ball,
\(\chi_{\mathrm{cut}}=\mathbf 1_{K^{\mathrm{mean}}_{7r_{\mathrm{mean}}/2}}*\varrho^{\mathrm{sm}}_{r_{\mathrm{mean}}/4}\)
equals \(1\) on \(K^{\mathrm{mean}}_{3r_{\mathrm{mean}}}\) and is supported in
\(K^{\mathrm{mean}}_{4r_{\mathrm{mean}}}\). The comparison with the true value uses the inner
mean collar \(K^{\mathrm{mean}}_{r_{\mathrm{mean}}}\) and the outer mean collar
\(K^{\mathrm{mean}}_{2r_{\mathrm{mean}}}\).

\paragraph{Explicit saturations.}
Fix a standard smooth cutoff \(\chi_0:\R\to[0,1]\) with \(\chi_0=1\) on
\((-\infty,0]\) and \(\chi_0=0\) on \([1,\infty)\). For \(R>0\), the map
\(\operatorname{Sat}_R(t)=\int_0^t\chi_0((z^2-R^2)/(3R^2))\,dz\) is smooth, equals \(t\) on \([-R,R]\), and is bounded by \(2R\) in absolute
value. Choose \(C_{\bar x}^{\mathrm{sat}}\) so large that \(X\) lies strictly
inside \([-C_{\bar x}^{\mathrm{sat}}+1,C_{\bar x}^{\mathrm{sat}}-1]\), and
choose mean-coordinate saturation radii large enough to contain
\(K^{\mathrm{mean}}_{4r_{\mathrm{mean}}}\) in their identity regions. For the
log-precision cap, put \(F_{\mathrm{cap}}(t)=\int_0^t\chi_0(z)\,dz\),
\(f_{\mathrm{cap}}=e^{F_{\mathrm{cap}}}\) and
\(\operatorname{Cap}_\epsilon^{\mathrm{var}}(\ell_V)
=\epsilon_{\mathrm{sc}}f_{\mathrm{cap}}(\ell_V-\log\epsilon_{\mathrm{sc}})\).
Then \(\operatorname{Cap}_\epsilon^{\mathrm{var}}(\ell_V)=e^{\ell_V}\) for
\(\ell_V\le\log\epsilon_{\mathrm{sc}}\), while
\(\operatorname{Cap}_\epsilon^{\mathrm{var}}\) and every fixed-order
\(\ell_V\)-derivative are \(O(\epsilon_{\mathrm{sc}})\), uniformly on the whole
real line.

Use the fixed-scale coordinates
\(z_\epsilon=(a,b,u_\epsilon,\bar x,\ell_V)\), with
\(u=\epsilon_{\mathrm{sc}}u_\epsilon\). Set
\(u_\epsilon^{\mathrm{sat}}=\operatorname{Sat}_2(u_\epsilon)\),
\(u^{\mathrm{sat}}=\epsilon_{\mathrm{sc}}u_\epsilon^{\mathrm{sat}}\) and
\(\bar x^{\mathrm{sat}}=\operatorname{Sat}_{C_{\bar x}^{\mathrm{sat}}}(\bar x)\),
let \(a^{\mathrm{sat}},b^{\mathrm{sat}}\) be the mean saturations, and define
\(D^{\mathrm{sat}}=x-\bar x^{\mathrm{sat}}\) and
\({\sigma_I^{\mathrm{sat}}}^2=1+u^{\mathrm{sat}}+\operatorname{Cap}_\epsilon^{\mathrm{var}}(\ell_V){D^{\mathrm{sat}}}^2\).
For sufficiently small \(\epsilon_{\mathrm{sc}}\), this variance is bounded
uniformly away from zero.

\paragraph{Artificial reward.} Define
\begin{equation}
g_\epsilon^{\mathrm{aux}}(z_\epsilon,e_{\mathrm{ctr}})
=\chi_{\mathrm{cut}}(a,b)\,
E[U(x,a^{\mathrm{sat}}+b^{\mathrm{sat}}x+\sigma_I^{\mathrm{sat}}Z)]
-(1-\chi_{\mathrm{cut}}(a,b))\frac{C_{\mathrm{curv}}^{\mathrm{stat}}}{4}e_{\mathrm{ctr}}^2.
\label{S1-eq-3-7}
\end{equation}
On the support of \(\chi_{\mathrm{cut}}\), the static objective has curvature
at most \(-C_{\mathrm{curv}}^{\mathrm{stat}}\) on the action interval and
endpoint derivatives of the appropriate signs. The exact Gaussian reward is
\(G_{\mathrm{var}}(x,a^{\mathrm{sat}}+b^{\mathrm{sat}}x,\tau_{\mathrm{var}}^{\mathrm{sat}})\)
with
\(\tau_{\mathrm{var}}^{\mathrm{sat}}={\sigma_I^{\mathrm{sat}}}^2-1
=u^{\mathrm{sat}}+\operatorname{Cap}_\epsilon^{\mathrm{var}}(\ell_V){D^{\mathrm{sat}}}^2\),
and the action \(x=x_{\mathrm{ctr}}(a,b)+e_{\mathrm{ctr}}\) lies in
\(\operatorname{int}X\). Since \(\tau_{\mathrm{var}}^{\mathrm{sat}}\) and its derivatives
are \(O(\epsilon_{\mathrm{sc}})\), the bounds of
Lemma~\ref{S1-lem:smoothing}\ref{S1-lem:smoothing:a} on the derivatives of
\(G_{\mathrm{var}}\) through order three make the exact Gaussian reward
\(O(\epsilon_{\mathrm{sc}})\)-close to the static objective in \(C^2\). No
derivative of \(U\) in the outcome is used. Hence, for small enough
\(\epsilon_{\mathrm{sc}}\), both components in \eqref{S1-eq-3-7} satisfy
\[
{g_\epsilon^{\mathrm{aux}}}_{e_{\mathrm{ctr}}e_{\mathrm{ctr}}}\le-C_{\mathrm{curv}}^{\mathrm{stat}}/2,\ \
{g_\epsilon^{\mathrm{aux}}}_{e_{\mathrm{ctr}}}(z_\epsilon,-h_x)\ge C_{\mathrm{curv}}^{\mathrm{stat}}h_x/2,\ \
{g_\epsilon^{\mathrm{aux}}}_{e_{\mathrm{ctr}}}(z_\epsilon,h_x)\le-C_{\mathrm{curv}}^{\mathrm{stat}}h_x/2,
\]
and the state-dependent blend preserves these inequalities. All derivatives of
\(g_\epsilon^{\mathrm{aux}}\) through order \(k_{\mathrm{reg}}\) are uniformly
bounded, by the chain rule and
Lemma~\ref{S1-lem:smoothing}\ref{S1-lem:smoothing:a}. The action stays in
\(\operatorname{int}X\), the mean and variance arguments of
\(G_{\mathrm{var}}\) and their derivatives are bounded, and the variance
argument stays in a compact subset of \((-1,\infty)\).

\paragraph{Full artificial successor map.} For every
\(z_\epsilon\in\R^5\) and every \(Z\in\R\), define
\begin{align*}
a^+&=a+\frac{u^{\mathrm{sat}}-\bar x^{\mathrm{sat}}\operatorname{Cap}_\epsilon^{\mathrm{var}}(\ell_V)D^{\mathrm{sat}}}
                    {\sigma_I^{\mathrm{sat}}}Z,&
b^+&=b+\frac{\operatorname{Cap}_\epsilon^{\mathrm{var}}(\ell_V)D^{\mathrm{sat}}}{\sigma_I^{\mathrm{sat}}}Z,\\
u_\epsilon^+&=u_\epsilon-\epsilon_{\mathrm{sc}}
       \frac{{u_\epsilon^{\mathrm{sat}}}^{\,2}}{1+\epsilon_{\mathrm{sc}}u_\epsilon^{\mathrm{sat}}},&\qquad
\bar x^+&=\bar x+\frac{u^{\mathrm{sat}}D^{\mathrm{sat}}}{1+u^{\mathrm{sat}}},\\
\ell_V^+&=\ell_V-
       \log\left(1+\frac{\operatorname{Cap}_\epsilon^{\mathrm{var}}(\ell_V){D^{\mathrm{sat}}}^2}{1+u^{\mathrm{sat}}}\right).
\end{align*}
Only the increments are saturated; the successor means themselves are not.
Every increment and every derivative of an increment through order
\(k_{\mathrm{reg}}\), in \((z_\epsilon,e_{\mathrm{ctr}})\), is bounded by
\(C\epsilon_{\mathrm{sc}}(1+\lvert Z\rvert)\). In the log increment every
differentiated term retains a factor
\(\operatorname{Cap}_\epsilon^{\mathrm{var}}(\ell_V)=O(\epsilon_{\mathrm{sc}})\),
the mean increments are linear in \(Z\), and the rest follows from the
bounded saturations and the denominator bound. Thus
Lemma~\ref{S1-lem:fixed-point} applies with
\(K_{\mathrm{der}}(Z)=C(1+\lvert Z\rvert)\). Let
\(J_{\delta,\epsilon}^{\mathrm{aux}}\) be its global bounded value, and let
\(\varphi_\epsilon(z_\epsilon)=x_{\mathrm{ctr}}(a,b)+e_{\mathrm{ctr},\epsilon}^\star(z_\epsilon)\)
be its action in the original coordinates. On
\((a,b)\in K^{\mathrm{mean}}_{3r_{\mathrm{mean}}}\),
\(\lvert u_\epsilon\rvert\le2\), \(\lvert\bar x\rvert\le C_{\bar x}^{\mathrm{sat}}\),
\(\ell_V\le\log\epsilon_{\mathrm{sc}}\), the reward and transition are exact for local actions, for \textit{every}
Gaussian realization.

\subsection{Global-action regret and tracking}\label{S1-section-4}
For fixed discount and scale, \(J^{\mathrm{aux},[\ell]}\),
\(\varphi^{[\ell]}\), \(J_\delta^{\mathrm{aux}}\) and \(\varphi\) are the maps
of Lemma~\ref{B-lem:virtual}\ref{B-virtual:regularity}. The \textit{outer
matching collar} consists of the feasible states with
\(\operatorname{dist}((a,b),K^{\mathrm{mean}})<2r_{\mathrm{mean}}\),
\(0<A^{-1}\le\epsilon_{\mathrm{sc}}\), \(V^{-1}\le\epsilon_{\mathrm{sc}}\) and
\(\bar x\in[-C_{\bar x}^{\mathrm{sat}}+1,C_{\bar x}^{\mathrm{sat}}-1]\). It lies
in the construction's matching domain and is the outer collar of the
comparison below.

\paragraph{The artificial residual is negative for every action.}
On the outer matching collar, for a uniform \(c_0>0\),
\begin{equation}
g_s(x)+\delta E[J_\delta^{\mathrm{aux}}(s^+)]-J_\delta^{\mathrm{aux}}(s)
\le-c_0\lvert x-\varphi(s)\rvert^2\qquad(x\in X).
\label{S1-eq-4-1}
\end{equation}
For \(x\) in the artificial action interval, this follows from uniform strict
concavity and the Bellman identity; outside it, use the static global gap.
On the interval, the \(x\)-derivative of the artificial objective differs
from \(\partial_xg^{\mathrm{full}}(x;(a,b))\) by \(O(A^{-1}+V^{-1})\), because
the predictive variance is \(1+A^{-1}+V^{-1}D^2\) with \(D\) bounded, the
derivatives of \(G_{\mathrm{var}}\) are bounded
(Lemma~\ref{S1-lem:smoothing}\ref{S1-lem:smoothing:a}), the
action derivative of the successor chart is
\(O((A^{-1}+V^{-1})(1+\lvert Z\rvert))\) with zero \(u\)-block, and
\(J_{\delta,\epsilon}^{\mathrm{aux}}\) has bounded derivatives. Since the
artificial derivative vanishes at the interior maximizer \(\varphi(s)\), the
artificial curvature is at most \(-C_{\mathrm{curv}}^{\mathrm{fp}}/2\) on the
interval, which contains \(x^\star(a,b)\), and
\(\partial_xg^{\mathrm{full}}(x^\star(a,b);(a,b))=0\), we get
\(\lvert\varphi-x^\star(a,b)\rvert\le C(A^{-1}+V^{-1})\). At a matching
current state,
\(\lvert\mathcal T^{[\epsilon]}(\iota_\epsilon(s),x,Z)-\mathcal T^{[\epsilon]}(\iota_\epsilon(s),x',Z)\rvert
\le C\epsilon_{\mathrm{sc}}(1+\lvert Z\rvert)\lvert x-x'\rvert\), so the global
Lipschitz bound on \(J_{\delta,\epsilon}^{\mathrm{aux}}\) gives
\(\lvert E[J_\delta^{\mathrm{aux}}(s_x^+)]-E[J_\delta^{\mathrm{aux}}(s_{x'}^+)]\rvert
\le C\epsilon_{\mathrm{sc}}\lvert x-x'\rvert\). For every \(x\in X\),
endpoints included,
\(g_s(x)=G_{\mathrm{var}}(x,a+bx,A^{-1}+V^{-1}D^2)\), so
Lemma~\ref{S1-lem:smoothing}\ref{S1-lem:smoothing:c} gives
\(\lvert g_s(x)-g^{\mathrm{full}}(x;(a,b))\rvert\le C(A^{-1}+V^{-1})
\le C\epsilon_{\mathrm{sc}}\). For small enough
\(\epsilon_{\mathrm{sc}}\), these two bounds cannot erase half the static
global gap, and
boundedness of \(X\) converts the remaining fixed gap into
\eqref{S1-eq-4-1}. This argument concerns the artificial problem only.

\paragraph{Lyapunov function.}\phantomsection\label{S1-section-lyapunov}
Write \(\epsilon=\epsilon_{\mathrm{sc}}\),
\(J_{\mathrm{in}}=[-C_{\bar x}^{\mathrm{sat}}+1,C_{\bar x}^{\mathrm{sat}}-1]\), and
let \(F_\epsilon\) be the set of feasible states with \(A^{-1}\le\epsilon\),
\(V^{-1}\le\epsilon\) and \(\bar x\in J_{\mathrm{in}}\). It is invariant under every
action in \(X\), because \(A^{-1}\) and \(V^{-1}\) decrease and \(\bar x^+\) is a
convex combination of \(\bar x\) and the action. Put
\(d(a,b)=(\operatorname{dist}((a,b),K^{\mathrm{mean}})-r_{\mathrm{mean}})_+\),
which is \(1\)-Lipschitz, zero on \(K^{\mathrm{mean}}_{r_{\mathrm{mean}}}\), at
least \(r_{\mathrm{mean}}\) at states of \(F_\epsilon\) off the outer matching
collar and bounded on that collar, and
\[
 \Phi(s)=\frac{A^{-1}}{\epsilon}+\lvert\varphi(s)-\bar x\rvert+d(a,b),
 \qquad e(s,x)=x-\varphi(s).
\]
On \(F_\epsilon\) and for \(x\in X\), the exact recursion bounds the increment of
the scaled coordinates by \(C\epsilon(1+\lvert Z\rvert)\psi\), with
\(\psi=A^{-1}/\epsilon+\lvert x-\bar x\rvert\). Indeed, the mean gain is at most
\(C(A^{-1}+V^{-1}\lvert D\rvert)\), the prediction error is
\(\sigma_IZ\) with \(\sigma_I\) bounded, and the \(u_\epsilon\), \(\bar x\) and log
increments are at most \(\epsilon u_\epsilon\), \(\epsilon\lvert D\rvert\) and
\(C\epsilon\lvert D\rvert\). Since \(\varphi=\varphi_\epsilon\circ z_\epsilon\) and
\(\varphi_\epsilon\) is Lipschitz in \(z_\epsilon\), uniformly over the small-scale
family (Lemma~\ref{B-lem:virtual}\ref{B-virtual:regularity}),
\[
 \Phi(\mathcal T(s,x,Z))\le\Phi(s)+\varkappa(1+\lvert Z\rvert)\psi,
 \qquad \psi\le\min\bigl\{\Psi,\ \Phi(s)+\lvert e(s,x)\rvert\bigr\},
\]
with \(\varkappa=C\epsilon\), \(C\) and \(\Psi\) fixed before \(\epsilon\). The
bound is absolute, since \(\psi\) involves neither the posterior mean nor \(d\).
Let \(\varpi=2\lambda\) and
\(\mathcal V_\epsilon(s)=\Phi(s)^4e^{\varpi\Phi(s)}\). With
\(G_0=1+\lvert Z\rvert\) and \(\varsigma=\varpi\varkappa\Psi\le1\), the convexity
inequality \((p+q)^4\le(1+\eta)^3p^4+(1+\eta^{-1})^3q^4\),
\(\psi^4\le8(\Phi(s)^4+e(s,x)^4)\), \(Ee^{\varsigma G_0}\le1+\varsigma E[G_0e^{G_0}]\) and
\(E[G_0^4e^{\varsigma G_0}]\le E[G_0^4e^{G_0}]\) give, for \(\eta>0\) with
\(\delta(1+\eta)^4\le\theta\coloneqq(1+\delta)/2\) and small \(\epsilon\),
\begin{equation}
 \delta E_Z[\mathcal V_\epsilon(\mathcal T(s,x,Z))]
 \le\theta\mathcal V_\epsilon(s)+K_e\varkappa^4e^{\varpi\Phi(s)}e(s,x)^4
 \qquad(s\in F_\epsilon,\ x\in X),
 \label{S1-eq-4-3}
\end{equation}
where \(K_e\) is fixed before \(\epsilon\). On the outer matching collar
\(\Phi\) is at most a constant \(\Phi_{\mathrm{out}}\), because the collar's means
lie in the bounded set \(K^{\mathrm{mean}}_{2r_{\mathrm{mean}}}\); off it,
\(\Phi\ge d\ge r_{\mathrm{mean}}\).

\paragraph{Comparison with the true value.}
On \(F_\epsilon\) the posterior covariance is bounded and
\(\max\{\lvert a\rvert,\lvert b\rvert\}\le R_0+d\) for a constant \(R_0\), so
\eqref{S1-eq:weight-bounds} gives
\(\mathsf W_\lambda(s)\le Ce^{\varpi d}\le Cr_{\mathrm{mean}}^{-4}\mathcal V_\epsilon(s)\)
off the outer collar; there \(\lvert g_s\rvert\), \(\lvert J_\delta^{\mathrm{aux}}\rvert\)
and the error term of \eqref{S1-eq-4-3} are at most \(C\mathcal V_\epsilon(s)\),
\(C\) and
\(K_e\varkappa^4r_{\mathrm{mean}}^{-4}\operatorname{diam}(X)^4\mathcal V_\epsilon(s)\).
Put \(h_\pm=J_\delta^{\mathrm{aux}}\pm L\mathcal V_\epsilon\) with \(L\) fixed
before \(\epsilon\). On the outer collar, \eqref{S1-eq-4-1}, \eqref{S1-eq-4-3},
\(\Phi\le\Phi_{\mathrm{out}}\) and \(e(s,x)^4\le\operatorname{diam}(X)^2e(s,x)^2\) give, for
small \(\epsilon\),
\begin{equation}
 g_s(x)+\delta E_Z[h_+(\mathcal T(s,x,Z))]\le h_+(s)-\tfrac{c_0}{2}\lvert x-\varphi(s)\rvert^2
 \qquad(x\in X),
 \label{S1-eq-4-8}
\end{equation}
and at \(x=\varphi(s)\) the hybrid identity gives
\(h_-(s)\le g_s(\varphi(s))+\delta E_Z[h_-(\mathcal T(s,\varphi(s),Z))]\). Off
the outer collar both inequalities hold for every action, because
\(\mathcal V_\epsilon\) dominates the reward, the auxiliary value and the error
term there. On \(F_\epsilon\), \(\Phi\le C+\lvert a\rvert+\lvert b\rvert\), since
\(d\) is \(1\)-Lipschitz, and \(q^4\le24e^q\) for \(q\ge0\); with the lower bound in
\eqref{S1-eq:weight-bounds} this gives
\(\mathcal V_\epsilon\le24e^{(\varpi+1)\Phi}\le C\mathsf W_{\varpi+1}\), so
Lemma~\ref{S1-lem:comparison} with \(\nu=\varpi+1\) and \(F=F_\epsilon\) gives
\(h_-\le J_\delta\le h_+\) on \(F_\epsilon\). On the inner collar, \(d=0\),
\(\Phi\) is bounded and \(\Phi\le(A^{-1}+\lvert\varphi-\bar x\rvert)/\epsilon\), so
\(\mathcal V_\epsilon\le C(A^{-1}+\lvert\varphi-\bar x\rvert)^4\) for the fixed
scale. If \(x\) maximizes \(Q_\delta(s,\cdot)\), then
\(h_-(s)\le J_\delta(s)=Q_\delta(s,x)\le h_+(s)-\tfrac{c_0}{2}\lvert x-\varphi(s)\rvert^2\)
by \eqref{S1-eq-4-8}. Consequently, for every exact maximizing action,
\[
\lvert J_\delta-J_\delta^{\mathrm{aux}}\rvert\le C(A^{-1}+\lvert\varphi-\bar x\rvert)^4,\qquad
\lvert x-\varphi\rvert\le C(A^{-1}+\lvert\varphi-\bar x\rvert)^2.
\]
All successor evaluations in this proof use the \textit{global artificial
value}, at every Gaussian innovation and at successors inside and outside the
collar.

\begin{proof}[Proof of Lemma~\ref{B-lem:virtual}]
\phantomsection\label{S1-lemma-discharge}
We identify the three outputs of Lemma~\ref{B-lem:virtual}. Choose
\(U_{\mathrm{in}}=\{(a,b):\operatorname{dist}((a,b),K^{\mathrm{mean}})<r_{\mathrm{mean}}\}\)
and
\(U_{\mathrm{out}}=\{(a,b):\operatorname{dist}((a,b),K^{\mathrm{mean}})<2r_{\mathrm{mean}}\}\).
Their closures lie in the neighborhood chosen in
Section~\ref{S1-section-3}, on which the maximizer is unique and interior, with strictly negative curvature, and in \(\Theta^I\) when
\(K^{\mathrm{mean}}\subset\Theta^I\). Take
\(J_{\mathrm{in}}=[-C_{\bar x}^{\mathrm{sat}}+1,C_{\bar x}^{\mathrm{sat}}-1]\),
\(J_{\mathrm{out}}=[-C_{\bar x}^{\mathrm{sat}},C_{\bar x}^{\mathrm{sat}}]\) and
\(\mathcal X(a,b)=x_{\mathrm{ctr}}(a,b)+[-h_x,h_x]\). The action-center bound
places \(x^\star(a,b)\) strictly inside this interval, and the lemma's matching
and feasible domains lie in the matching and initial domains established
above. Lemma~\ref{S1-lem:fixed-point} and the explicit global construction give the
bounded value and the \(C^{2,1}\) action map, with scaled norms uniform in
\(\epsilon_{\mathrm{sc}}\). The smooth center has bounded derivatives through
order four, so adding it to the selector preserves these bounds, and linear
pullback by \(S_\epsilon\) gives the stated fixed-scale derivative conversions
and the global extension in part~\ref{B-virtual:regularity}. On matching current states and for every local action, the constructed
reward and full transition agree with the true-form expressions for every
Gaussian realization, so the fixed-point Bellman identity gives the hybrid
identity. The objective-curvature bound
\(Q^{\mathrm{aux}}_{e_{\mathrm{ctr}}e_{\mathrm{ctr}}}\le-C_{\mathrm{curv}}^{\mathrm{fp}}/2\)
of Section~\ref{S1-fixed-point-proof} gives part~\ref{B-virtual:hybrid} with
\(c_Q=C_{\mathrm{curv}}^{\mathrm{fp}}/2\), since differentiation in the
physical action equals differentiation in its translated coordinate. Only the
current state is restricted, and the value at all successors is the global
auxiliary value. Finally, the comparison above gives
part~\ref{B-virtual:tracking} on \(\mathcal I_\epsilon\).
\end{proof}

\subsection{Covariance derivative calculations}\label{S1-section-5}\label{S1-covariance-proof}
\begin{proof}[Proof of Lemma~\ref{B-lem:covariance-bounds}]
Use the independent unscaled coordinates \eqref{B-eq:coordinates} and the
transition \eqref{B-eq:transition}, with derivatives taken at fixed physical
action \(x\), on the domain of the lemma; \(Q^{[\ell]}\) is the hybrid
objective of Lemma~\ref{B-lem:virtual}\ref{B-virtual:hybrid}, and that lemma
supplies global value regularity and hybrid action curvature. We first bound
\((Q^{[\ell]})_{x\bar x}\) and \((Q^{[\ell]})_{x\ell_V}\), with constants
independent of a lower \(\ell_V\)-cutoff, and then the policy derivatives by
implicit differentiation. Only derivatives of \(J^{\mathrm{aux},[\ell]}\) in
the mean, \(\bar x\) and \(\ell_V\) coordinates enter, so the constants are uniform in the
anisotropic construction, that is, independent of the scale conversion in
\(u\); only two derivatives of \(J^{\mathrm{aux},[\ell]}\) and a Lipschitz
modulus of the second are needed, and
Lemma~\ref{B-lem:virtual}\ref{B-virtual:regularity} supplies them.

\emph{Transition derivatives.} Put \(\bar x^+=(\bar x+ux)/(1+u)\) and
\(\boldsymbol n(\bar x^+)=(-\bar x^+,1)\). A direct computation gives the
cancellations
\[
{\boldsymbol k_I}_x=\frac{e^{\ell_V}(1+u)}{\sigma_I^3}\boldsymbol n(\bar x^+),
\quad
{\boldsymbol k_I}_{x\ell_V}
={\boldsymbol k_I}_x\left(1-\frac{3e^{\ell_V}D^2}{2\sigma_I^2}\right),
\quad
{\boldsymbol k_I}_{\bar x}=-{\boldsymbol k_I}_x-\frac{e^{\ell_V}D}{\sigma_I}\boldsymbol e_a,
\]
and in particular
\begin{equation}
\lvert\boldsymbol k_I\rvert\le C(\lvert u\rvert+e^{\ell_V}\lvert D\rvert),\quad
\lvert{\boldsymbol k_I}_x\rvert\le Ce^{\ell_V},\quad
\lvert{\boldsymbol k_I}_{\ell_V}\rvert\le Ce^{\ell_V}\lvert D\rvert,\quad
\lvert{\boldsymbol k_I}_{x\ell_V}\rvert\le Ce^{\ell_V}.
\label{S1-eq-5-4}
\end{equation}
In the blocks (mean, \(\bar x\), \(\ell_V\)), the \(u\)- or
\(u_\epsilon\)-blocks being zero,
\[
\begin{gathered}
\mathcal T^{[\ell]}_x=\left({\boldsymbol k_I}_xZ,\frac u{1+u},-\frac{2e^{\ell_V}D}{\sigma_I^2}\right),\quad
\mathcal T^{[\ell]}_{\ell_V}=\left({\boldsymbol k_I}_{\ell_V}Z,0,\frac{1+u}{\sigma_I^2}\right),\\
\mathcal T^{[\ell]}_{x\ell_V}=\left({\boldsymbol k_I}_{x\ell_V}Z,0,-\frac{2e^{\ell_V}(1+u)D}{\sigma_I^4}\right).
\end{gathered}
\]

\emph{The \((Q^{[\ell]})_{x\bar x}\) estimate.} Differentiating the hybrid
objective,
\[
(Q^{[\ell]})_{x\bar x}
=\partial_{x\bar x}g_{z_\ell}
+\delta E_Z\bigl[\mathrm D^2J^{\mathrm{aux},[\ell]}[\mathcal T^{[\ell]}_x,\mathcal T^{[\ell]}_{\bar x}]+\mathrm DJ^{\mathrm{aux},[\ell]}\,\mathcal T^{[\ell]}_{x\bar x}\bigr].
\]
The transition formulas give
\(\lvert\mathcal T^{[\ell]}_x\rvert\le C(\lvert u\rvert+e^{\ell_V})(1+\lvert Z\rvert)\),
\(\lvert\mathcal T^{[\ell]}_{\bar x}\rvert\le C(1+e^{\ell_V}\lvert Z\rvert)\) and
\(\lvert\mathcal T^{[\ell]}_{x\bar x}\rvert\le Ce^{\ell_V}(1+\lvert Z\rvert)\),
and \(\partial_{x\bar x}g_{z_\ell}=O(e^{\ell_V})\), because \(\bar x\) enters
\(g_{z_\ell}(x)=G_{\mathrm{var}}(x,a+bx,u+e^{\ell_V}D^2)\) only through the
variance argument. Here and below, the derivatives of \(G_{\mathrm{var}}\) are
bounded on this domain by
Lemma~\ref{S1-lem:smoothing}\ref{S1-lem:smoothing:a}, since local actions lie
in \(\operatorname{int}X\), the mean argument is bounded and the variance
argument \(u+e^{\ell_V}D^2\) lies in a compact subset of \((-1,\infty)\),
because \(D\) is bounded and \(u>-2\epsilon_{\mathrm{sc}}\ge-1/2\), since the
construction scale satisfies \(\epsilon_{\mathrm{sc}}\le\epsilon_*\le1/4\). Hence
\(\lvert(Q^{[\ell]})_{x\bar x}\rvert\le C(\lvert u\rvert+e^{\ell_V})\).

\emph{The \((Q^{[\ell]})_{x\ell_V}\) estimate.} Direct differentiation of
\(g_{z_\ell}(x)=G_{\mathrm{var}}(x,a+bx,u+e^{\ell_V}D^2)\)
gives
\[
\partial_{x\ell_V}g_{z_\ell}
=e^{\ell_V}D^2(G_{\mathrm{var},x\tau_{\mathrm{var}}}+bG_{\mathrm{var},m\tau_{\mathrm{var}}})
 +2e^{\ell_V}DG_{\mathrm{var},\tau_{\mathrm{var}}}
 +2{e^{\ell_V}}^2D^3G_{\mathrm{var},\tau_{\mathrm{var}}\tau_{\mathrm{var}}},
\]
so \(\lvert\partial_{x\ell_V}g_{z_\ell}\rvert\le Ce^{\ell_V}\lvert D\rvert\). For
the continuation,
\[
(Q^{[\ell]})_{x\ell_V}-\partial_{x\ell_V}g_{z_\ell}
=\delta E_Z\bigl[\mathrm D^2J^{\mathrm{aux},[\ell]}[\mathcal T^{[\ell]}_x,\mathcal T^{[\ell]}_{\ell_V}]+\mathrm DJ^{\mathrm{aux},[\ell]}\,\mathcal T^{[\ell]}_{x\ell_V}\bigr].
\]
Most terms are immediately \(O(\lvert u\rvert+e^{\ell_V}\lvert D\rvert)\). The
terms for which a crude bound gives only \(O(e^{\ell_V})\) are
\[
E_Z\left[
 Z\,\frac{1+u}{\sigma_I^2}{\boldsymbol k_I}_x\cdot J^{\mathrm{aux},[\ell]}_{\mathrm{mean},\ell_V}(\mathcal T^{[\ell]})
 +Z\,{\boldsymbol k_I}_{x\ell_V}\cdot J^{\mathrm{aux},[\ell]}_{\mathrm{mean}}(\mathcal T^{[\ell]})
\right].
\]
The successor mean is its deterministic mean plus \(\boldsymbol k_IZ\), its
other coordinates are deterministic given the current state and action, and
\((1+u)/\sigma_I^2\le1\). Hence, for
\(f(Z)=\frac{1+u}{\sigma_I^2}{\boldsymbol k_I}_x\cdot J^{\mathrm{aux},[\ell]}_{\mathrm{mean},\ell_V}(\mathcal T^{[\ell]}(z_\ell,x,Z))\),
since \(E_Z[Z]=0\) and
\(\lvert\mathcal T^{[\ell]}(z_\ell,x,Z)-\mathcal T^{[\ell]}(z_\ell,x,0)\rvert=\lvert\boldsymbol k_I\rvert\lvert Z\rvert\),
\[
\lvert E_Z[Zf(Z)]\rvert
=\lvert E_Z[Z(f(Z)-f(0))]\rvert
\le\lvert{\boldsymbol k_I}_x\rvert\operatorname{Lip}(J^{\mathrm{aux},[\ell]}_{\mathrm{mean},\ell_V})\lvert\boldsymbol k_I\rvert E_Z[Z^2],
\]
where \(\operatorname{Lip}(\cdot)\) is the Lipschitz constant of the second
derivative in the mean directions, uniform by
Lemma~\ref{B-lem:virtual}\ref{B-virtual:regularity}; the same bound holds
with \({\boldsymbol k_I}_{x\ell_V}\) and \(J^{\mathrm{aux},[\ell]}_{\mathrm{mean}}\).
By \eqref{S1-eq-5-4}, both terms are at most
\(Ce^{\ell_V}(\lvert u\rvert+e^{\ell_V}\lvert D\rvert)\). The term with
\(Z^2{\boldsymbol k_I}_x^\top J^{\mathrm{aux},[\ell]}_{\mathrm{mean},\mathrm{mean}}{\boldsymbol k_I}_{\ell_V}\)
is \(O({e^{\ell_V}}^2\lvert D\rvert)\). Terms with the deterministic
\(\bar x\)-component of \(\mathcal T^{[\ell]}_x\) are \(O(\lvert u\rvert)\);
those with its deterministic \(\ell_V\)-component or the deterministic
component of \(\mathcal T^{[\ell]}_{x\ell_V}\) are
\(O(e^{\ell_V}\lvert D\rvert)\). Consequently
\(\lvert(Q^{[\ell]})_{x\ell_V}\rvert\le C(\lvert u\rvert+e^{\ell_V}\lvert D\rvert)\).

Implicit differentiation at \(x=\varphi^{[\ell]}(z_\ell)\), using uniform
negative action curvature, now yields
\[
\lvert\varphi^{[\ell]}_{\bar x}\rvert\le C(\lvert u\rvert+e^{\ell_V}),\qquad
\lvert\varphi^{[\ell]}_{\ell_V}\rvert\le C(\lvert u\rvert+e^{\ell_V}\lvert D^\varphi\rvert).
\]
These are the asserted bounds.
\end{proof}

\section{Infinite patience}\label{supp:infinite}

This section proves Theorem~\ref{thm:infinite}. Throughout, \(X\subset\R\) is
a nonempty compact set, not necessarily an interval; \(U\) is Borel on
\(X\times\R\) with \(\lvert U(x,y)\rvert\le Ce^{c_U\lvert y\rvert}\) for finite constants \(C\)
and \(c_U\ge0\); \(G\) is continuous on \(X\times\R\); and, for every
\(\vartheta=(\alpha,\beta)\), \(x\mapsto E\lvert U(x,\alpha+\beta x+Z)\rvert\) is
continuous on \(X\), where \(Z\sim N(0,1)\).
Assumption~\ref{ass:fl-payoff} on \(X=[\underline x,\overline x]\)
implies these conditions, the first two by Appendix~\ref{app:value} and the
third by the dominated-convergence argument given there for \(G\), applied to
\(\lvert U\rvert\) with \(m=\alpha+\beta x\). Every \(U\) that is continuous
on \(X\times\R\) with the growth bound satisfies them as well, by dominated
convergence, so a finite set of actions is covered. Fix the true
parameter \(\vartheta=(\alpha,\beta)\) and a policy \(\pi\). The arguments use only that
the action \(x_{t+1}\) is measurable with respect to
\[
 \mathcal F_t=\sigma\left(x_1,y_1,\dots,x_t,y_t\right)\vee
 \sigma(\zeta_1,\dots,\zeta_{t+1}),
\]
the observed history through date \(t\) together with the randomization seeds
through date \(t+1\); every policy of Section~\ref{sec:policies}, with values
in \(X\), has this property. Write
\[
 g^{\mathrm{full}}_*(\vartheta)=\max_{x\in X}g^{\mathrm{full}}(x;\vartheta),
 \qquad
 \mathfrak r_t^\pi(\vartheta)
 =E_\vartheta^\pi\bigl[g^{\mathrm{full}}_*(\vartheta)
 -g^{\mathrm{full}}(x_{t+1};\vartheta)\bigr],
\]
so that \(R_T^\pi(\vartheta)=\sum_{t=0}^{T-1}\mathfrak r_t^\pi(\vartheta)\), and
write \(U_{t+1}=U(x_{t+1},y_{t+1})\). As in Section~\ref{sec:infinite}, the
normalized discounted payoff under the fixed truth is
\[
 U_{\delta,\vartheta}^\pi=(1-\delta)E_\vartheta^\pi
 \left[\sum_{t\ge0}\delta^tU_{t+1}\right],
 \qquad 0\le\delta<1,
\]
and \(U_{1,\vartheta}^\pi=\liminf_{\delta\uparrow1}U_{\delta,\vartheta}^\pi\) and
\(U_{\mathrm{av},\vartheta}^\pi=\liminf_{T\to\infty}T^{-1}E_\vartheta^\pi
\sum_{t=0}^{T-1}U_{t+1}\). The normalization by \(1-\delta\) preserves the
ordering of policies at every fixed \(\delta<1\) and keeps the payoff scale
finite as \(\delta\uparrow1\).

\begin{lemma}[Payoff primitives]\label{S2-lem:primitives}
Let \(X\subset\R\) be nonempty and compact, and let \(U\) be Borel on
\(X\times\R\) with \(\lvert U(x,y)\rvert\le Ce^{c_U\lvert y\rvert}\), \(c_U\ge0\), such
that \(G\) is continuous on \(X\times\R\) and, for every \(\vartheta\),
\(x\mapsto E\lvert U(x,\alpha+\beta x+Z)\rvert\) is continuous on \(X\). Then
\(g^{\mathrm{full}}(x;\vartheta)\) is jointly continuous in
\((x,\alpha,\beta)\in X\times\R^2\), so \(g^{\mathrm{full}}_*(\vartheta)\) is
attained. For every policy \(\pi\) and every \(t\ge0\),
\begin{equation}\label{S2-eq:conditional-payoff}
 E_\vartheta^\pi[U_{t+1}\mid\mathcal F_t]=g^{\mathrm{full}}(x_{t+1};\vartheta),
\end{equation}
and
\begin{equation}\label{S2-eq:envelope}
 \sup_{t\ge0}E_\vartheta^\pi\lvert U_{t+1}\rvert
 \le\max_{x\in X}E\lvert U(x,\alpha+\beta x+Z)\rvert<\infty,
 \qquad Z\sim N(0,1).
\end{equation}
\end{lemma}

\begin{proof}
Since \(g^{\mathrm{full}}(x;\vartheta)=G(x,\alpha+\beta x)\) and \(G\) is
continuous, \(g^{\mathrm{full}}\) is jointly continuous, and compactness of
\(X\) gives the maximum. Under \(P_\vartheta^\pi\),
\(\varepsilon_{t+1}\sim N(0,1)\) is independent of \(\mathcal F_t\), which is
generated by earlier noise and the seeds, while \(x_{t+1}\) is
\(\mathcal F_t\)-measurable. The map
\((x,z)\mapsto U(x,\alpha+\beta x+z)\) is Borel on \(X\times\R\), and, since
\(X\) is bounded, it is dominated by a multiple of \(e^{c_U\lvert z\rvert}\)
uniformly in \(x\in X\). Conditioning on \(\mathcal F_t\) therefore gives
\eqref{S2-eq:conditional-payoff} and
\(E_\vartheta^\pi\lvert U_{t+1}\rvert=E_\vartheta^\pi[h(x_{t+1})]\) with
\(h(x)=E\lvert U(x,\alpha+\beta x+Z)\rvert\). The function \(h\) is
continuous on the compact set \(X\), so it attains its maximum, which is
finite by the growth bound; this gives \eqref{S2-eq:envelope}.
\end{proof}

\begin{proposition}[Ces\`aro and Abel criteria]\label{S2-prop:patient-limit}
Let \(X\) and \(U\) be as in Lemma~\ref{S2-lem:primitives}, and fix
\(\vartheta\) and a policy \(\pi\). Then:
\begin{enumerate}[label=\textup{(\roman*)},ref=(\roman*)]
\item For every \(T\ge1\),\label{S2-prop:patient-limit:i}
\begin{equation}\label{S2-eq:cesaro-identity}
 \frac1TE_\vartheta^\pi\sum_{t=0}^{T-1}U_{t+1}
 =g^{\mathrm{full}}_*(\vartheta)-\frac{R_T^\pi(\vartheta)}{T},
 \qquad\text{so}\qquad
 U_{\mathrm{av},\vartheta}^\pi=g^{\mathrm{full}}_*(\vartheta)
 -\limsup_{T\to\infty}\frac{R_T^\pi(\vartheta)}{T}.
\end{equation}
\item For every \(0\le\delta<1\),\label{S2-prop:patient-limit:ii}
\begin{equation}\label{S2-eq:abel-identity}
 g^{\mathrm{full}}_*(\vartheta)-U_{\delta,\vartheta}^\pi
 =(1-\delta)\sum_{t\ge0}\delta^t\mathfrak r_t^\pi(\vartheta)
 =(1-\delta)^2\sum_{T\ge1}\delta^{T-1}R_T^\pi(\vartheta).
\end{equation}
\item The following are equivalent:\label{S2-prop:patient-limit:iii}
\begin{equation}\label{S2-eq:equivalence}
 R_T^\pi(\vartheta)=o(T),
 \qquad
 \lim_{\delta\uparrow1}U_{\delta,\vartheta}^\pi=g^{\mathrm{full}}_*(\vartheta),
 \qquad
 U_{1,\vartheta}^\pi=g^{\mathrm{full}}_*(\vartheta),
 \qquad
 U_{\mathrm{av},\vartheta}^\pi=g^{\mathrm{full}}_*(\vartheta).
\end{equation}
\item If at least one policy has sublinear cumulative expected regret at\label{S2-prop:patient-limit:iv}
\(\vartheta\), then the optimal values of \(U_{\mathrm{av},\vartheta}^\pi\) and
\(U_{1,\vartheta}^\pi\) over policies both equal
\(g^{\mathrm{full}}_*(\vartheta)\), and every policy with
\(R_T^\pi(\vartheta)=o(T)\) attains them.
\end{enumerate}
\end{proposition}

\begin{proof}
Suppress \(\pi\) and \(\vartheta\) in \(\mathfrak r_t\), \(R_T\),
\(U_\delta\), \(U_{\mathrm{av}}\) and \(g^{\mathrm{full}}_*\). Taking
expectations in \eqref{S2-eq:conditional-payoff} and summing gives
\(E\sum_{t=0}^{T-1}U_{t+1}=\sum_{t=0}^{T-1}Eg^{\mathrm{full}}(x_{t+1};\vartheta)
=Tg^{\mathrm{full}}_*-R_T\), which is the first identity in
\eqref{S2-eq:cesaro-identity}. Taking the lower limit and using
\(\liminf_T(c-a_T)=c-\limsup_Ta_T\) gives the second.

The envelope \eqref{S2-eq:envelope} permits interchange of expectation and
the discounted series, and \((1-\delta)\sum_{t\ge0}\delta^t=1\) gives the
first equality in \eqref{S2-eq:abel-identity}. Since
\(g^{\mathrm{full}}(\cdot;\vartheta)\) is continuous on the compact set \(X\),
\(0\le\mathfrak r_t\le\bar r<\infty\) and \(0\le R_T\le\bar rT\), so every
series converges absolutely, and summation by parts with
\(\mathfrak r_t=R_{t+1}-R_t\), \(R_0=0\), gives the second.

For \ref{S2-prop:patient-limit:iii}, let \(R_T=o(T)\), fix \(\epsilon>0\), and choose \(T_\epsilon\)
with \(R_T\le\epsilon T\) for \(T\ge T_\epsilon\). Splitting the last sum in
\eqref{S2-eq:abel-identity} at \(T_\epsilon\) and using
\(\sum_{T\ge1}T\delta^{T-1}=(1-\delta)^{-2}\),
\[
 0\le g^{\mathrm{full}}_*-U_\delta
 \le(1-\delta)^2\sum_{T=1}^{T_\epsilon-1}R_T+\epsilon\longrightarrow\epsilon
 \qquad(\delta\uparrow1).
\]
Conversely, let \(U_\delta\to g^{\mathrm{full}}_*\), and put
\(\delta_T=1-1/T\) for \(T\ge2\). Keeping only the terms \(t\le T-1\) of
\eqref{S2-eq:abel-identity} at \(\delta=\delta_T\),
\[
 g^{\mathrm{full}}_*-U_{\delta_T}
 \ge(1-\delta_T)\delta_T^{T-1}\sum_{t=0}^{T-1}\mathfrak r_t
 =\left(1-\frac1T\right)^{T-1}\frac{R_T}{T},
\]
and \((1-1/T)^{T-1}\to e^{-1}>0\), so \(R_T/T\to0\). Since
\(U_\delta\le g^{\mathrm{full}}_*\) for every \(\delta\), by
\eqref{S2-eq:abel-identity}, \(U_1=\liminf_{\delta\uparrow1}U_\delta\) equals
\(g^{\mathrm{full}}_*\) if and only if \(U_\delta\to g^{\mathrm{full}}_*\).
Finally, \(R_T\ge0\), so \(\limsup_TR_T/T=0\) if and only if \(R_T=o(T)\), and
\eqref{S2-eq:cesaro-identity} shows that this is
\(U_{\mathrm{av}}=g^{\mathrm{full}}_*\).

For \ref{S2-prop:patient-limit:iv}, the regret terms in \eqref{S2-eq:cesaro-identity} and
\eqref{S2-eq:abel-identity} are nonnegative, so \(g^{\mathrm{full}}_*\)
bounds the long-run average payoff and, for every \(\delta\), the normalized
discounted payoff of every policy, hence also \(U_{1,\vartheta}^\pi\). By
\ref{S2-prop:patient-limit:iii}, every policy with sublinear regret attains this bound under both
criteria.
\end{proof}

\begin{proposition}[Consistent estimation and asymptotic exploitation]
\label{S2-prop:exploitation}
Let \(X\) and \(U\) be as in Lemma~\ref{S2-lem:primitives}, and fix
\(\vartheta\) and a policy \(\pi\). Let \(\widehat\vartheta_t\) be random parameters with
values in \(\R^2\) such that \(\widehat\vartheta_t\to\vartheta\) in
\(P_\vartheta^\pi\)-probability, and suppose that
\begin{equation}\label{S2-eq:exploitation}
 \max_{x\in X}g^{\mathrm{full}}(x;\widehat\vartheta_t)
 -g^{\mathrm{full}}(x_{t+1};\widehat\vartheta_t)\longrightarrow0
 \quad\text{in }P_\vartheta^\pi\text{-probability}.
\end{equation}
Then \(\mathfrak r_t^\pi(\vartheta)\to0\) and \(R_T^\pi(\vartheta)=o(T)\).
\end{proposition}

\begin{proof}
By Lemma~\ref{S2-lem:primitives}, \(g^{\mathrm{full}}\) is jointly
continuous, so by compactness of \(X\) (the tube lemma)
\(\sup_{x\in X}\lvert g^{\mathrm{full}}(x;\theta)-g^{\mathrm{full}}(x;\vartheta)\rvert\to0\)
as \(\theta\to\vartheta\); the supremum, over a countable dense subset of
\(X\), is measurable at \(\theta=\widehat\vartheta_t\), and it tends to zero
in probability. The realized payoff gap obeys
\[
\begin{aligned}
 0&\le g^{\mathrm{full}}_*(\vartheta)-g^{\mathrm{full}}(x_{t+1};\vartheta)\\
 &\le2\sup_{x\in X}\bigl\lvert g^{\mathrm{full}}(x;\widehat\vartheta_t)
 -g^{\mathrm{full}}(x;\vartheta)\bigr\rvert
 +\max_{x\in X}g^{\mathrm{full}}(x;\widehat\vartheta_t)
 -g^{\mathrm{full}}(x_{t+1};\widehat\vartheta_t),
\end{aligned}
\]
and the right side converges to zero in probability. The left side is
bounded, because \(g^{\mathrm{full}}(\cdot;\vartheta)\) is continuous on the
compact set \(X\), so its convergence to zero in probability implies
convergence in expectation. Thus \(\mathfrak r_t^\pi(\vartheta)\to0\), and
Ces\`aro summation gives \(R_T^\pi(\vartheta)/T\to0\).
\end{proof}

\begin{proof}[Proof of Theorem~\ref{thm:infinite}]
The equivalence of \ref{thm:infinite:i}, \ref{thm:infinite:ii} and \ref{thm:infinite:iii} is part of
Proposition~\ref{S2-prop:patient-limit}\ref{S2-prop:patient-limit:iii}. A policy with
consistent estimates \(\widehat\vartheta_t\) and vanishing maximization error
\eqref{S2-eq:exploitation} has \(R_T^\pi(\vartheta)=o(T)\) by
Proposition~\ref{S2-prop:exploitation}.

For the last statement, suppose that, under \(P_\vartheta^\pi\), the actions
converge in probability to a random action \(x_\infty\) with
\(g^{\mathrm{full}}(x_\infty;\vartheta)=g^{\mathrm{full}}_*(\vartheta)\) almost
surely. This is the case \(\widehat\vartheta_t=\vartheta\) of
Proposition~\ref{S2-prop:exploitation}. By continuity of
\(g^{\mathrm{full}}(\cdot;\vartheta)\),
\(g^{\mathrm{full}}_*(\vartheta)-g^{\mathrm{full}}(x_{t+1};\vartheta)\to0\) in
probability. Hence \(R_T^\pi(\vartheta)=o(T)\), and the policy attains
\(U_{1,\vartheta}^\pi=U_{\mathrm{av},\vartheta}^\pi=g^{\mathrm{full}}_*(\vartheta)\).
By Proposition~\ref{S2-prop:patient-limit}\ref{S2-prop:patient-limit:iv}, these are the optimal
infinite-patience and long-run average payoffs.
\end{proof}

\section{Application proofs}\label{supp:applications}

This section verifies, for the three examples of Section~\ref{sec:equilibria},
the hypotheses of Theorems~\ref{thm:fl-local} and~\ref{thm:fl-global}. In each
example \(X=[\underline x,\overline x]\) is a compact interval with
\(\underline x<\overline x\). Three facts are checked.
\begin{enumerate}[label=\textup{(\alph*)}]
\item The payoff \(U\) satisfies Assumption~\ref{ass:fl-payoff}.
\item A selected triple \((a^e,b^e,x^e)\) with \(b^e\ne\beta\) is an interior
regular Berk--Nash equilibrium: it satisfies the no-surprise condition
\eqref{eq:bne-ns}; \(x^e\in\operatorname{int}X\) is the unique maximizer of
\(z\mapsto G(z,a^e+b^ez)\) over \(X\), which gives \eqref{eq:bne-ic}; the second
derivative of this function at \(x^e\),
\(G_{xx}+2b^eG_{xm}+(b^e)^2G_{mm}\), is negative; and
\(G_m(x^e,a^e+b^ex^e)\ne0\). Theorem~\ref{thm:fl-local} then applies to
\((a^e,b^e,x^e)\) for every \(\delta\in[0,1)\), and since \(b^e\ne\beta\),
persistence entails incorrect learning.
\item The complement of \(\Theta^R\) is Lebesgue-null. A nondegenerate Gaussian
prior has a density, so \(\mu(\Theta^R)=1\) for every such prior.
Theorem~\ref{thm:fl-global} then gives, for every nondegenerate Gaussian prior
\(\mu\), every \(\delta\in[0,1)\) and every optimal policy \(\pi_\delta\),
\(P^{\mu,\pi_\delta}(\mathcal W,\ \vartheta\in B)=\mu(B)\) for every Borel
\(B\subseteq\Theta^R\), hence \(P^{\mu,\pi_\delta}(\mathcal W)=1\), and
\(P^{\pi_\delta}_\vartheta(\mathcal W)=1\) for Lebesgue-almost every
\(\vartheta\in\Theta^R\), hence for Lebesgue-almost every \(\vartheta\in\R^2\),
with an exceptional null set of truths for the given prior and policy.
\end{enumerate}
Throughout, the curvature and outcome-relevance coefficients at a truth
\(\vartheta\in\Theta^I\) are
\(\partial_{xx}g^{\mathrm{full}}(x^\star(\vartheta);\vartheta)
=G_{xx}+2\beta G_{xm}+\beta^2G_{mm}\) and
\(G_m(x^\star(\vartheta),\alpha+\beta x^\star(\vartheta))\), evaluated at
\((x^\star(\vartheta),\alpha+\beta x^\star(\vartheta))\).

\subsection{Advertising}\label{S3-sec:advertising}

\begin{proposition}[Advertising]\label{S3-prop:advertising}
Let \(U(x,y)=y-x^2/2\) on a compact interval \(X\) with
\(\underline x<\overline x\).
\begin{enumerate}[label=\textup{(\roman*)},ref=(\roman*)]
\item \(U\) satisfies Assumption~\ref{ass:fl-payoff} with \(c_U=1\).\label{S3-prop:advertising:i}
\item For every truth \(\vartheta\) and every \(b^e\in\operatorname{int}X\)\label{S3-prop:advertising:ii}
with \(b^e\ne\beta\), put \(x^e=b^e\) and \(a^e=\alpha+(\beta-b^e)b^e\). Then
\((a^e,b^e,x^e)\) is an interior regular Berk--Nash equilibrium with an incorrect
slope, and Theorem~\ref{thm:fl-local} applies to it.
\item Every truth with \(\beta\in\operatorname{int}X\) is in \(\Theta^I\), with\label{S3-prop:advertising:iii}
\(x^\star(\vartheta)=\beta\), and every truth with \(\beta\notin X\) is in
\(\Theta^B\). Hence \(\R^2\setminus\Theta^R\subseteq
\{\beta\in\{\underline x,\overline x\}\}\), \(\mu(\Theta^R)=1\) for every
nondegenerate Gaussian prior \(\mu\), and Theorem~\ref{thm:fl-global}
applies.
\end{enumerate}
\end{proposition}

\begin{proof}
\ref{S3-prop:advertising:i} For every \(y\), \(x\mapsto U(x,y)\) is a
polynomial, with \(\partial_xU=-x\), \(\partial_x^2U=-1\) and higher
derivatives zero, and for every \(x\), \(y\mapsto U(x,y)\) is continuous,
hence Borel. On \(X\times\R\),
\(\lvert U\rvert\le(1+R_X^2/2)(1+\lvert y\rvert)\) and
\(\lvert\partial_xU\rvert\le R_X\), where
\(R_X=\max\{\lvert\underline x\rvert,\lvert\overline x\rvert\}\); so
Assumption~\ref{ass:fl-payoff} holds with \(c_U=1\), since \(1+t\le e^t\). Gaussian
integration gives \(G(x,m)=m-x^2/2\), so \(G_{xx}=-1\), \(G_{xm}=G_{mm}=0\) and
\(G_m=1\).

\parstartref{S3-prop:advertising:ii} The definition of \(a^e\) is \eqref{eq:bne-ns}. Under the belief
\((a^e,b^e)\), \(a^e+b^ez-z^2/2\) is strictly concave with maximizer
\(z=b^e=x^e\in\operatorname{int}X\), second derivative \(-1\), and
\(G_m=1\ne0\).

\parstartref{S3-prop:advertising:iii} The complete-information payoff is
\(g^{\mathrm{full}}(x;\vartheta)=\alpha+\beta x-x^2/2\), with
\(g^{\mathrm{full}}_x(x;\vartheta)=\beta-x\). For
\(\beta\in\operatorname{int}X\), the strictly concave
\(g^{\mathrm{full}}(\cdot;\vartheta)\) has the unique maximizer
\(x^\star(\vartheta)=\beta\in\operatorname{int}X\), with
\(G_{xx}+2\beta G_{xm}+\beta^2G_{mm}=-1<0\) and \(G_m=1\ne0\); so
\(\vartheta\in\Theta^I\). For \(\beta>\overline x\),
\(\beta-x\ge\beta-\overline x>0\) on \(X\), so the unique maximizer is
\(\overline x\), with \(g^{\mathrm{full}}_x(\overline x;\vartheta)>0\); for
\(\beta<\underline x\), symmetrically, it is \(\underline x\), with
\(g^{\mathrm{full}}_x(\underline x;\vartheta)<0\). These truths are in
\(\Theta^B\). The two omitted horizontal lines are Lebesgue-null, so
\(\mu(\Theta^R)=1\) under every nondegenerate Gaussian prior.
\end{proof}

For \(b^e=\beta\in\operatorname{int}X\), the triple of part~\ref{S3-prop:advertising:ii} is
\((\alpha,\beta,\beta)\), the complete-information optimum, which is interior
regular by part~\ref{S3-prop:advertising:iii}.

\subsection{Pricing}\label{S3-sec:pricing}

\begin{proposition}[Pricing]\label{S3-prop:pricing}
Let \(U(x,y)=(x-k)y\) with \(k\in\R\), and let \(X\) be a compact interval
with \(\underline x<\overline x\); in the economic application, \(k\ge0\) is
a marginal cost and \(X\Subset(0,\infty)\) is a set of positive prices.
\begin{enumerate}[label=\textup{(\roman*)},ref=(\roman*)]
\item \(U\) satisfies Assumption~\ref{ass:fl-payoff} with \(c_U=1\).\label{S3-prop:pricing:i}
\item Fix a truth with \(\beta<0\) and \(\alpha+\beta k>0\). For every\label{S3-prop:pricing:ii}
\(b^e<0\) with \(b^e\ne\beta\), put
\[
 x^e=\frac{kb^e-\alpha}{b^e+\beta},
 \qquad
 a^e=\alpha+(\beta-b^e)x^e.
\]
Then \(x^e-k=-(\alpha+\beta k)/(b^e+\beta)>0\). If
\(x^e\in\operatorname{int}X\), then \((a^e,b^e,x^e)\) is an interior regular
Berk--Nash equilibrium with an incorrect slope, and Theorem~\ref{thm:fl-local}
applies to it.
\item \(\R^2\setminus\Theta^R\) is contained in the union of the five lines\label{S3-prop:pricing:iii}
\[
\begin{aligned}
 &L_1=\{\alpha+\beta(2\underline x-k)=0\},\quad
 L_2=\{\alpha+\beta(2\overline x-k)=0\},\\
 &L_3=\{\alpha+\beta k=0\},\quad
 L_4=\{\alpha+\beta(\underline x+\overline x-k)=0\},\quad
 L_5=\{\beta=0\}.
\end{aligned}
\]
Hence \(\mu(\Theta^R)=1\) for every nondegenerate Gaussian prior \(\mu\), and
Theorem~\ref{thm:fl-global} applies.
\end{enumerate}
\end{proposition}

\begin{proof}
\ref{S3-prop:pricing:i} For every \(y\), \(x\mapsto U(x,y)\) is affine, with
\(\partial_xU=y\) and higher derivatives zero, and for every \(x\),
\(y\mapsto U(x,y)\) is continuous, hence Borel. Since
\(\lvert x-k\rvert\le\lvert\underline x\rvert+\lvert\overline x\rvert+\lvert k\rvert\) on \(X\),
\(\lvert\partial_x^jU\rvert\le(1+\lvert\underline x\rvert+\lvert\overline x\rvert+\lvert k\rvert)(1+\lvert y\rvert)\)
on \(X\times\R\) for \(j\le4\), the bound of Assumption~\ref{ass:fl-payoff}
with \(c_U=1\), since \(1+t\le e^t\). Gaussian integration gives \(G(x,m)=(x-k)m\), so \(G_{xx}=G_{mm}=0\),
\(G_{xm}=1\) and \(G_m=x-k\).

\parstartref{S3-prop:pricing:ii} Since \(b^e+\beta<0\), \(x^e-k=-(\alpha+\beta k)/(b^e+\beta)>0\). The
definition of \(a^e\) is \eqref{eq:bne-ns}. Under the belief \((a^e,b^e)\),
perceived profit \((z-k)(a^e+b^ez)\) is strictly concave, with second
derivative \(2b^e<0\), and its derivative at \(x^e\) is
\(\alpha+\beta x^e+b^e(x^e-k)=0\) by the definition of \(x^e\). So the
interior \(x^e\) is the unique maximizer over \(X\), and
\(G_m(x^e,a^e+b^ex^e)=x^e-k\ne0\).

\parstartref{S3-prop:pricing:iii} Write \(g(x)=g^{\mathrm{full}}(x;\vartheta)=(x-k)(\alpha+\beta x)\) and
\(m(x)=g'(x)=\alpha+\beta(2x-k)\), so \(g''=2\beta\).

\emph{Negative slopes.} For \(\beta<0\), \(m\) is strictly decreasing. If
\(m(\overline x)>0\), then \(m>0\) on \(X\) and \(\overline x\) is the unique
maximizer, with \(g_x(\overline x)>0\); if \(m(\underline x)<0\), then
\(\underline x\) is the unique maximizer, with \(g_x(\underline x)<0\). Both
truths are in \(\Theta^B\). If \(m(\underline x)>0>m(\overline x)\), strict
concavity gives the unique maximizer
\(x^\star(\vartheta)=(\beta k-\alpha)/(2\beta)\in\operatorname{int}X\), at which
\(G_{xx}+2\beta G_{xm}+\beta^2G_{mm}=2\beta<0\) and
\(G_m=x^\star(\vartheta)-k=-(\alpha+\beta k)/(2\beta)\); so
\(\vartheta\in\Theta^I\) when \(\alpha+\beta k\ne0\). Every other negative-slope
truth lies on \(L_1\), \(L_2\) or \(L_3\).

\emph{Positive slopes.} For \(\beta>0\), \(g\) is strictly convex. For
\(x=(1-\lambda)\underline x+\lambda\overline x\) with \(\lambda\in(0,1)\),
\(g(x)<(1-\lambda)g(\underline x)+\lambda g(\overline x)
\le\max\{g(\underline x),g(\overline x)\}\), so every maximizer is an endpoint,
and it is unique exactly when the two endpoint profits differ. Moreover,
\[
 g(\overline x)-g(\underline x)
 =\beta(\overline x^2-\underline x^2)+(\alpha-\beta k)(\overline x-\underline x)
 =(\overline x-\underline x)\bar m,
 \qquad
 \bar m=\alpha+\beta(\underline x+\overline x-k),
\]
so the endpoint profits tie exactly on \(L_4\). Since \(m\) is affine with slope
\(2\beta\) and \(\bar m=m((\underline x+\overline x)/2)\),
\(m(\overline x)=\bar m+\beta(\overline x-\underline x)\) and
\(m(\underline x)=\bar m-\beta(\overline x-\underline x)\). If \(\bar m>0\), the
unique maximizer is \(\overline x\) and \(m(\overline x)>0\); if \(\bar m<0\), it
is \(\underline x\) and \(m(\underline x)<0\). Hence every positive-slope truth
off \(L_4\) is in \(\Theta^B\).

\emph{Null sets.} Thus \(\R^2\setminus\Theta^R\subseteq L_1\cup\dots\cup L_5\),
five proper affine lines, each Lebesgue-null by Fubini's theorem and hence
null under every nondegenerate Gaussian prior, so \(\mu(\Theta^R)=1\).
\end{proof}

For \(b^e=\beta\), with \(\beta<0\) and \(\alpha+\beta k>0\), the formulas of
part~\ref{S3-prop:pricing:ii} give \(x^e=(\beta k-\alpha)/(2\beta)\) and \(a^e=\alpha\), so the
triple is \((\alpha,\beta,x^\star(\vartheta))\), the complete-information
optimum; if \(x^e\in\operatorname{int}X\), it is interior regular by
part~\ref{S3-prop:pricing:iii}, and Theorem~\ref{thm:fl-local} applies to it.

The local application in part~\ref{S3-prop:pricing:ii} concerns truths with \(\beta<0\); this
restricts the truth for that calculation, and the prior remains a
nondegenerate Gaussian. Part~\ref{S3-prop:pricing:iii} covers every sign of the slope. The
identity of Theorem~\ref{thm:fl-global} at \(B=\Theta^R\cap\{\beta<0\}\) is its
restriction to negative slopes.

\subsection{Inflation}\label{S3-sec:inflation}

\begin{proposition}[Inflation]\label{S3-prop:inflation}
Let \(U(x,y)=-[(y-y^*)^2+\omega x^2]\) with \(y^*\in\R\) and \(\omega>0\), and let
\(X=[\underline x,\overline x]\) be a compact interval with
\(\underline x<\overline x\).
\begin{enumerate}[label=\textup{(\roman*)},ref=(\roman*)]
\item \(U\) satisfies Assumption~\ref{ass:fl-payoff} with \(c_U=2\).\label{S3-prop:inflation:i}
\item Fix a truth with \(\alpha>y^*\) and \(\beta<0\). For every \(b^e<0\) with\label{S3-prop:inflation:ii}
\(b^e\ne\beta\), put \(K^e=\omega+b^e\beta>0\) and
\[
 x^e=\frac{b^e(y^*-\alpha)}{K^e}>0,
 \qquad
 a^e=\alpha+(\beta-b^e)x^e.
\]
If \(x^e\in\operatorname{int}X\), then \((a^e,b^e,x^e)\) is an interior regular
Berk--Nash equilibrium with an incorrect slope, and Theorem~\ref{thm:fl-local}
applies to it.
\item Let \(S_\vartheta(x)=\beta(\alpha+\beta x-y^*)+\omega x\) and\label{S3-prop:inflation:iii}
\(\mathcal N=\{\vartheta:S_\vartheta(\underline x)S_\vartheta(\overline x)=0\}\).
Then \(\R^2\setminus\Theta^R\subseteq\mathcal N\cup\{\vartheta:\alpha=y^*\}\),
and \(\R^2\setminus\Theta^R\subseteq\mathcal N\) if
\(0\notin\operatorname{int}X\), in particular if \(X\Subset(0,\infty)\). The
sets \(\mathcal N\) and \(\{\vartheta:\alpha=y^*\}\) are Lebesgue-null,
\(\mu(\Theta^R)=1\) for every nondegenerate Gaussian prior \(\mu\), and
Theorem~\ref{thm:fl-global} applies.
\end{enumerate}
\end{proposition}

\begin{proof}
\ref{S3-prop:inflation:i} For every \(y\), \(x\mapsto U(x,y)\) is a quadratic
polynomial, with \(\partial_xU=-2\omega x\), \(\partial_x^2U=-2\omega\) and
higher derivatives zero, and for every \(x\), \(y\mapsto U(x,y)\) is
continuous, hence Borel. Since \(X\) is bounded, \(U\) and its
\(x\)-derivatives through order four are bounded on \(X\times\R\) by
\(C(1+\lvert y\rvert)^2\) for a finite \(C\) depending on \(y^*\), \(\omega\)
and \(\max\{\lvert\underline x\rvert,\lvert\overline x\rvert\}\), hence by
\(Ce^{2\lvert y\rvert}\).

Gaussian integration gives \(G(x,m)=-[(m-y^*)^2+1+\omega x^2]\), so
\[
 G_{xx}=-2\omega,\qquad G_{xm}=0,\qquad G_{mm}=-2,\qquad G_m=-2(m-y^*).
\]

\parstartref{S3-prop:inflation:ii} Put \(\Delta=y^*-\alpha<0\). Since \(\omega>0\) and \(b^e,\beta<0\),
\(K^e>0\), and \(x^e=b^e\Delta/K^e>0\). The definition of \(a^e\) gives
\(a^e+b^ex^e=\alpha+\beta x^e\), which is \eqref{eq:bne-ns}. Under the belief
\((a^e,b^e)\), the perceived reward is
\(-[(a^e+b^ez-y^*)^2+1+\omega z^2]\). Its first-order condition at \(x^e\) is
\[
 b^e(\alpha+\beta x^e-y^*)+\omega x^e=-b^e\Delta+K^ex^e=0,
\]
and its second derivative is \(-2((b^e)^2+\omega)<0\), so \(x^e\) is the unique
unconstrained maximizer and, being interior, the unique maximizer over \(X\).
The second derivative is
\(G_{xx}+2b^eG_{xm}+(b^e)^2G_{mm}=-2(\omega+(b^e)^2)<0\). At the target,
\[
 \alpha+\beta x^e-y^*=-\frac{\omega\Delta}{K^e}>0,
 \qquad\text{so}\qquad
 G_m(x^e,a^e+b^ex^e)=-2(\alpha+\beta x^e-y^*)=\frac{2\omega\Delta}{K^e}\ne0.
\]

\parstartref{S3-prop:inflation:iii} Write \(g(x)=g^{\mathrm{full}}(x;\vartheta)=
-[(\alpha+\beta x-y^*)^2+1+\omega x^2]\). Then \(g'=-2S_\vartheta\) and
\(g''=-2(\beta^2+\omega)<0\), so \(g\) is a strictly concave quadratic, and
\(S_\vartheta\) is strictly increasing because \(\omega>0\). Partition \(\R^2\)
into \(\mathcal N\) and
\[
 \mathcal R_L=\{S_\vartheta(\underline x)>0\},\qquad
 \mathcal R_I=\{S_\vartheta(\underline x)<0<S_\vartheta(\overline x)\},\qquad
 \mathcal R_R=\{S_\vartheta(\overline x)<0\}.
\]
On \(\mathcal R_L\), \(S_\vartheta>0\) on \(X\), so \(g\) is strictly decreasing,
with unique maximizer \(\underline x\) and \(g'(\underline x)<0\); on
\(\mathcal R_R\), \(g\) is strictly increasing, with unique maximizer
\(\overline x\) and \(g'(\overline x)>0\). Hence
\(\mathcal R_L\cup\mathcal R_R\subseteq\Theta^B\). On \(\mathcal R_I\),
\(S_\vartheta\) has a unique zero
\(x^\star(\vartheta)=-\beta(\alpha-y^*)/(\beta^2+\omega)\in\operatorname{int}X\),
the unique maximizer of \(g\) by strict concavity, at which
\[
 G_{xx}+2\beta G_{xm}+\beta^2G_{mm}=-2(\beta^2+\omega)<0,
 \qquad
 G_m=-2(\alpha+\beta x^\star(\vartheta)-y^*).
\]
Here \(G_m\ne0\) unless \(\alpha=y^*\), because \(S_\vartheta(x^\star)=0\) together with
\(\alpha+\beta x^\star-y^*=0\) would force \(\omega x^\star=0\), so
\(x^\star=0\); since \(x^\star\in\operatorname{int}X\) on \(\mathcal R_I\),
this needs \(0\in\operatorname{int}X\), and then
\(\alpha=\alpha+\beta x^\star=y^*\). Hence
\(\mathcal R_I\setminus\{\alpha=y^*\}\subseteq\Theta^I\), and
\(\mathcal R_I\subseteq\Theta^I\) when \(0\notin\operatorname{int}X\); this
gives the two inclusions.
For an endpoint \(e\in\{\underline x,\overline x\}\) and fixed \(\beta\ne0\), the
equation \(S_\vartheta(e)=0\) is affine in \(\alpha\) with slope \(\beta\) and has
one solution. Every \(\beta\)-section of the closed set \(\mathcal N\) with
\(\beta\ne0\) therefore has at most two points, so \(\mathcal N\) is
Lebesgue-null by Fubini's theorem; the section at \(\beta=0\) may be all of
\(\R\) when an endpoint is zero, but the single value \(\beta=0\) is a
Lebesgue-null set of slopes. The line
\(\{\alpha=y^*\}\) is null, so \(\mu(\R^2\setminus\Theta^R)=0\) for every
nondegenerate Gaussian prior \(\mu\).
\end{proof}

For a positive instrument, \(X\Subset(0,\infty)\), the exceptional set is
\(\mathcal N\). Part~\ref{S3-prop:inflation:iii} does not need this. In
general, the only additional irregular parameters have \(\alpha=y^*\) and
complete-information optimum \(x^\star=0\), where inflation is on target at a
zero instrument and the mean outcome is not payoff relevant.

The conditions \(\alpha>y^*\) and \(\beta<0\) in part~\ref{S3-prop:inflation:ii}, that inflation would
exceed target at a zero instrument and that a higher instrument lowers
inflation, restrict the truth for the local calculation; part~\ref{S3-prop:inflation:iii} imposes neither of
these sign restrictions. Supplement~\ref{supp:technologies} gives the same
probability-one conclusion for pricing and inflation under two alternative
action technologies, a finite set of economically available actions and a fixed
cost of changing the action (Proposition~\ref{S4-prop:applications}).

\section{Alternative action technologies}\label{supp:technologies}

This section replaces the main text's continuous action interval without
adjustment cost by one of two economically different technologies: finitely
many economically available actions, such as price ticks, policy increments or
budget units; or a nonempty compact set of real actions with a fixed cost of
every change of the action. Under the conditions below, in both cases every
exact Bellman-optimal selection
eventually holds its action constant, and incorrect learning follows with
prior-predictive probability one. With a fixed cost, the same holds for
approximately optimal and myopic selections, and more generally for every rule
that revises its action only for a fixed gain. These technologies change the economic
model. They differ from randomization in the continuous model, and a finite
economic action set differs from a computational grid used to solve the
continuous model.

\subsection{Setup}\label{S4-sec:setup}

The observation equation, the parameter \(\vartheta=(\alpha,\beta)\) and the
nondegenerate Gaussian prior \(\mu\) are as in Section~\ref{sec:model}, except
that the action set \(X\subset\R\) is a nonempty compact set, the leading cases
being a finite set and a compact interval. We write the posterior state as
\(s=(m,\Sigma)\) with \(m=(a,b)^\top\), in the closed state space
\(\mathcal S\) of pairs \((m,\Sigma)\in\R^2\times\R^{2\times2}\) with
\(\Sigma=\Sigma^\top\succeq0\); it contains the singular states, including
complete information \(\Sigma=0\). For \(d(x)=(1,x)^\top\), the subjective
predictive transition has, on all of \(\mathcal S\), the common-normal
representation
\begin{equation}\label{S4-eq:transition}
 m^+=m+\frac{\Sigma d(x)}{\sqrt{1+d(x)^\top\Sigma d(x)}}Z,
 \qquad
 \Sigma^+=\Sigma-\frac{\Sigma d(x)d(x)^\top\Sigma}{1+d(x)^\top\Sigma d(x)},
 \qquad Z\sim N(0,1).
\end{equation}
Write \(g_s(x)=E_s[g^{\mathrm{full}}(x;\tilde\vartheta)]\), with
\(\tilde\vartheta\sim N(m,\Sigma)\), and fix \(\delta\in[0,1)\), with
\(\delta^0=1\). Instead of Assumption~\ref{ass:fl-payoff}, we impose the
primitive envelope
\begin{equation}\label{S4-eq:envelope}
 \begin{gathered}
 U\text{ is Borel on }X\times\R,
 \qquad
 (\vartheta,x)\longmapsto g^{\mathrm{full}}(x;\vartheta)\text{ is continuous},
 \\
 \sup_{x\in X}E_\varepsilon\lvert U(x,\alpha+\beta x+\varepsilon)\rvert
 \le C(1+\lVert\vartheta\rVert^p)
 \end{gathered}
\end{equation}
for finite constants \(C\) and \(p\). The expectations of
\(U(x_{t+1},y_{t+1})\), whose action \(x_{t+1}\) is random, presuppose the
first condition. For a finite \(X\), the first condition is equivalent to the
Borel measurability of each \(U(x,\cdot)\), which the other two conditions
already presuppose. The envelope makes every discounted return
absolutely integrable and \(g_s(x)\) continuous on \(\mathcal S\times X\),
singular states included.

Induction on \eqref{eq:design}, which uses the action set only through the
realized actions, gives the prior-augmented centered-sum identity
\begin{equation}\label{S4-eq:centered-sum}
 V_t=V_0+A_0(\bar x_0-\bar x_t)^2+\sum_{j=1}^t(x_j-\bar x_t)^2,
\end{equation}
and, since \(\bar x_{t+1}\) is a convex combination of \(\bar x_t\) and
\(x_{t+1}\),
\[
 \sup_t\lvert\bar x_t\rvert\le\max\left\{\lvert\bar x_0\rvert,\max_{x\in X}\lvert x\rvert\right\}<\infty.
\]
The selections below are deterministic Borel functions of the observed history
with values in \(X\), hence Borel predictable policies in the sense of
\PartIRemPredictable. So \PartILemPosterior{} and \PartILemDisintegration{}
apply to them, \PartILemEventualConstant{} holds pathwise, and, by
\PartILemPosterior(a) and~(c), under \(P^{\mu,\pi}\) the posterior means
converge almost surely, \(s_t\to(\vartheta,0)\) almost surely on
\(\{V_\infty=\infty\}\), and \(P^{\mu,\pi}(V_\infty<\infty,\ b_\infty=\beta)=0\).
In case~\textup{\ref{cond:G}} below, \(x_{t+1}=\pi(s_t)\) is moreover a decision rule in
the sense of \citet{kky2026parti} once \(X\) is embedded in
\([\min X,\max X]\), or in \([x_0-1,x_0+1]\) if \(X=\{x_0\}\), so
\PartILemComplete{} applies to it directly.

\paragraph{Finite actions.} When \(X\) is finite and there is no adjustment
cost, let \(J^X_\delta(s)\) be the unnormalized infinite-horizon Bellman value
and \(Q^X_\delta(s,x)=g_s(x)+\delta E_s[J^X_\delta(s^+)\mid x]\) the value of
choosing \(x\) first. By Lemma~\ref{S4-lem:grid-continuity} below, these are
continuous, so the conditional expectation is well defined and a Borel
selection of \(\operatorname*{argmax}_{x\in X}Q^X_\delta(s,x)\) exists. An
exact Borel Bellman-optimal selection \(\pi\) chooses \(x_{t+1}=\pi(s_t)\) in
this argmax. The complete-information tie set is
\[
 \mathcal T_X\coloneqq\left\{\vartheta:\
 \operatorname*{argmax}_{x\in X}g^{\mathrm{full}}(x;\vartheta)
 \text{ is not a singleton}\right\}.
\]

\paragraph{Fixed adjustment cost.} For a nonempty compact \(X\), let
\(x_t^-\in X\) be the inherited action at date \(t\), with a given
\(x_0^-\in X\). Choosing \(x_{t+1}\) incurs the cost
\begin{equation}\label{S4-eq:fixed-cost}
 K\1\{x_{t+1}\ne x_t^-\},
 \qquad K>0,
 \qquad x_{t+1}^-=x_{t+1}.
\end{equation}
The Bellman value is
\begin{equation}\label{S4-eq:fixed-bellman}
 J_{\delta,K}(s,x^-)=\max_{x\in X}
 \bigl\{g_s(x)-K\1\{x\ne x^-\}+\delta E_s[J_{\delta,K}(s^+,x)\mid x]\bigr\},
\end{equation}
with pre-cost return and its maximum
\[
 Q^{\mathrm{pre}}_{\delta,K}(s,x)=g_s(x)+\delta E_s[J_{\delta,K}(s^+,x)\mid x],
 \qquad
 M_{\delta,K}(s)=\max_{x\in X}Q^{\mathrm{pre}}_{\delta,K}(s,x).
\]
An exact
Borel Bellman-optimal selection \(\pi\) chooses \(x_{t+1}=\pi(s_t,x_t^-)\)
attaining the maximum in \eqref{S4-eq:fixed-bellman}. It is Markov in the
augmented state \((s,x^-)\), not in the posterior alone.

\subsection{Bellman lemmas}\label{S4-sec:lemmas}

\begin{lemma}[Commitment and clairvoyance]\label{S4-lem:evpi}
Suppose \(X\) is finite. Put
\(g^{\mathrm{full}}_*(\vartheta)=\max_{z\in X}g^{\mathrm{full}}(z;\vartheta)\) and
\[
 \operatorname{EVPI}(s)=E_s[g^{\mathrm{full}}_*(\tilde\vartheta)]
 -\max_{z\in X}g_s(z),\qquad\tilde\vartheta\sim N(m,\Sigma).
\]
Then, for every \(x,z\in X\),
\begin{equation}\label{S4-eq:evpi-gap}
 \bigl\lvert[Q^X_\delta(s,x)-Q^X_\delta(s,z)]-[g_s(x)-g_s(z)]\bigr\rvert
 \le\frac{\delta}{1-\delta}\operatorname{EVPI}(s).
\end{equation}
If \(s_n=(m_n,\Sigma_n)\to(\vartheta,0)\), then \(\operatorname{EVPI}(s_n)\to0\)
and \(g_{s_n}(x)\to g^{\mathrm{full}}(x;\vartheta)\) uniformly over \(x\in X\).
\end{lemma}

\begin{proof}
After a forced first action \(x\), the decision maker may commit forever to a
posterior expected-payoff maximizer chosen after the first observation, and
cannot do better than if the truth were revealed. Convexity of the finite
maximum and posterior martingality give
\[
 g_s(x)+\frac\delta{1-\delta}\max_zg_s(z)\le Q^X_\delta(s,x)
 \le g_s(x)+\frac\delta{1-\delta}E_s[g^{\mathrm{full}}_*(\tilde\vartheta)].
\]
The continuation parts of every two action values lie in this common interval,
whose length is the right side of \eqref{S4-eq:evpi-gap}. Continuity,
\eqref{S4-eq:envelope} and Gaussian uniform integrability give the limits.
\end{proof}

\begin{lemma}[Finite-action Bellman continuity]\label{S4-lem:grid-continuity}
Suppose \(X\) is finite. Under \eqref{S4-eq:envelope}, \(J^X_\delta\) and every
\(Q^X_\delta(\cdot,x)\) are continuous on \(\mathcal S\), including at singular
covariance matrices.
\end{lemma}

\begin{proof}
A finite \(X\) is compact, and the proof of
Lemma~\ref{S4-lem:continuous-pre-cost} below uses the cost only through
\(K\ge0\). With \(K=0\), its identity~\eqref{S4-eq:finite-horizon} reads
\(J_{n+1}(s,x^-)=M_n(s)\), which does not depend on \(x^-\), so its value
iteration is the finite-action value iteration \(J_0=0\),
\(Q_n(s,x)=g_s(x)+\delta E_s[J_n(s^+)\mid x]\), \(J_{n+1}=\max_{x\in X}Q_n\).
That proof therefore shows that the continuous \(J_n\) and \(Q_n(\cdot,x)\)
converge locally uniformly on \(\mathcal S\) to \(J^X_\delta\) and
\(Q^X_\delta(\cdot,x)\).
\end{proof}

\begin{lemma}[Continuous pre-cost return]\label{S4-lem:continuous-pre-cost}
Let \(X\subset\R\) be nonempty and compact. Under \eqref{S4-eq:envelope}, the
fixed-cost Bellman problem has a finite continuous value
\(J_{\delta,K}(s,x^-)\), the pre-cost return \(Q^{\mathrm{pre}}_{\delta,K}\) is
jointly continuous on \(\mathcal S\times X\), a Borel Bellman-optimal selection
exists, and
\begin{equation}\label{S4-eq:max-form}
 J_{\delta,K}(s,x^-)=\max\bigl\{Q^{\mathrm{pre}}_{\delta,K}(s,x^-),\
 M_{\delta,K}(s)-K\bigr\}.
\end{equation}
\end{lemma}

\begin{proof}
Use backward induction from zero terminal value. If the finite-horizon
pre-cost return \(Q_n\) is continuous, Berge's maximum theorem makes
\(M_n(s)=\max_xQ_n(s,x)\) continuous, and
\begin{equation}\label{S4-eq:finite-horizon}
 J_{n+1}(s,x^-)=\max\{Q_n(s,x^-),\ M_n(s)-K\}.
\end{equation}
The Bellman equation has \(\sup_{x\ne x^-}Q_n(s,x)-K\) in place of
\(M_n(s)-K\), and if this supremum is below \(M_n(s)\), then \(x^-\) is the
unique maximizer, so both sides equal \(M_n(s)\) because \(K\ge0\). The same
case distinction gives attainment, and \eqref{S4-eq:transition}, local
dominated convergence and \eqref{S4-eq:envelope} continue the induction. On
compact sets of states, \eqref{S4-eq:envelope} bounds the tail uniformly over
policies,
\[
 \sup_\pi E_s\sum_{j\ge n}\delta^j\bigl(\lvert U_{j+1}\rvert+K\bigr)
 \le\frac{\delta^n}{1-\delta}\bigl(C\,E_s[1+\lVert\tilde\vartheta\rVert^p]+K\bigr)
 \longrightarrow0
\]
locally uniformly in \(s\). So the finite-horizon values and pre-cost returns
converge locally uniformly to continuous limits \(J_{\delta,K}\) and
\(Q^{\mathrm{pre}}_{\delta,K}\), and \eqref{S4-eq:finite-horizon} passes to the
limit as \eqref{S4-eq:max-form}.

For the selector, \(\widehat x(s)=\min\operatorname*{argmax}_{x\in X}
Q^{\mathrm{pre}}_{\delta,K}(s,x)\) exists and is lower semicontinuous. If
\(s_n\to s\), compactness of \(X\) and joint continuity make every limit point
of \(\widehat x(s_n)\) a maximizer at \(s\), so
\(\{s:\widehat x(s)\le c\}\) is closed for every \(c\). Hence \(\widehat x\)
and \(M_{\delta,K}(s)=Q^{\mathrm{pre}}_{\delta,K}(s,\widehat x(s))\) are Borel,
and so is the selector \(\pi(s,x^-)=x^-\) if
\(Q^{\mathrm{pre}}_{\delta,K}(s,x^-)\ge M_{\delta,K}(s)-K\) and
\(\pi(s,x^-)=\widehat x(s)\) otherwise. By \eqref{S4-eq:max-form} it is exact.
In the first case \(x^-\) is costless and attains \(J_{\delta,K}(s,x^-)\); in
the second, \(Q^{\mathrm{pre}}_{\delta,K}(s,\widehat x(s))
>Q^{\mathrm{pre}}_{\delta,K}(s,x^-)+K\) forces \(\widehat x(s)\ne x^-\), and
\(\widehat x(s)\) attains \(M_{\delta,K}(s)-K=J_{\delta,K}(s,x^-)\).
\end{proof}

\subsection{Eventual constancy and incorrect learning}\label{S4-sec:theorem}

In the fixed-cost case the argument uses optimality only to rule out small
revisions. We therefore state that case for every deterministic policy that
revises its action only for a fixed gain in some continuous index. Such a
policy is given by Borel maps \(\pi_t\) with
\(x_{t+1}=\pi_t(y_1,\ldots,y_t)\in X\), \(t\ge0\), for the fixed initial
state; it may depend on the whole history.

\begin{theorem}[Granular revision]\label{S4-thm:granular}
Assume the model of Section~\ref{S4-sec:setup} and a nondegenerate Gaussian
prior \(\mu\). Suppose that one of the following holds:
\begin{enumerate}
\item[\textup{\condtarget{cond:G}{(G)}}] \(X\) is finite, there is no adjustment cost,
\(\delta\in[0,1)\), the envelope \eqref{S4-eq:envelope} holds,
\(\mu(\mathcal T_X)=0\), and \(\pi\) is a fixed exact Borel Bellman-optimal
selection;
\item[\textup{\condtarget{cond:F}{(F)}}] \(X\) is a nonempty compact set, \(\pi\) is a fixed
deterministic Borel policy, possibly history dependent, and there are a
jointly continuous \(F:\mathcal S\times X\to\R\) and \(K'>0\) such that, for
every sequence of outcomes and every \(t\ge1\),
\[
 x_{t+1}\ne x_t\quad\Longrightarrow\quad
 F(s_t,x_{t+1})\ge F(s_t,x_t)+K';
\]
no condition on \(U\) or \(\delta\) is needed.
\end{enumerate}
Put \(\mathcal R_*=\R^2\setminus\mathcal T_X\) in case~\textup{\ref{cond:G}} and
\(\mathcal R_*=\R^2\) in case~\textup{\ref{cond:F}}. Then, for every Borel
\(B\subseteq\mathcal R_*\),
\[
 P^{\mu,\pi}(\mathcal W,\ \vartheta\in B)=\mu(B),
 \qquad
 \mathcal W=\{V_\infty<\infty,\ b_t\to b_\infty\ne\beta\}.
\]
Since \(\mu(\mathcal R_*)=1\) in either case,
\begin{equation}\label{S4-eq:prior-predictive}
 P^{\mu,\pi}(\mathcal W)=1.
\end{equation}
In case~\textup{\ref{cond:F}}, the number of action changes is finite almost surely. In
case~\textup{\ref{cond:G}}, the action is also eventually constant almost surely.
Consequently, for Lebesgue-almost every truth \(\vartheta\),
\begin{equation}\label{S4-eq:fixed-truth}
 P^\pi_\vartheta(\mathcal W)=1.
\end{equation}
\end{theorem}

\begin{proof}
Each case uses \PartILemEventualConstant. An eventually constant action gives
\(\sum_t\kappa_tD_t^2<\infty\), that is, \(V_\infty<\infty\) by
\eqref{eq:design}.

\emph{Finite actions.} In case~\textup{\ref{cond:G}}, on
\(\{V_\infty=\infty,\ \vartheta\notin\mathcal T_X\}\), \PartILemComplete{}
gives \(s_t\to(\vartheta,0)\) almost surely. If \(X\) is a singleton, the
action is constant. Otherwise the unique complete-information maximizer
\(x^*(\vartheta)\) beats every other action of the finite set \(X\) by a
positive margin, and by Lemma~\ref{S4-lem:evpi} every Bellman gap
\(Q^X_\delta(s_t,x)-Q^X_\delta(s_t,z)\) converges to the corresponding
complete-information payoff gap. So \(x^*(\vartheta)\) is the unique Bellman
action after a finite random date, and \PartILemEventualConstant{} contradicts
\(V_\infty=\infty\). As \(\mu(\mathcal T_X)=0\),
\(P^{\mu,\pi}(V_\infty<\infty)=1\). Eventual constancy of the action now
follows from finite slope precision. If distinct \(x_i,x_j\in X\) have occurred
\(n_i(t)\) and \(n_j(t)\) times by date \(t\), one of them lies at distance at
least \(\lvert x_i-x_j\rvert/2\) from \(\bar x_t\), so by \eqref{S4-eq:centered-sum}
\[
 V_t-V_0\ge\tfrac14\min\{n_i(t),n_j(t)\}\lvert x_i-x_j\rvert^2.
\]
On \(\{V_\infty<\infty\}\) no two distinct actions occur infinitely often, and
since \(X\) is finite, exactly one does.

\emph{Revision for a fixed gain.} In case~\textup{\ref{cond:F}}, the posterior means
converge almost surely by \PartILemPosterior(c), and \(\Sigma_t\) converges
because the covariance update in \eqref{S4-eq:transition} is nonincreasing in
the positive-semidefinite order and bounded below by zero. So
\(s_t\to s_\infty\in\mathcal S\), and by joint continuity of \(F\) and
compactness of \(X\),
\begin{equation}\label{S4-eq:uniform}
 e_t=\sup_{x\in X}\bigl\lvert F(s_t,x)-F(s_\infty,x)\bigr\rvert\longrightarrow0.
\end{equation}
At a revision date \(t\ge1\), with \(x_{t+1}\ne x_t\), the hypothesis
of~\textup{\ref{cond:F}} and \eqref{S4-eq:uniform} yield
\[
 F(s_\infty,x_{t+1})\ge F(s_\infty,x_t)+K'-2e_t.
\]
For all sufficiently large \(t\), \(e_t\le K'/4\), so every revision after
that date raises the bounded function \(F(s_\infty,\cdot)\) by at least
\(K'/2\), and the action is unchanged between revisions. So revisions are
finitely many, and \(P^{\mu,\pi}(V_\infty<\infty)=1\).

\emph{Conclusion.} In either case \(P^{\mu,\pi}(V_\infty<\infty)=1\), while
\(P^{\mu,\pi}(V_\infty<\infty,\ b_\infty=\beta)=0\) by \PartILemPosterior(c) and
\PartIRemPredictable. Hence \(P^{\mu,\pi}(\mathcal W)=1\) and
\(P^{\mu,\pi}(\mathcal W,\ \vartheta\in B)=\mu(B)\) for every Borel
\(B\subseteq\mathcal R_*\). As \(\mathcal W\) is a joint Borel event,
\PartILemDisintegration{} with \PartIRemPredictable{} gives
\(P^\pi_\vartheta(\mathcal W)=1\) for \(\mu\)-almost every, equivalently
Lebesgue-almost every, \(\vartheta\).
\end{proof}

Case~\textup{\ref{cond:F}} covers the following fixed-cost policies, under
\eqref{S4-eq:envelope}, \(\delta\in[0,1)\) and the cost \eqref{S4-eq:fixed-cost}
with \(K>0\). Each chooses \(x_{t+1}=\pi(s_t,x_t^-)\) for a Borel \(\pi\), so it
is a deterministic Borel policy, and \(x_t^-=x_t\) for \(t\ge1\).
\begin{enumerate}[label=\textup{(\alph*)}]
\item An exact Borel Bellman-optimal selection of
\eqref{S4-eq:fixed-bellman}. At a revision, optimality against inaction gives
\(Q^{\mathrm{pre}}_{\delta,K}(s_t,x_{t+1})-K\ge Q^{\mathrm{pre}}_{\delta,K}(s_t,x_t)\),
so \textup{\ref{cond:F}} holds with \(F=Q^{\mathrm{pre}}_{\delta,K}\), which is jointly
continuous by Lemma~\ref{S4-lem:continuous-pre-cost}, and \(K'=K\).
\item An \(\varepsilon\)-optimal Borel selection with \(0\le\varepsilon<K\),
that is, one whose value of
\(Q^{\mathrm{pre}}_{\delta,K}(s,x)-K\1\{x\ne x^-\}\) is within \(\varepsilon\)
of the maximum at every \((s,x^-)\). Then \textup{\ref{cond:F}} holds with
\(F=Q^{\mathrm{pre}}_{\delta,K}\) and \(K'=K-\varepsilon\).
\item A myopic decision maker who pays the fixed cost, that is, a Borel
selection whose value of \(g_s(x)-K\1\{x\ne x^-\}\) is within
\(\varepsilon\in[0,K)\) of the maximum at every \((s,x^-)\). Then \textup{\ref{cond:F}}
holds with \(F(s,x)=g_s(x)\), which is jointly continuous on
\(\mathcal S\times X\) by \eqref{S4-eq:envelope}, and \(K'=K-\varepsilon\).
\end{enumerate}

\begin{remark}[The tie condition]\label{S4-rem:tie}
Without \(\mu(\mathcal T_X)=0\), the conclusion of case~\textup{\ref{cond:G}} need not
hold for every exact Borel selection. If every action has identically zero
payoff, every policy is Bellman-optimal, and a Borel selection can use the
effective sample size \(A_t\) to alternate between two distinct actions
forever, so that \(V_\infty=\infty\). Thus finite action cardinality alone is
insufficient. The fixed-cost case needs no such genericity, because every
revision must overcome the positive cost.
\end{remark}

\subsection{Pricing and inflation}\label{S4-sec:applications}

\begin{proposition}[Pricing and inflation]\label{S4-prop:applications}
Let \(X\subset\R\) be a nonempty compact set. Theorem~\ref{S4-thm:granular}
applies to the payoffs below, in case~\textup{\ref{cond:F}} to every policy listed
after the theorem; the payoffs are
\[
 U(x,y)=(x-k)y
 \qquad\text{and}\qquad
 U(x,y)=-[(y-y^*)^2+\omega x^2],
\]
with \(\omega\in\R\); in the economic application, \(\omega\ge0\) is a weight
on the instrument.
Their complete-information expected payoffs are
\begin{equation}\label{S4-eq:application-payoffs}
 g^{\mathrm{full}}(x;\vartheta)=(x-k)(\alpha+\beta x)
 \qquad\text{and}\qquad
 g^{\mathrm{full}}(x;\vartheta)=-[(\alpha+\beta x-y^*)^2+\omega x^2+1].
\end{equation}
For a finite action set, their complete-information tie sets are null under
every nondegenerate Gaussian prior. For a nonempty compact action set with a
fixed action-change cost \(K>0\), no tie or sign restriction is required.
Hence both pricing and inflation satisfy \eqref{S4-eq:prior-predictive} and
\eqref{S4-eq:fixed-truth} under either action technology.
\end{proposition}

\begin{proof}
Both payoffs \(U\) are continuous, hence Borel, on \(X\times\R\).
Gaussian integration gives \eqref{S4-eq:application-payoffs}, including
\(E[\varepsilon^2]=1\) in the inflation payoff; both are continuous in
\((\vartheta,x)\). Uniformly over the bounded \(X\), the expected absolute
payoffs are at most \(C_P(1+\lVert\vartheta\rVert)\) for pricing and
\((\alpha+\beta x-y^*)^2+1+\lvert\omega\rvert x^2\le C_I(1+\lVert\vartheta\rVert^2)\) for inflation,
so \eqref{S4-eq:envelope} holds. For distinct \(x_i,x_j\), a direct computation
shows that the pricing payoffs tie exactly on the proper affine line
\(\alpha+\beta(x_i+x_j-k)=0\), and the inflation payoffs exactly on the zero set
of the nonzero polynomial \(2\beta(\alpha-y^*)+(\beta^2+\omega)(x_i+x_j)\). A
finite union of these sets is Lebesgue-null, hence null under a nondegenerate
Gaussian prior; this verifies case~\textup{\ref{cond:G}}. The same checks with
compactness of \(X\) verify the envelope for case~\textup{\ref{cond:F}}, which then
applies to every policy listed after Theorem~\ref{S4-thm:granular}. Both arguments cover every sign
of \(\beta\). Compactness gives a pricing maximum even when expected profit is
convex.
\end{proof}

The conclusions are for a fixed finite action set and a fixed cost \(K>0\).
A smooth quadratic adjustment cost has zero marginal cost at inaction and can
generate infinitely many gradual adjustments; an absolute adjustment cost
creates an inaction band, but an optimizer may track its moving boundary.
Theorem~\ref{S4-thm:granular} covers the technologies \ref{cond:G} and \ref{cond:F} only. In
case~\ref{cond:G} it concerns exact Bellman-optimal selections, in case~\ref{cond:F} policies
that revise only for a fixed gain, and its Lebesgue-null exceptional set of
truths is the one for the given prior and policy.

\end{document}